\documentclass[a4paper,11pt]{article}
\pdfoutput=1

\usepackage{jheppub} 
\usepackage[T1]{fontenc}
\usepackage[utf8]{inputenc}
\usepackage[english]{babel}
\usepackage{amsmath,bbm}
\usepackage{amsfonts}
\usepackage{amssymb}
\allowdisplaybreaks[1]
\usepackage{graphicx}
\usepackage[normalem]{ulem}

\usepackage[usenames,dvipsnames]{xcolor}
\usepackage[compat=1.1.0]{tikz-feynman}
\usepackage{hyperref}

\hypersetup{unicode,psdextra}

\newcommand{\msbar}{$\overline{\mathrm{MS}}$~}

\newcommand*{\eweakgroup}{\mbox{$\mathit{SU}(2)_L \times U(1)_Y$}}
\newcommand*{\emgroup}{\mbox{$U(1)_{em}$}}

\newcommand*{\matheweakgroup}{\mathit{SU}(2)_L \times U(1)_Y}
\newcommand*{\mathemgroup}{U(1)_{em}}

\newcommand*{\unitmatrix}{\mathbbm{1}}

\newcommand*{\twomat}[1]{\underline{#1}}             
\newcommand*{\tvec}[1]{\boldsymbol{#1}}              
\newcommand*{\im}[1]{\text{Im} {#1}}                        
\newcommand*{\re}[1]{\text{Re} {#1}}                        

\newcommand*{\trans}{\mathrm{T}}                     

\newcommand{\KT}{\widetilde{\tvec{K}}}
\newcommand{\xiT}{\widetilde{\tvec{\xi}}}
\newcommand{\ET}{\widetilde{E}}
\newcommand{\gT}{\widetilde{g}}
\newcommand{\gK}{\big( \gT \KT \big)}

\DeclareMathOperator{\diag}{diag}		
\DeclareMathOperator{\tr}{Tr}		

\title{The Two-Higgs Doublet Model beyond tree-level:\\
A gauge-invariant formalism}

\author[b]{T. Guerandel}
\author[a]{M. Maniatis}
\author[b]{L. Sartore}
\author[b]{I. Schienbein}

\affiliation[a]{Centro de Ciencias Exactas,  Universidad del B\'io-B\'io,  Casilla  447,  Chill\'{a}n, Chile}
\affiliation[b]{Laboratoire de Physique Subatomique et de Cosmologie, Universit\'e Grenoble-Alpes, CNRS/IN2P3, 53 Avenue des Martyrs, 38026 Grenoble, France}

\emailAdd{thomas.guerandel@lpsc.in2p3.fr}
\emailAdd{maniatis8@gmail.com}
\emailAdd{lohan.sartore@lpsc.in2p3.fr}
\emailAdd{ingo.schienbein@lpsc.in2p3.fr}

\abstract{
Employing the gauge-invariant formalism in the two-Higgs-doublet model (THDM) offers profound insights into the model's fundamental structure. A specific set of gauge-invariant bilinear combinations, constructed from the Higgs doublets, establishes a one-to-one correspondence between the components of the doublet fields and real-valued bilinears. This formalism provides a compact and consistent framework to study various aspects of the THDM, including stability, electroweak symmetry breaking, basis transformations, and general symmetries of the Higgs potential.

Recently, the bilinear formalism has been extended beyond the Higgs potential to encompass the full THDM, including the gauge and Yukawa sectors, all in gauge-invariant terms. In this work, we advance the formalism further by incorporating quantum corrections. Specifically, we show how bilinears, combined with the $\hbar$-expansion, can be used to compute one-loop corrections. We provide concise, gauge-invariant expressions for these corrections, which are directly applicable to the THDM.
}

\begin{document} 
\maketitle
\flushbottom

\section{Introduction}
\label{sec:introduction}

Extending the Standard Model~\cite{ParticleDataGroup:2022pth}, which features a single Higgs doublet, to the two-Higgs-doublet model (THDM) is motivated by several theoretical and phenomenological considerations. One of the original motivations, proposed by T.~D.~Lee~\cite{Lee:1973iz}, remains valid today: the THDM offers an additional, potentially strong, source of CP violation.
Such a source is among the necessary conditions to dynamically generate the observed baryon–antibaryon asymmetry in the Universe~\cite{Sakharov:1967dj}.
Supersymmetric extensions of the Standard Model provide another motivation for the THDM, as they require at least two Higgs doublets; see, for instance, the review~\cite{Martin:1997ns}.
Moreover, the THDM features a richer scalar spectrum, with five physical Higgs bosons compared to a single one in the Standard Model. These include three neutral scalars and a pair of charged Higgs bosons, which are degenerate in mass.
For a review of the THDM we refer to~\cite{Branco:2011iw}. 

For any phenomenologically viable model, one of the primary tasks is to analyze stability and electroweak symmetry breaking.
In the Standard Model with one complex Higgs-boson doublet
\begin{equation}
\varphi(x) = \begin{pmatrix} \varphi^+(x)\\ \varphi^0(x) \end{pmatrix}\,,
\end{equation}
the most general renormalizable, tree-level potential we can write down reads 
\begin{equation}
V_{\text{SM}}^{(0)} = - m^2 \varphi^\dagger \varphi + \lambda (\varphi^\dagger \varphi)^2\,.
\end{equation}
The potential involves one dimensionful parameter, $m^2$, with units of mass squared, and one dimensionless coupling, $\lambda$.
Stability—that is, the potential being bounded from below—and the correct spontaneous symmetry breaking require $m^2 > 0$ and $\lambda > 0$.

Conventionally, to study a stable model with correct electroweak symmetry breaking that yields an electrically neutral vacuum, a gauge transformation is performed such that the Higgs doublet takes the form:
\begin{equation}
\varphi(x) = \frac{1}{\sqrt{2}} \begin{pmatrix} 0 \\ v + \phi(x) \end{pmatrix},
\end{equation}
where $v$ is the real vacuum expectation value and $\phi(x)$ is the real-valued excitation about the vacuum.
As is well known, the drawback, performing this specific gauge rotation, is that the gauge symmetry is no longer manifest. In contrast, keeping the gauge symmetry manifest turns out to have many advantages. 
For instance, the potential of the Standard Model can easily be written manifestly gauge invariant if we introduce one {\em bilinear}
$K = \varphi^\dagger \varphi$, resulting in a simple, real potential, 
$V_{\text{SM}}^{(0)} = - m^2 K + \lambda K^2$ of quadratic order.

In contrast to the Standard Model, the most general tree-level potential of the THDM, which includes two Higgs doublets $\varphi_1$ and $\varphi_2$, contains 14 real parameters, making the analysis of stability and electroweak symmetry breaking more complex. Moreover, in the Yukawa sector of the THDM, fermions can couple to both doublets in different ways. For instance, in the so-called type~I THDM, all fermions couple exclusively to one of the two doublets.

A gauge-invariant formulation of the THDM can be accomplished by introducing four bilinear combinations of the original doublet fields -- as has been shown in detail in~\cite{Nagel:2004sw, Maniatis:2006fs, Nishi:2006tg}. 
For the convenience of the reader, we recall the bilinear formalism in Sec.~\ref{sec:bilinear}.
As is shown, these bilinears arise naturally from the matrix constructed from the two Higgs-boson doublets in the form of a $2 \times 2$ matrix,
\begin{equation} \label{inphi}
\psi = \begin{pmatrix} \varphi_1^\trans\\ \varphi_2^\trans \end{pmatrix}.
\end{equation}
Note that the fields implicitly depend on a space-time argument.
We obtain the bilinear matrix of all  gauge-invariant scalar products as the outer product of $\psi$, explicitly,
\begin{equation} \label{intwomat}
\twomat{K} \equiv \psi \psi^\dagger = 
\begin{pmatrix} \varphi_1^\dagger \varphi_1 & \varphi_2^\dagger \varphi_1\\ \varphi_1^\dagger \varphi_2 & \varphi_2^\dagger \varphi_2 \end{pmatrix}.
\end{equation}
It has been shown that stability, electroweak symmetry breaking and the scalar mass matrices for any THDM tree-level potential can be 
studied concisely with the help of this bilinear matrix~\cite{Maniatis:2006fs}. Since the gauge symmetry is not obscured, symmetries of the model turn out to be very transparent and appear in subgroups of $O(3)$ in terms of bilinears~\cite{Ivanov:2007de, Maniatis:2011qu, Ferreira:2020ana, Bento:2020jei}.  Let us briefly demonstrate how electroweak-symmetry breaking can be understood in terms of bilinears:
We see that the $2 \times 2$ matrix $\psi$ in~\eqref{inphi} trivially has rank less than or equal to two, a fact which holds by construction also directly for the $2 \times 2$ matrix~$\twomat{K}$ in~\eqref{intwomat}. 
A vacuum with both, electrically charged and neutral components non-vanishing, corresponds therefore to 
a matrix $\twomat{K}$ with full rank~2. In this case the electroweak symmetry is completely broken.
A vacuum with electroweak symmetry breaking $\eweakgroup \to \emgroup$ leads to a matrix $\twomat{K}$ with rank-1
and the case with rank-0 corresponds to an unbroken electroweak symmetry.

Let us note that the bilinear formalism has been extended to the potential of the three Higgs-boson doublet model~\cite{Maniatis:2015kma} and the general N-Higgs-doublet model~\cite{Maniatis:2015gma} and has also been used to study the vacuum structure of the general THDM 
at large field values \cite{Maniatis:2020wfz} in the context of the principle of multiple point criticality \cite{Bennett:1993pj,Bennett:1996hx,Bennett:1996vy}.

Until recently, only the THDM potential has been considered in a manifestly gauge-invariant form whereas other sectors of the
Lagrangian could not be treated that way.
A long-standing open question was therefore how bilinears apply in a similar way to the gauge and Yukawa sectors.
In any viable model 
it is clear that all physical, i.e.~observable, quantities should be gauge independent. 
On the other hand, the Yukawa couplings, for instance for an electron coupled to the
Higgs-boson doublet $\varphi_1$, read
\begin{equation} \label{inYuk}
- {\cal L}_{\text{Yuk, e}} = y_e \left(\bar{L} \varphi_1 \right) e_R + h. c.\, ,
\end{equation}
with the left-handed electron doublet denoted by $L$, the right-handed singlet by $e_R$, and $y_e$ a generic coupling constant -- usually fixed by the mass of the fermion. Since these kind of Yukawa couplings are linear in the Higgs-boson doublet, it is not immediately clear how to express them in terms of gauge-invariant bilinears. 
However, recently it has been shown that the bilinear formalism can indeed be applied to the {\em complete} THDM \cite{Sartore:2022sxh},
due to the fact that the bilinear matrix~$\twomat{K}$, defined in~\eqref{intwomat} must have rank~1 
for the correct electroweak symmetry breaking. 
Therefore, this matrix can be decomposed as
\begin{equation}
\twomat{K}^\alpha_\beta = \kappa^\alpha \kappa_\beta^*,
\qquad
\alpha, \beta \in \{1,2\}\;,
\end{equation}
that is, in terms of one complex, two-dimensional, vector~$\kappa$.
Consequently, it becomes possible to express the Yukawa couplings in terms of this gauge-invariant vector~$\kappa$, which is linear in the doublet fields.
Furthermore, in the scalar sector it is possible to derive the expressions for the scalar squared mass matrix in a general way, without resorting to the actual potential. 
Therefore, the results obtained in \cite{Sartore:2022sxh} are valid in any THDM - to any perturbation order and even in effective THDM's.
In particular, it has been shown that the masses of the charged Higgs boson pair are determined by the vacuum in a particularly simple way.\\
Let us mention that the
beta functions in the THDM in terms of bilinears have been studied in~\cite{Ma:2009ax} and \cite{Maniatis:2020wfz}. The thermal one loop corrections of the effective potential have been studied gauge invariantly very recently in~\cite{Cao:2023kgq}.

In this work we want to apply the gauge-invariant formalism based on bilinears to the THDM beyond tree-level order.
Motivated by precision measurements as that of the electroweak $\rho$ parameter, defined as
\begin{equation}
\rho \equiv \frac{M_W^2}{M_Z^2 \cos^2(\theta_W)}
\end{equation}
which is measured as $\rho = 1.01019 \pm 0.00009$~\cite{ParticleDataGroup:2022pth},
we aim to confront these observations with equally precise theoretical predictions.
In the literature we can find accordingly a large number of calculations on quantum corrections in the THDM; 
see for instance~\cite{Xiao:2003ya,Krawczyk:2004na,Lopez-Val:2009xtx,Lopez-Val:2010asi,Diaz-Cruz:2014aga,Jenniches:2018zlb,Su:2019ibd}. Let us also mention in
this context the computation of the radiative corrections to the $\rho$ parameter, 
that is, more generally, the electroweak oblique parameters,
for any multi-Higgs doublet model~\cite{Grimus:2007if}.

The main purpose of this work is to derive one-loop corrections to the scalar vacuum in the THDM, as well as to the mass spectrum. In particular, we present a formulation that avoid any gauge-dependent quantities.  The vacuum structure is studied perturbatively using the so-called $\hbar$-expansion~\cite{Brodsky:2010zk}, allowing for an iterative solution of the stationary point equations starting from the tree-level solution. With this method, loop corrections to the scalar potential can be evaluated at the tree-level vacuum, providing simple analytic expressions for 
both the stationary point equations and the quantum corrections
to the scalar spectrum. Several typically problematic features inherent to the effective potential are avoided, such as explicit gauge dependence and infrared divergences in the Goldstone sector. Combining the $\hbar$-expansion method with the bilinear formalism, we are able to compute explicitly the one-loop corrections to the scalar spectrum. This is achieved in a concise manner, avoiding mixing between the Goldstone and physical sectors, while ensuring that the Goldstone modes remain massless.

Let us briefly outline the paper:
To make this work self-contained, we briefly recall the bilinear formalism and its relevance to electroweak symmetry breaking in Sec.~\ref{sec:bilinear}. We recall in Sec.~\ref{sec:gaugeInvariantFormalism} the computation of the scalar squared mass matrix to all orders from~\cite{Sartore:2022sxh}. 
In Sec.~\ref{sec:hbarExpansion}, we discuss the $\hbar$~expansion of the potential and the vacuum in the bilinear formalism. In Sec.~\ref{sec:oneLoopPotential}, we present the one-loop effective potential formulation. In this section we express the mass matrices for the gauge, fermionic, and the scalar contributions. In Sec.~\ref{sec:oneLoopMasses} we derive the corrections at one-loop order to the scalar squared mass matrices in detail. 
Readers primarily interested in the result of the $\hbar$ expansion without the technical derivations may refer directly to Sec.~\ref{sec:hbar_nuts}, where all results are summarized. Finally, in Sec.~\ref{sec:application}, we apply the formalism to a specific model: a CP-conserving THDM.
\section{Review of the bilinear formalism}
\label{sec:bilinear}

For the reader's convenience, we briefly review the bilinear formalism in the THDM~\cite{Nagel:2004sw, Nishi:2006tg, Maniatis:2006fs}. We adopt the convention of positive hypercharge $y=+1/2$, where the upper components of the doublets are electrically charged.
\begin{equation} \label{eq:doublets}
\begin{split}
&\varphi_1(x) = \begin{pmatrix} \varphi_1^{+}(x)\\ \varphi_1^{0}(x) \end{pmatrix} =
\frac{1}{\sqrt{2}} 
\begin{pmatrix} \pi_1^1(x)+i \sigma_1^1(x)\\ \pi_1^2(x)+i \sigma_1^2(x) \end{pmatrix}
,\\
&\varphi_2(x) = \begin{pmatrix} \varphi_2^{+}(x)\\ \varphi_2^{0}(x) \end{pmatrix} =
\frac{1}{\sqrt{2}} 
\begin{pmatrix} \pi_2^1(x)+i \sigma_2^1(x)\\ \pi_2^2(x)+i \sigma_2^2(x) \end{pmatrix}.
\end{split}
\end{equation}
Next, the complex fields are decomposed into their  real and imaginary components.
In what follows,
we omit the space-time argument of the fields. 
We form a new vector~$\phi$ out of all eight real and imaginary upper and lower components of the two doublets,
\begin{equation} \label{eq:phi}
    \phi = \begin{pmatrix}
    \pi_1^1, \pi_1^2, \sigma_1^1, \sigma_1^2, \pi_2^1, \pi_2^2, \sigma_2^1, \sigma_2^2\end{pmatrix}^\trans
    , \quad \text{with }
    \phi \in \mathbb{R}^8\,. 
\end{equation}

The most general, tree-level, renormalizable and gauge-invariant potential with two Higgs-boson doublets 
reads \cite{Gunion:1989we,Gunion:1992hs}
\begin{equation}
\label{eq:Vconv}
\begin{split}
V^{(0)}_{\text{THDM}} (\varphi_1, \varphi_2) =~& 
m_{11}^2 (\varphi_1^\dagger \varphi_1) +
m_{22}^2 (\varphi_2^\dagger \varphi_2) -
m_{12}^2 (\varphi_1^\dagger \varphi_2) -
(m_{12}^2)^* (\varphi_2^\dagger \varphi_1)\\
& +\frac{1}{2} \lambda_1 (\varphi_1^\dagger \varphi_1)^2 
+ \frac{1}{2} \lambda_2 (\varphi_2^\dagger \varphi_2)^2 
+ \lambda_3 (\varphi_1^\dagger \varphi_1)(\varphi_2^\dagger \varphi_2) \\ 
&+ \lambda_4 (\varphi_1^\dagger \varphi_2)(\varphi_2^\dagger \varphi_1)
+ \frac{1}{2} \Big[ \lambda_5 (\varphi_1^\dagger \varphi_2)^2 + \lambda_5^* 
(\varphi_2^\dagger \varphi_1)^2 \Big] \\ 
&+ \Big[ \lambda_6 (\varphi_1^\dagger \varphi_2) + \lambda_6^* 
(\varphi_2^\dagger \varphi_1) \Big] (\varphi_1^\dagger \varphi_1) + \Big[ \lambda_7 (\varphi_1^\dagger 
\varphi_2) + \lambda_7^* (\varphi_2^\dagger \varphi_1) \Big] (\varphi_2^\dagger \varphi_2).
\end{split}
\end{equation}
The potential contains two real quadratic parameters~$m_{11}^2$, $m_{22}^2$, one complex quadratic parameter~$m_{12}^2$ and seven quartic parameters $\lambda_1, \dots, \lambda_7$.
Among the quartic parameters the first four are real, while $\lambda_5$, $\lambda_6$, $\lambda_7$ are complex. 
In total, the potential contains 14 real parameters.

We now outline how the bilinears are introduced; for further details, see~\cite{Maniatis:2006fs}.
In a first step, the $2 \times 2$ matrix of the two doublets is formed:
\begin{equation} \label{eq:defphi}
    \psi = \begin{pmatrix} \varphi_1^\trans \\ \varphi_2^\trans \end{pmatrix} 
    = \begin{pmatrix} \varphi_1^+& \varphi_1^0 \\ \varphi_2^+& \varphi_2^0 \end{pmatrix} \,.
\end{equation}
Taking the outer product of this matrix yields the bilinear matrix
\begin{equation} \label{eq:Kbar}
\twomat{K} = {\psi} {\psi}^\dagger =
\begin{pmatrix}
 \varphi_1^\dagger \varphi_1 &  ~\varphi_2^\dagger \varphi_1\\
 \varphi_1^\dagger \varphi_2 &  ~\varphi_2^\dagger \varphi_2
 \end{pmatrix}.
\end{equation}
We see that all possible invariant scalar products of the two doublets $\varphi_1$ and $\varphi_2$  appear in this matrix.

This  $2 \times 2$ matrix~$\twomat{K}$ is by construction hermitian and can be decomposed into a basis of the unit matrix and the Pauli matrices,
\begin{equation}
 \twomat{K} = \frac{1}{2} \bigg( K_0 \unitmatrix_2 + K_a \sigma_a \bigg), \qquad a=1,2,3, \label{eq:KbarToBilinears}
\end{equation}
with the usual convention to sum over repeated indices. The four real coefficients $K_0$, $K_a$ are called bilinears. 
From traces on both sides of this equation, also with products of Pauli matrices, we get the four real bilinears explicitly,
\begin{align}
\label{eq:Kphi}
&K_0 = \varphi_1^\dagger \varphi_1 + \varphi_2^\dagger \varphi_2,
&&K_1 = \varphi_1^\dagger \varphi_2 + \varphi_2^\dagger \varphi_1, \nonumber \\
&K_2 = i\big( \varphi_2^\dagger \varphi_1 - \varphi_1^\dagger \varphi_2 \big), 
&&K_3 = \varphi_1^\dagger \varphi_1 - \varphi_2^\dagger \varphi_2,
\end{align}
which justifies the term {\em bilinear}. Inverting these relations shows how any THDM potential can be expressed in terms of bilinears,
\begin{align}
\label{eq:phiK}  
& \varphi_1^\dagger \varphi_1 = \frac{1}{2}\left( K_0 + K_3\right),
&& \varphi_1^\dagger \varphi_2 = \frac{1}{2}\left( K_1 + i K_2\right), \nonumber\\
& \varphi_2^\dagger \varphi_1 = \frac{1}{2}\left( K_1 - i K_2\right),
&& \varphi_2^\dagger \varphi_2 = \frac{1}{2}\left( K_0 - K_3\right)\,.
\end{align}
The matrix $\twomat{K}$ is positive semi-definite. Moreover, 
$K_0 = \tr (\twomat{K})$  and $\det(\twomat{K}) = \tfrac{1}{4} (K_0^2 -K_a K_a)$
and
\begin{equation} \label{eq:Kdom}
K_0 \ge 0, \qquad K_0^2 - K_a K_a \ge 0\,.
\end{equation}
As has been shown in~\cite{Maniatis:2006fs} there is a one-to-one correspondence between the original doublet fields and the bilinears - modulo unphysical gauge-degrees of freedom. 
Since constant terms can be omitted, the most general THDM potential formulated in terms of bilinears
is
\begin{equation}
\label{eq:pot}
V^{(0)}_{\text{THDM}} (K_0, K_a)=
 \xi_0 K_0 + \xi_a K_a + \eta_{00} K_0^2 + 2 K_0 \eta_a K_a + K_a E_{ab} K_b\,.
\end{equation}
All parameters $\xi_0$, $\xi_a$, $\eta_{00}$, $\eta_a$, $E_{ab} = E_{ba}$, $a,b \in \{1,2,3\}$ are real
and are expressed in terms of the conventional parameters in \eqref{eq:Vconv} as follows:
\begin{align}
\xi_0 &= \frac{1}{2}\left(m_{11}^2+m_{22}^2\right) ,
\quad
\tvec{\xi} = (\xi_\alpha)=\frac{1}{2}
\begin{pmatrix}
- 2 \re(m_{12}^2), &
 2 \im(m_{12}^2), &
 m_{11}^2-m_{22}^2
\end{pmatrix}^\trans,
\nonumber
\\
\eta_{00} &= \frac{1}{8}
(\lambda_1 + \lambda_2) + \frac{1}{4} \lambda_3  ,
\quad
\tvec{\eta} = (\eta_{a})=\frac{1}{4}
\begin{pmatrix}
\re(\lambda_6+\lambda_7), & 
-\im(\lambda_6+\lambda_7), & 
\frac{1}{2}(\lambda_1 - \lambda_2)
\end{pmatrix}^\trans, 
\nonumber \\
E &= (E_{ab})= \frac{1}{4}
\begin{pmatrix}
\lambda_4 + \re(\lambda_5) & 
-\im(\lambda_5) & \re(\lambda_6-\lambda_7) 
\\ 
-\im(\lambda_5) & \lambda_4 - \re(\lambda_5) & 
-\im(\lambda_6-\lambda_7) \\ 
\re(\lambda_6-\lambda_7) & 
-\im(\lambda_6 -\lambda_7) & 
\frac{1}{2}(\lambda_1 + \lambda_2) - \lambda_3
\end{pmatrix}.
\label{eq:para3}
\end{align}

We can write~\eqref{eq:KbarToBilinears} shortly
\begin{equation}
    \label{eq:KbarToKM}
   \twomat{K} = \frac{1}{2} 
   \sigma_\mu K_\mu, \qquad
   \text{where }
   \sigma_0 \equiv \unitmatrix_2,
\qquad \mu=0,1,2,3\;,
\end{equation}
with the usual summation convention of repeated indices, 
with upper and lower indices used interchangeably.
The Pauli matrices satisfy
\begin{equation}
\label{eq:Pauli}
\tr(\sigma_\mu \sigma_\nu)
= 2 \delta_{\mu\nu}
\end{equation}
and we denote the components of the Pauli matrices as $(\sigma_\mu)^a_{\;b}$.
It is natural to define a four-component vector
from the bilinears~\cite{Maniatis:2006fs}, 
\begin{equation}
    \KT=\begin{pmatrix} K_0\\ \tvec{K}\end{pmatrix}, \quad
    \text{with } \tvec{K}=
    \begin{pmatrix} K_1\\ K_2\\ K_3 \end{pmatrix}.
\end{equation}
If we now express the parameters of the potential as
\begin{equation} \label{eq:para4}
    \tilde{\tvec{\xi}}=\begin{pmatrix} \xi_0 \\ \tvec{\xi}\end{pmatrix},
    \qquad
    \tilde{E}=\begin{pmatrix} \eta_{00} & \tvec{\eta}^\trans\\
    \tvec{\eta} & E \end{pmatrix},
\end{equation}
we see that the tree-level potential~\eqref{eq:pot} can be written
\begin{equation} \label{eq:potK4}
    V^{(0)}_{\text{THDM}}(\KT) =  \KT^\trans \tilde{\tvec{\xi}} + 
     \KT^\trans \tilde{E}  \KT\,.
\end{equation}
If we introduce the constant matrix 
\begin{equation}
    \tilde{g} = \diag(1, -\unitmatrix_3)\,,
\end{equation}
we may write the conditions~\eqref{eq:Kdom} shortly
\begin{equation} \label{eq:dom4}
    K_0 \ge 0, \qquad \KT^\trans \tilde{g} \KT \ge 0\;.
\end{equation}

Let us also recall the unitary mixing of the two doublets,
\begin{equation} \label{eq:basistf}
\varphi'_i = U_{ij} \varphi_j, \quad \text{with } U = (U_{ij}),  \quad U^\dagger U = \unitmatrix_2\,.
\end{equation}
These basis transformations give in terms of bilinears~\cite{Maniatis:2006fs}
\begin{equation} \label{eq:b1}
K_0' = K_0, \quad K_a' = R_{ab}(U) K_b \;,
\end{equation}
with $R_{ab}(U)$ defined by
\begin{equation} \label{eq:bilinearRtransfo}
U^\dagger \sigma^a U = R_{ab}(U) \sigma^b \;.
\end{equation}
The transformations $R(U) \in SO(3)$ are proper rotations in three dimensions. The potential~\eqref{eq:pot} 
is invariant under a change of basis
\eqref{eq:b1} by a simultaneous transformation of the parameters
\begin{equation} \label{eq:b2}
\xi'_0 = \xi_0, \quad
\xi_a' = R_{ab} \xi_b, \quad
\eta_{00}' = \eta_{00}, \quad
\eta_{a}' = R_{ab} \eta_{b}, \quad
E_{cd}' = R_{ca}  E_{ab} R_{bd}^\trans \, .
\end{equation}
We may, by a change of basis,  diagonalize the real symmetric matrix~$E$. 

Let us recall from~\cite{Maniatis:2006fs} the electroweak symmetry breaking in the bilinear formalism.
We assume to have a stable potential, that is, one that is bounded from below.
At a minimum of the potential~\eqref{eq:defphi} becomes, 
\begin{equation} \label{eq:phivev}
    \langle \psi \rangle 
    = \begin{pmatrix} v_1^+& v_1^0 \\ v_2^+& v_2^0 \end{pmatrix}\,,
\end{equation}
with the vacuum expectation values of the components of the doublets denoted by 
$v^+_{1/2} = \langle \varphi^+_{1/2} \rangle$ and
$v^0_{1/2} = \langle \varphi^0_{1/2} \rangle$. 

We can now distinguish different cases of the vacuum of $\twomat{K}$.
We have a charge-breaking (CB) minimum if the matrix $\langle \psi \rangle$, and therefore $\langle \twomat{K} \rangle = \langle \psi \rangle \langle \psi \rangle^\dagger$ has full rank giving the condition for the bilinear fields
\begin{equation} \label{bEW}
\text{CB:} \qquad K_0>0, \qquad K_0^2 -\tvec{K}^\trans \tvec{K}  = \KT^\trans \tilde{g} \KT > 0\;.
\end{equation}

For the case of the matrix $\langle \psi \rangle$ together with $\langle \twomat{K} \rangle$ of rank one
we have a charge-conserving (CC) vacuum leading to the condition
\begin{equation} \label{EW}
\text{CC:} \qquad K_0>0, \qquad K_0^2 -\tvec{K}^\trans \tvec{K}  = \KT^\trans \tilde{g} \KT = 0\;.
\end{equation}
Note that this requirement yields a light-like Minkowski-type structure; see~\cite{Maniatis:2006fs}.

Finally, a vacuum with
\begin{equation}
    K_0 = 0
\end{equation}
corresponds to an unbroken electroweak symmetry resulting 
in~$\KT=0$. 

For a charge-conserving minimum we can, by a change of basis, achieve that only the component~$\varphi_1^0$ gets a non-vanishing real vacuum-expectation value, in particular we can write, imposing a conventional factor $1/\sqrt{2}$,
\begin{equation} \label{eq:v10beta}
    \langle {\varphi_1}'\rangle = \frac{1}{\sqrt{2}}
    \begin{pmatrix} 0 \\ v \end{pmatrix},
    \qquad
    \langle {\varphi_2}'\rangle = \begin{pmatrix} 0 \\ 0 \end{pmatrix}.
  \end{equation}
This basis, in which only the neutral component of $\varphi_1$ acquires a non-zero vacuum expectation value, is sometimes referred to as the Higgs basis.
The bilinears in this basis read $\KT=\begin{pmatrix} v^2/2,& 0, & 0,& v^2/2 \end{pmatrix}^\trans$.
Alternatively, we may for a charge-conserving minimum, by a change of basis with mixing angle~$\beta$, achieve the form
\begin{equation} \label{eq:v10beta2}
    \langle {\varphi_1}'\rangle = \frac{1}{\sqrt{2}}
    \begin{pmatrix} 0 \\ v \cos(\beta) \end{pmatrix},
    \qquad
    \langle {\varphi_2}'\rangle = \frac{1}{\sqrt{2}}
    \begin{pmatrix} 0 \\ v \sin(\beta) \end{pmatrix}\,.
\end{equation}

If the potential is stable, the minima can be found from the gradient of the potential.
The conditions for a charge-breaking minimum read
\begin{equation} \label{eqbEW}
\text{CB:} \qquad 
\partial_\mu V \equiv \frac{\partial V}{\partial K^\mu} = 0
\end{equation}
on the domain~\eqref{bEW}. 
Unbroken minima follow from
vanishing doublets, corresponding to $K_0=0$. As we see, for the most general potential we have in this case 
a vanishing potential at the position of the minimum.
A charge-conserving electroweak symmetry-breaking minimum can be found by introducing a Lagrange multiplier~$u$ in order to satisfy the second equation in~\eqref{EW}. That is, the conditions for  a minimum with the correct electroweak symmetry breaking read 
\begin{equation} \label{eqEW}
\text{CC:} \qquad 
\partial_\mu V =2 u  (\tilde{g} \KT)_\mu
\end{equation}
on the hypersurface~\eqref{EW}.
Hence these gradient equations together with the domain condition $\KT^\trans \tilde{g} \KT=0$ 
determine all stationary points of the potential 
in terms of $u$ and $\KT$ corresponding to the charge-conserving case
\footnote{Alternatively, the constraint \eqref{EW} could be used to eliminate, say $K_0$, in $V(K_0, K_1, K_2, K_3)$ leading to a potential
$\tilde V(K_1, K_2, K_3)$ for which one would have to calculate the gradient $\partial_{\mu=1,2,3} \tilde V = 0$ to find the stationary points
without introducing a Lagrange multiplier.}.

Later we will also need dimensionless bilinears for the case $K_0>0$. The case $K_0=0$ is trivial and simply 
gives a vanishing potential. Therefore, we define for $K_0 > 0$~\cite{Maniatis:2006fs}
\begin{equation}
\label{eq-ksde}
k_a = \frac{K_a}{K_0}, \quad a \in \{1,2,3\} \quad \text{and} \quad 
\tvec{k} = 
\begin{pmatrix} k_1, & k_2, & k_3 \end{pmatrix}^\trans \;.
\end{equation}
Using these dimensionless bilinears, the domain~\eqref{eq:Kdom} becomes
\begin{equation} \label{domk}
|\tvec{k}| \leq 1 \;.
\end{equation}

We will follow the convention to use Greek indices $\mu$, $\nu,  \ldots \in \{0, \ldots, 3\}$ for the Minkowski-type four-vectors, for instance we write $K_\mu$. The two doublets themselves are distinguished by Latin indices $i$, $j, \ldots \in \{1,2\}$. For the component fields 
of the two doublets~\eqref{eq:phi} we also use Latin indices $i$, $j, \ldots \in \{1, \ldots, 8\}$. Let us note that
in our notation, upper and lower indices will be used interchangeably.

\section{Gauge-invariant scalar mass matrices}
\label{sec:gaugeInvariantFormalism}

We briefly recall the derivation of the scalar squared mass matrices from~\cite{Sartore:2022sxh}.
The principal idea is to express the mass matrices in a gauge-invariant form. Since the mass matrices are defined as the second derivative of the Lagrangian with respect to the fields, the first step is to establish a connection between the component fields of the doublets~\eqref{eq:phi}, $\phi^i$, $i \in \{1, \ldots, 8\}$, and the bilinear fields $K^\mu$, $\mu \in \{0, \ldots 3\}$.

The gauge-invariant bilinear fields can be written in terms of the components of the doublets~\eqref{eq:phi}
\begin{equation}\label{eq:KphiDef}
	K^\mu \equiv \frac{1}{2} \Delta^\mu_{ij} \phi^i \phi^j\,,
	\qquad i,j \in \{1,\ldots,8\}\,.
\end{equation}
In the basis~\eqref{eq:phi}, we find explicitly the four real and symmetric $8 \times 8$ matrices $\Delta^\mu_{ij}$:
\begin{align}
\label{eq:DeltaMatrices}
	&\Delta^0 = \begin{pmatrix}
		\phantom{+}\unitmatrix_2 & & & \\
		& \phantom{+}\unitmatrix_2 & & \\
		& &\phantom{+}\unitmatrix_2 &  \\
		& & &\phantom{+}\unitmatrix_2
	\end{pmatrix}, \qquad
	&&\Delta^1 = \begin{pmatrix}
		& & \phantom{+}\unitmatrix_2 & \\
		& & & \phantom{+}\unitmatrix_2 \\
		\phantom{+}\unitmatrix_2 & & &  \\
		& \phantom{+}\unitmatrix_2 & &
	\end{pmatrix}, \nonumber\\
	&\Delta^2 = \begin{pmatrix}
		& & & \phantom{+}\unitmatrix_2 \\
		& & -\unitmatrix_2 & \\
		& -\unitmatrix_2 & &  \\
		\phantom{+}\unitmatrix_2 & & &
	\end{pmatrix}, \qquad
	&&\Delta^3 = \begin{pmatrix}
		\phantom{+}\unitmatrix_2 & & & \\
		& \phantom{+}\unitmatrix_2 & & \\
		& &-\unitmatrix_2 &  \\
		& & &-\unitmatrix_2
	\end{pmatrix}\, .
\end{align}
Here, $\unitmatrix_2$ denotes the $2 \times 2$ identity matrix and the blank entries represent zeros.

Then the connection between the bilinears and the component fields are defined as the $8 \times 4$ matrix $\Gamma$,
\begin{equation} \label{eq:defgamma}
   \Gamma^\mu_i \equiv \frac{\partial K^\mu}{\partial \phi^i} = \partial_i K^\mu = \Delta^\mu_{ij} \phi^j  \,.
\end{equation}
We emphasize that  the Greek indices refer to the gauge invariants $K^\mu$, while Latin indices label the component fields~$\phi^i$.

The next step is to derive the algebra of the
matrices $\Delta_{ij}^\mu$. We have
\begin{align}\label{eq:ga21}
    \left(\Gamma^2\right)^{\mu\nu} = \left(\Gamma^\trans \Gamma\right)^{\mu\nu} = \Gamma^\mu_i \Gamma^\nu_i = \Delta^\mu_{ij} \Delta^\nu_{ik}\, \phi^j \phi^k  
    = \frac{1}{2}\left\{\Delta^\mu, \Delta^\nu\right\}_{jk} \phi^j \phi^k\, ,
\end{align}
with $\{A, B\} = A B + B A$ denoting, as usual, the anti-commutator.
 We see from~\eqref{eq:ga21} that $\Gamma^2$ is a gauge-independent expression, carrying only Greek bilinear-field indices and that it depends quadratically on the fields. A fully symmetric rank-3 tensor $T_\lambda^{\mu \nu}$ can be defined by
\begin{equation}\label{eq:ga22}
    \left(\Gamma^2\right)^{\mu\nu} = T_\lambda^{\mu\nu} K^\lambda\,.
\end{equation}
We then get from~\eqref{eq:KphiDef}, \eqref{eq:ga21} and \eqref{eq:ga22},
\begin{equation}\label{eq:anti-comm}
    \left\{\Delta^\mu, \Delta^\nu\right\}_{ij} =  T^{\mu\nu}_\lambda \Delta^\lambda_{ij}\,.
\end{equation}
The symmetric matrices
$\Delta^\mu$
satisfy the closure relation
\begin{equation}
    \Delta^\mu_{ij} \Delta^\nu_{ij} = \Delta^\mu_{ij} \Delta^\nu_{ji} =\tr (\Delta^\mu \Delta^\nu) = 8 \delta^{\mu\nu}\,.
\end{equation}
Explicitly we have
for $T^{\mu\nu}_\lambda$:
\footnote{$T^{\mu\nu}_\lambda$ can also be expressed in terms of the Pauli matrices as 
$T^{\mu\nu}_\lambda = \frac{1}{2}\mathrm{Tr}\,\big(\{\sigma^\mu, \sigma^\nu\}\,\sigma_\lambda\big)$.}
\begin{equation} \label{eq:Texp}
    T^{\mu\nu}_\lambda = \frac{1}{8}\mathrm{Tr}\left(\left\{\Delta^\mu, \Delta^\nu\right\} \Delta_\lambda\right)\,.
\end{equation}
We obtain 
\begin{equation}\label{eq:Gamma2expr}
    \Gamma^2 = 
     2 \begin{pmatrix}
       K_0 & \tvec{K}^\trans\\ \tvec{K} & K_0 \unitmatrix_3
    \end{pmatrix}\,.
\end{equation}

The field-dependent scalar mass matrix can now be expressed as
\begin{equation} \label{eq:msExpr}
    \left(M_s^2\right)_{ij} = \partial_i \partial_j V = \partial_i \left(\Gamma^\mu_j \partial_\mu V \right) = \Delta^\mu_{ij} \partial_\mu V + \Gamma^\mu_i \Gamma^\nu_j \partial_\mu \partial_\nu V \,.
\end{equation}
With the definition
\begin{equation} \label{eq:Mmunu}
    \mathcal{M}_{\mu\nu} = \partial_\mu \partial_\nu V\,
\end{equation}
and writing $\mathcal{M} = (\mathcal{M}_{\mu\nu})$
we express~\eqref{eq:msExpr} in matrix form:
\begin{equation} \label{eq:msExprMat}
    M_s^2 = \Delta^\mu \partial_\mu V + \Gamma \mathcal{M} \Gamma^\trans.
\end{equation}

We now introduce an orthogonal rotation of the component fields,
\begin{equation}
    \widehat{\phi}^i = U^{ij} \phi^j \quad 
    \text{with} \quad U^{ij} U^{kj} = \delta^{ik}, \quad
    i,j,k \in \{1,\ldots,8\}\,.
\end{equation}
In matrix notation we write simply
$\widehat{\phi}= U \phi$ with $U^\trans U = \unitmatrix_8$.
By an orthogonal rotation - only applied to the component fields, leaving the gauge invariants $\KT$ unchanged - we can always bring the 
$8\times4$ matrix $\Gamma$ into a form where
the first four rows vanish. This comes from the fact that the matrix $\Gamma$ can be understood as four columns of 8-component vectors, and by an orthogonal transformation, say, considering only the last column we may rotate this vector pointing to the eighth direction. Then by applying a transformation that only 
affects the upper seven components, the next-to-last column can be rotated into the plane spanned by directions 7-8, and so on. In this way we see that 
we can achieve a form of $\Gamma$ such that at most the four lower rows are non-vanishing.
\begin{equation} \label{eq:canonicalGamma}
    \widehat{\Gamma} = U_c\, \Gamma = \begin{pmatrix}
    0_{4\times4} \\ \gamma
    \end{pmatrix}\,,
\end{equation}
and we will call~$U_c$ a canonical rotation and the 
corresponding bases, 
where $\widehat{\Gamma}$ has this form, canonical bases. 
The matrix $\gamma$ is of dimension  $4\times4$. 
We find
\begin{equation}
    \Gamma^2 = \Gamma^\trans \Gamma = \widehat{\Gamma}^\trans \widehat{\Gamma} = 
    \gamma^\trans \gamma
    \,,
\end{equation}
meaning that the $\gamma$ matrix can be obtained from a Cholesky-like decomposition
of $\Gamma^2$, and that the resulting expression only depends on gauge invariants, that is, the bilinear fields. Requiring $\gamma$ to be a lower-triangular uniquely determines its components, 
\begin{equation} \label{eq:gamma}
    \gamma = \sqrt{2 K_0} \begin{pmatrix}
       \sqrt{1 - \tvec{k}^\trans \tvec{k}} & \tvec{0}^\trans \\
       \tvec{k} & \unitmatrix_3
    \end{pmatrix}\,.
\end{equation}
In a canonical basis we can always express $\widehat{\Gamma}$ in terms of the field components $\widehat{\phi}=U_c \phi$:
\begin{equation}
    \widehat{\Gamma} = U_c \Gamma = U_c \Delta \phi = U_c \Delta U_c^\trans \widehat{\phi} = 
    \widehat{\Delta} \widehat{\phi}\, ,
\end{equation}
where $\widehat{\Delta}= U_c \Delta U_c^\trans$ is simply obtained from the canonical orthogonal rotation $U_c$.
Note that all basis-dependent quantities in the canonical basis are denoted with a hat.

Transforming~\eqref{eq:msExprMat} to the canonical basis, using for the case of a charge-conserving 
minimum~\eqref{eqEW}, we have
\begin{equation} \label{eq:MsqCC}
    \widehat{M}_s^2  \stackrel{CC}{=}  2u  \, \big(\gT \KT\big)_\mu \widehat{\Delta}^\mu + \widehat{\Gamma} \mathcal{M} \widehat{\Gamma}^\trans\,.
\end{equation}
The explicit form of this squared scalar mass matrix for a charge-conserving minimum
has been derived in \cite{Sartore:2022sxh}:
\begin{equation} \label{eq:Deltahat}
    \widehat{\Delta}^\mu = \begin{pmatrix}
        A_{55}^\mu & C_{53}^\mu \\[.15cm]
        (C_{53}^\mu)^\trans & B_{33}^\mu
    \end{pmatrix}\,,
\end{equation}
with
\begin{align}
    A_{55}^0 &= \unitmatrix_5, & A_{55}^a &= \begin{pmatrix}
        k_a & \tvec{0}_{1\times 3} & 0\\
        \;\;\tvec{0}_{3\times 1}\;\; & k_a \unitmatrix_3 - 2 k_a \tvec{k} \tvec{k}^\trans\!\! + \!\! \left(\left[\tvec{e}_a\times\tvec{k}\right] \tvec{k}^\trans + \tvec{k} \left[\tvec{e}_a\times\tvec{k}\right]^\trans\right)\;
        & \tvec{e}_a - k_a \tvec{k} \\
        0 & \tvec{e}_a^\trans - k_a \tvec{k}^\trans & - k_a
    \end{pmatrix},\\[.2cm] 
    B_{33}^0 &= \unitmatrix_3, & B_{33}^a &= - k_a \unitmatrix_3 + \tvec{e}_a \tvec{k}^\trans + \tvec{k} \tvec{e}_a^\trans,
    \label{eq:Ahatcomp}\\[.2cm]
    C_{53}^0 &= 0_{5\times3}, & C_{53}^a &= \begin{pmatrix}
        \left[\tvec{e}_a \times \tvec{k}\right]^\trans \\[.15cm]
    	0_{4\times 3}
    \end{pmatrix}\,.
\end{align}
This gives the following expressions for the terms appearing in~\eqref{eq:MsqCC} 
\begin{equation}
\big(\gT \KT\big)_\mu \widehat{\Delta}^\mu  =
\begin{pmatrix}
        A_{55} & 0_{5\times3} \\
        0_{3\times5} & B_{33} \\
    \end{pmatrix}
\end{equation}
and
\begin{equation} \label{eq:GMG}
\widehat{\Gamma} \mathcal{M} \widehat{\Gamma}^\trans  =
    \begin{pmatrix}
        0_{5\times5} & 0_{5\times3} \\
        0_{3\times5} & \gamma_3 \mathcal{M} \gamma_3^\trans
    \end{pmatrix}\, .
\end{equation}
The matrices $A_{55}$ and $B_{33}$ read explicitly
\begin{equation}
    A_{55} = 2 K_0 \begin{pmatrix}
        0 & \tvec{0}^\trans & 0 \\
        \tvec{0} & \tvec{k} \tvec{k}^\trans & \tvec{0}\\
        0 & \tvec{0}^\trans & 1\\
    \end{pmatrix}\, ,\qquad 
    B_{33} = 2 K_0 \left(\unitmatrix_3 - \tvec{k} \tvec{k}^\trans\right) = - \gamma_3 \gT \gamma_3^\trans\, , \qquad
\end{equation}
where the matrix $\gamma_3$ is defined to be the non-vanishing, lower $3\times4$ block of the matrix~$\gamma$ evaluated at a charge-conserving minimum where 
$1-\tvec{k}^\trans \tvec{k}=0$,
\begin{equation} \label{eq:ga3def}
    \gamma \stackrel{CC}{=} \sqrt{2 K_0} \begin{pmatrix}
        0 & \tvec{0}^\trans\\
        \tvec{k} & \unitmatrix_3
    \end{pmatrix} \equiv \begin{pmatrix}
        0_{1\times4}\\
        \gamma_3
    \end{pmatrix}\,.
\end{equation}
We can therefore write
\begin{equation}
\widehat{M}_s^2  \stackrel{CC}{=} 
   \begin{pmatrix}
        \widehat{\mathcal{M}}^2_\mathrm{CC} & 0_{5\times3}\\
        0_{3\times5} &  \widehat{\mathcal{M}}^2_\mathrm{neutral}
    \end{pmatrix}\,.
    \label{eq:ms2cc}
\end{equation}
The $5 \times 5$ block $\widehat{\mathcal{M}}^2_\mathrm{CC} = 2 u A_{55}$ 
accounts for the Goldstone sector as well as the two massive charged Higgs fields.

By a further transformation from one canonical basis to another we can completely
disentangle the massless Goldstone sector, the 
electrically charged as well the electrically neutral contributions,
\begin{equation}\label{eq:massMatPhysical}
    \widehat{M}_s^2 \stackrel{CC}{=} \begin{pmatrix}
        0_{3\times3} & & \\
        & \widehat{\mathcal{M}}^2_\mathrm{charged} & \\
        & & \widehat{\mathcal{M}}^2_\mathrm{neutral}
    \end{pmatrix}\,.
\end{equation}
In~\cite{Sartore:2022sxh} it is shown that the charged mass matrix can be diagonalized,
\begin{equation} \label{eq:allOrderMcharged}
    \overline{\mathcal{M}}^2_\mathrm{charged} = \diag\begin{pmatrix} m_{H^\pm}^2, & m_{H^\pm}^2 \end{pmatrix} = \diag\begin{pmatrix} 4 u K_0, & 4 u K_0 \end{pmatrix}\,,
\end{equation}
and the neutral scalar mass matrix reads
\begin{equation}\label{eq:allOrderMneutral}
    \widehat{\mathcal{M}}^{2}_\mathrm{neutral} = \gamma_3 \left(\mathcal{M} - 2 u \gT\right) \gamma_3^\trans\,.
\end{equation}

To get the scalar squared masses we have to diagonalize the 
neutral part~\eqref{eq:allOrderMneutral} by a similarity transformation,
defined as 
\begin{equation}\label{eq:diagonalBasisNeutral}
    \overline{\mathcal{M}}^2_\mathrm{neutral} = R\widehat{\mathcal{M}}^2_\mathrm{neutral}R^\trans = \diag\begin{pmatrix}m_1^2, & m_2^2, & m_3^2\end{pmatrix}.
\end{equation}
With this diagonalization we eventually get, in the {\em mass basis},
\begin{equation}\label{eq:diagonalMs}
    \overline{M}_s^2 = \bar{U} \widehat{M}_s^2 \bar{U}^\trans =
    \diag\begin{pmatrix}
        0, & 0, & 0, & m_{H^\pm}^2, & m_{H^\pm}^2, & m_1^2, & m_2^2, & m_3^2
    \end{pmatrix}\,.
\end{equation}
Here the $(8 \times 8)$ matrix
\begin{equation} \label{eq:Ubar}
\bar{U} = \diag(1, R_H, 1, R)\;,
\end{equation}
where $R$ is the basis transformation~\eqref{eq:diagonalBasisNeutral}  and $R_H$ is defined as the basis transformations to the Higgs basis (see also Appendix~A in~\cite{Sartore:2022sxh}),
\begin{equation} \label{eq:RH}
R_H\, \tvec{k} = \begin{pmatrix} 0, 0, 1\end{pmatrix}^\trans.
\end{equation}
This transformation with the matrix~$R$ corresponds to a basis change in terms of bilinears with the orthogonal 
three dimensional rotation~$R$,
\begin{equation} \label{bilbasis}
    K_0 \to K_0, \quad
    \tvec{K} \to \overline{\tvec{K}} = R \tvec{K}\,,
    \quad
     \tvec{k} \to \bar{\tvec{k}} = R \tvec{k} \, .
\end{equation}
With respect to the Minkowski-type four vectors, indicated by a tilde symbol, we write 
\begin{equation}
 \overline{\KT} = \widetilde{R} \KT
    \quad \text{with }
    \widetilde{R} =
    \begin{pmatrix}
    1 & \tvec{0}^\trans\\
    \tvec{0} & R
    \end{pmatrix}, 
\end{equation}
that is, $R$ is a $3 \times 3$ matrix, whereas $\widetilde{R}$ is a $4 \times 4$ matrix. 

The transformations of the connection from the canonical basis $\widehat{\Gamma}$ to 
the mass basis, as well as $\widehat{\Delta}$, given in \eqref{eq:Deltahat}, read
\begin{equation} \label{Gamma_Delta_bar}
\bar{\Gamma}_i^\mu =
\bar{U}_{ij} \widehat{\Gamma}_j^\mu,\ \qquad
\bar{\Delta}_{ij}^\mu =
\bar{U}_{ia} \widehat{\Delta}_{ab}^\mu
\bar{U}_{bj}^\trans\;.
\end{equation}

\subsection{Tree-level scalar mass matrix}

We now compute the squared mass matrix for the tree-level THDM potential.
The charged scalar squared masses are in any potential given in~\eqref{eq:allOrderMcharged}:
\begin{equation} \label{mHpm}
m_{H^\pm}^2 = 4 u K_0\, .
\end{equation}
Recall that both $u$ and $K_0$ are determined from the stationarity equations; see~\eqref{eqEW}.
For the neutral $3 \times 3$  matrix~\eqref{eq:allOrderMneutral} we find in this case
\begin{equation} \label{eq:MneutralTreeLevel}
    \widehat{\mathcal{M}}_\mathrm{neutral}^2 = 
    4 K_0 \left[ \eta_{00} \tvec{k} \tvec{k}^\trans + \tvec{\eta} \tvec{k}^\trans + \tvec{k} \tvec{\eta}^\trans + E + u (\unitmatrix_3 -  \tvec{k} \tvec{k}^\trans) \right].
\end{equation}
This explicitly gauge invariant, real, symmetric matrix can be diagonalised
with the rotation matrix~$R$ given in \eqref{eq:diagonalBasisNeutral}.
Under this change of basis, the bilinears transform as
shown in~\eqref{bilbasis}. The parameters of the tree-level potential transform as, see \eqref{eq:b2},
\begin{equation} 
\label{eq:b2b}
\bar{\xi}_0 = \xi_0, \quad
\bar{\tvec{\xi}} = R \tvec{\xi}, \quad
\bar{\eta}_{00} = \eta_{00}, \quad
\bar{\tvec{\eta}} = R \tvec{\eta}, \quad
\bar{E} = R  E R^\trans \, .
\end{equation}
We get for the neutral squared mass matrix 
\begin{equation} \label{eq:Mbar0tree}
    \overline{\mathcal{M}}_\mathrm{neutral}^2 = 4 K_0 \left[ \eta_{00} \bar{\tvec{k}} \bar{\tvec{k}}^\trans + \bar{\tvec{\eta}} \bar{\tvec{k}}^\trans + \bar{\tvec{k}} \bar{\tvec{\eta}}^\trans + \bar{E} + u (\unitmatrix_3 -  \bar{\tvec{k}} \bar{\tvec{k}}^\trans) \right].
\end{equation}
Let us mention that we can use~\eqref{eq:Mbar0tree} and \eqref{mHpm} together with the parameters 
$\eta_{00}$ and $\tvec{\bar{\eta}}$ to fix the parameter matrix~$\bar{E}$ of the THDM in terms of the scalar masses.

\section{The $\hbar$-expansion}
\label{sec:hbarExpansion}

Once a minimum of the tree-level potential has been found, one can compute the 
$n$-loop corrections to the vacuum position and to quantities derived from the scalar potential, such as the scalar masses, using the so-called $\hbar$-expansion~\cite{Brodsky:2010zk}. The general idea is to combine the perturbative expansion of the scalar potential with a Taylor expansion around the tree-level vacuum. 
The perturbative expansion is then obtained by iteratively solving the corresponding stationarity conditions.
In the following, we assume that the quantum corrections to the potential, like the vacuum itself, can be expressed as a perturbative expansion around the tree-level approximation.
Considering $\kappa = 1/(16\pi^2)$ as the perturbative expansion parameter\footnote{Had we not been working with natural units, the actual expansion parameter would have been $\frac{\hbar}{16 \pi^2}$.} we define
\begin{gather}
    V = V^{(0)} +\kappa V^{(1)} + \kappa^2 V^{(2)} + \dots \label{eq:hbarV}\,,\\
    \phi = \phi^{(0)} +\kappa \phi^{(1)} + \kappa^2 \phi^{(2)} + \dots \label{eq:hbarPhi}\,.
\end{gather}
Let us emphasize that in~\eqref{eq:hbarPhi} we are referring to the vacuum expectation value of the fields, $\langle \phi \rangle$, 
and similarly for the radiative corrections $\langle \phi^{(n)} \rangle$, for which we shall omit the brackets to simplify the notation. 
Moreover, 
$\phi$, $\phi^{(0)}$, $\phi^{(1)}$,~$\ldots$ each represents a complete vector of components $\phi_i$, $\phi_i^{(0)}$, $\phi_i^{(1)}$,~$\ldots$, 
where $i=1,\ldots,8$.
The $\hbar$-expansions of the scalar potential and of its first derivatives read
\begin{align}
\begin{split}
    V(\phi) &= V^{(0)}(\phi^{(0)}) + \kappa\left[V^{(1)}(\phi^{(0)}) + \phi^{(1)}_i \partial_i V^{(0)}(\phi^{(0)})\right]\\
    &+ \kappa^2 \left[V^{(2)}(\phi^{(0)}) + \phi^{(1)}_i \partial_i V^{(1)}(\phi^{(0)}) + \phi^{(2)}_i \partial_i V^{(0)}(\phi^{(0)}) +  \frac{1}{2} \phi^{(1)}_i \phi^{(1)}_j \partial_i \partial_j V^{(0)}(\phi^{(0)}) \right] \\
    &+ \mathcal{O}\left(\kappa^3\right)\,,
\end{split}\\
\begin{split} \label{eq:hbarSPE}
    \partial_a V(\phi) &= \partial_a V^{(0)}(\phi^{(0)}) + \kappa\left[\partial_a  V^{(1)}(\phi^{(0)}) + \phi^{(1)}_i \partial_i \partial_a V^{(0)}(\phi^{(0)})\right]\\
    &+ \kappa^2 
        \left[\partial_a V^{(2)}(\phi^{(0)}) + \phi^{(1)}_i \partial_i \partial_a V^{(1)}(\phi^{(0)}) + \phi^{(2)}_i \partial_i \partial_a V^{(0)}(\phi^{(0)})  \right.
        \\ & \qquad 
        +   \left. \frac{1}{2} \phi^{(1)}_i \phi^{(1)}_j \partial_i \partial_j \partial_a V^{(0)}(\phi^{(0)}) \right] \\
    &+ \mathcal{O}\left(\kappa^3\right)\,.
\end{split}
\end{align}
The vacuum expectation values
$\phi^{(k)}$ are precisely defined such that the stationarity conditions derived from~\eqref{eq:hbarSPE} are satisfied order by order. With this method and the analytic expressions of $V^{(0)}, \dots, V^{(n)}$, the position of the minimum can be computed iteratively up to order~$n$, starting with a minimization of the polynomial tree-level scalar potential. We note that the sum 
$\phi^{(0)} + \dots + \kappa^n \phi^{(n)}$
is not an exact solution of the $n$-loop stationarity condition (\textit{i.e.}~it does not precisely minimize the $n$-loop scalar potential) but rather an order $\kappa^n$ approximation of it.

While the latter is probably sufficient for many applications, the former can still be approached to arbitrary precision by truncating the scalar potential to order $n$ and solving the stationary point equation~\eqref{eq:hbarSPE} iteratively up to order $N > n$.\\

One major advantage of using the $\hbar$-expansion method is that the loop-corrections to the scalar potential are evaluated at a tree-level minimum, where the scalar spectrum can be easily computed. In particular, in our effort to study the effective potential in a gauge-invariant way, we will be able to derive analytic expressions for $V^{(1)}$ at a minimum of the tree-level potential, making the $\hbar$-expansion method well suited to the purpose of this work.

We now apply the $\hbar$-expansion method within the context of the gauge-invariant bilinear formalism. 
First, we substitute the $\phi$-expansion~\eqref{eq:hbarPhi} into the definition of the bilinear fields~\eqref{eq:KphiDef}, yielding
\begin{align} \label{Khbarexp}
\begin{split}
    K_\mu &= \frac{1}{2} \Delta_\mu^{ij} \left[\phi^{(0)}_i \phi^{(0)}_j + \kappa \left( 2 \phi^{(0)}_i \phi^{(1)}_j \right) + \kappa^2 \left(\phi^{(1)}_i \phi^{(1)}_j + 2 \phi^{(0)}_i \phi^{(2)}_j\right) + \dots \right]\\
    &\equiv K_\mu^{(0)} + \kappa K_\mu^{(1)} + \kappa^2 K_\mu^{(2)} + \dots\;,
\end{split}
\end{align}
such that the bilinear fields $\KT^{(n)} = (K_\mu^{(n)})$ are uniquely defined. 
Then, an $\hbar$-expansion for the stationary point equation can be given in terms of gauge-invariants, hence allowing for a minimization of the scalar potential in terms of bilinear fields. The derivatives are decomposed as follows (all quantities on the right-hand side are evaluated at the tree-level vacuum~$\KT^{(0)}$):
\begin{align}
\begin{split} \label{eq:KhbarSPE}
    \partial_\alpha V(\KT) &= \partial_\alpha V^{(0)} + \kappa\left[\partial_\alpha V^{(1)} + K^{(1)}_\mu \partial_\mu \partial_\alpha V^{(0)}\right] \\
    &+ \kappa^2 \left[\partial_\alpha V^{(2)} + K^{(1)}_\mu \partial_\mu \partial_\alpha V^{(1)} + K^{(2)}_\mu \partial_\mu \partial_\alpha V^{(0)} +  \frac{1}{2} K^{(1)}_\mu K^{(1)}_\nu \partial_\mu \partial_\nu \partial_\alpha V^{(0)} \right] + \mathcal{O}\left(\kappa^3\right)\,.
\end{split}
\end{align}
Note that the last summand at order~$\kappa^2$ vanishes as the tree-level potential is at most quadratic 
in the bilinears.
With an expansion of the Lagrange multiplier 
\begin{equation} \label{uexp}
u = u^{(0)} + \kappa u^{(1)}  + \kappa^2 u^{(2)} + \ldots 
\end{equation}
the stationary point equations for a charge-conserving minimum are given by
\begin{align}
\begin{split}
    \partial_\alpha V (\KT) = 2 u \big(\gT\KT\big)_\alpha &= 2 u^{(0)} \big(\gT\KT^{(0)}\big)_\alpha 
    + \kappa \left[ 2 u^{(0)} \big(\gT\KT^{(1)}\big)_\alpha +  2 u^{(1)} \big(\gT\KT^{(0)}\big)_\alpha\right] \\
    &+ \kappa^2 \left[2 u^{(0)} \big(\gT\KT^{(2)}\big)_\alpha + 2 u^{(1)} \big(\gT\KT^{(1)}\big)_\alpha + 2 u^{(2)} \big(\gT\KT^{(0)}\big)_\alpha \right] + \mathcal{O}\left(\kappa^3\right)
\end{split}
\end{align}
with
\begin{equation} \label{eq:kgk}
    \KT^\trans\gT\KT = 0 = \KT^{(0)\, \trans} \gT \KT^{(0)} + \kappa \left[ 2 \KT^{(0)\, \trans} \gT \KT^{(1)} \right] + \kappa^2 \left[ 2 \KT^{(0)\, \trans} \gT \KT^{(2)} + \KT^{(1)\, \trans} \gT \KT^{(1)} \right] + \mathcal{O}\left(\kappa^3\right)
\end{equation}
which vanishes at each order. 
Up to order $\kappa^2$, we finally obtain the following set of stationary point equations:
\begin{align}
    \partial_\alpha V^{(0)} &= 2 u^{(0)} \big( \gT \KT^{(0)}\big)_\alpha\,, \label{eq:kSPE0} \\
    \partial_\alpha V^{(1)} + \left(\mathcal{M}^{(0)} - 2 u^{(0)}\gT\right)^{\alpha \mu} K^{(1)}_\mu &= 2 u^{(1)} \big( \gT \KT^{(0)}\big)_\alpha\,, \label{eq:kSPE1}\\
    \partial_\alpha V^{(2)} + \left(\mathcal{M}^{(1)} - 2 u^{(1)}\gT\right)^{\alpha \mu} K^{(1)}_\mu + \left(\mathcal{M}^{(0)} - 2 u^{(0)}\gT\right)^{\alpha \mu} K^{(2)}_\mu &= 2 u^{(2)} \big( \gT \KT^{(0)}\big)_\alpha \label{eq:kSPE2}\,,
\end{align}
where we use the shorthand notation
\begin{equation}
    \mathcal{M}^{(n)}_{\mu\nu} = \partial_\mu \partial_\nu V^{(n)}\, .
\end{equation}

As stated above, the stationary point equations together with the constraints in \eqref{eq:kgk} can be solved iteratively in order to estimate the position of the loop-corrected minimum: The first equation~\eqref{eq:kSPE0} corresponding to a tree-level minimization, together with the first term in \eqref{eq:kgk}, allows to determine $\KT^{(0)}$ and $u^{(0)}$. 
In the next step, \eqref{eq:kSPE1}, together with the second term in \eqref{eq:kgk}, is solved for $\KT^{(1)}$ and $u^{(1)}$, given that an expression for $\partial_\alpha V^{(1)}$ can be found. Subsequently, \eqref{eq:kSPE2} and the order $\kappa^2$ term in \eqref{eq:kgk} are 
solved for $\KT^{(2)}$ and $u^{(2)}$, given that expressions of $\partial_\alpha V^{(2)}$ and $\partial_\alpha\partial_\mu V^{(1)}$ 
are known, and so on.\\

Although the order-$n$ stationary point equations can be solved at once for $\KT^{(n)}$ and $u^{(n)}$, we propose an alternative 
two-step method that will greatly simplify the computations and provide better insight into the nature of the loop corrections 
to the various physical quantities. 
The first step is to multiply the stationary point equation from the left by the matrix 
$\gamma$~\eqref{eq:ga3def},
which is understood to be evaluated at the tree-level minimum.
For conciseness, we define 
\begin{equation}\label{eq:D-derivative}
    D^\mu \equiv \gamma^{\mu\nu} \partial_\nu\,, \quad \text{and} \quad D^a \equiv \gamma_3^{a\nu} \partial_\nu\,,
\end{equation}
and obtain, recognizing that $\gamma  \tilde{g} \KT^{(0)}=0$, at one-loop order the new set of stationary point equations,
\begin{equation}
    D^\mu V^{(1)} + \left[\gamma \left(\mathcal{M}^{(0)} - 2 u^{(0)} \tilde{g}\right) \KT^{(1)}\right]^\mu = 0\,.
\end{equation}

Then, we observe that~\eqref{eq:kgk} implies
\begin{equation}
    \KT^{(1)} = \begin{pmatrix}
        \frac{1}{K_0^{(0)}}\tvec{K}^{(0)\,\trans} \tvec{K}^{(1)} \\
        \tvec{K}^{(1)}
    \end{pmatrix} = \frac{1}{\sqrt{2 K_0^{(0)}}} \gamma^\trans \KT^{(1)}\,,
\end{equation}
and, using the expression of the neutral mass matrix in~\eqref{eq:allOrderMneutral}, the one-loop stationary point equations can eventually be put in the form 
\begin{equation} \label{eq:canonicalSPE}
D^0 V^{(1)} =0\, , \quad
    \sqrt{2 K_0^{(0)}} D^a V^{(1)} + \left[\left(\widehat{\mathcal{M}}^2_\mathrm{neutral}\right)^{(0)} \tvec{K}^{(1)}\right]^a = 0,\quad (a=1,2,3)\, ,
\end{equation}
where $D^0 V^{(1)} =0$ is trivially satisfied.
From this form it is straightforward to compute the one-loop shift $K^{(1)}$ from the $D$-derivatives of the one-loop potential. In fact, the last equation can be further simplified when it is written in the diagonal basis~\eqref{eq:diagonalBasisNeutral},
\begin{equation} \label{eq:oneLoopSPE}
    \sqrt{2 K_0^{(0)}} \bar{D}^a V^{(1)} + \left(m_a^2\right)^{(0)} {\bar{K}^{(1)a}} = 0\,,
\end{equation}
where
\begin{equation} \label{eq:diagonalD}
    \bar{D}^a = R^{ab} D_b
\end{equation}
expresses the $D$-derivative with respect to the rotated bilinear fields. Here, the tree level squared scalar masses are denoted by $\left(m_a^2\right)^{(0)}$.
In fact, recalling from~\eqref{eq:canonicalGamma} and~\eqref{eq:ga3def} that $\gamma_3$ is the lower $3\times4$ block of $\widehat{\Gamma}$ in the canonical basis, the operator $D^a$ defined in~\eqref{eq:D-derivative} expresses the derivatives with respect to the neutral scalar field components. Accordingly, $\bar{D}^a$ expresses the derivative with respect to the diagonalized neutral field components (\textit{i.e.}~the neutral mass eigenstates) $H^a$ ($a=1,2,3$). Hence, the stationary point equation~\eqref{eq:canonicalSPE} matches the conventional minimization condition with respect to the gauge-singlet vacuum expectation values, and~\eqref{eq:oneLoopSPE} is its analogue in the mass basis. Additionally, the neutral scalar mass matrix in the canonical basis can be simply expressed as
\begin{equation}
    \left(\widehat{\mathcal{M}}^2_\mathrm{neutral}\right)^{ab} = D^a D^b V\,,
\end{equation}
and similarly for its diagonalized counterpart $\overline{\mathcal{M}}_\mathrm{neutral}^2$, given by
\begin{equation} \label{eq:MneutralDab}
    \left(\overline{\mathcal{M}}_\mathrm{neutral}^2\right)^{ab} = \bar{D}^a \bar{D}^b V\,.
\end{equation}
The second step consists in determining the value of $u^{(1)}$ which previously dropped out from the stationary point equation. This is done by evaluating the~$\alpha=0$ component of~\eqref{eq:kSPE1}. With $\mathcal{M}^{(0)}_{\mu \nu} = \partial_\mu \partial_\nu V^{(0)}$ and the explicit form of the tree-level potential, \eqref{eq:potK4}, applying~\eqref{eq:kgk}, we get the 
following formula for $u^{(1)}$,
\begin{equation}\label{eq:u1aux}
    u^{(1)} = \frac{1}{2 K_0^{(0)}} \partial_0 V^{(1)} + \frac{1}{K_0^{(0)}}\left[(\eta_{00} - u^{(0)})\tvec{k}^{(0)\,\trans} + \tvec{\eta}^\trans\right]\tvec{K}^{(1)}\,,
\end{equation}
where
\begin{equation}
    \tvec{k}^{(0)} = \frac{\tvec{K}^{(0)}}{K_0^{(0)}}\,.
\end{equation}
Here and in the following, it is useful to further define, using~\eqref{mHpm},
\begin{align}
    \tvec{k}^{(1)} &= \frac{\tvec{K}^{(1)}}{K_0^{(0)}}\,, \label{eq:k1def}\\
    \tvec{f}_\pm &= 8 K_0^{(0)} \left[\left(\eta_{00} - u^{(0)}\right)\tvec{k}^{(0)} + \tvec{\eta}\right] = 8 K_0^{(0)} \left(\eta_{00}\tvec{k}^{(0)} + \tvec{\eta}\right) - 2 \left(m_{H^\pm}^2\right)^{(0)} \tvec{k}^{(0)} \label{eq:fpmdef}\,,
\end{align}
bringing~\eqref{eq:u1aux} to the form
\begin{equation} \label{eq:u1}
    u^{(1)} = \frac{1}{4 K_0^{(0)}}\left[ 2 \partial_0 V^{(1)} + \frac{1}{2} \tvec{f}_\pm^\trans \tvec{k}^{(1)} \right]\,.
\end{equation}
This concludes the discussion on the stationary point equations and their solutions at one-loop order. In the next section, we derive analytic expressions for all the relevant quantities needed to solve the one-loop stationary point equations.

\section{One-loop effective potential and its derivatives}
\label{sec:oneLoopPotential}

In this section we express the one-loop contributions to the scalar potential and their derivatives in terms of the bilinear fields. 
This will allow us to solve the first-order stationarity condition analytically and compute the first-order corrections to the scalar masses in Sec.~\ref{sec:oneLoopMasses}. In the Landau gauge, the one-loop contributions to the effective potential can be generally written~\cite{Coleman:1973jx} (note that the original factor of $1/(16 \pi^2)$ is our factor $\kappa$ in the definition~\eqref{eq:hbarV}):
\begin{align}\label{eq:V1}
	V^{(1)}\left(\mu; \phi_a\right) &= \frac{1}{4} \sum_{i=s,f,g} n_i \mathrm{Tr}\left\{ M_i^4\left(\phi_a\right) \left[\log \frac{M_i^2\left(\phi_a\right)}{\mu^2} - C_i\right] \right\}\,.
\end{align}
As usual, $M_i^2 (\phi_a)$ with $i=s,f,g$ with $s$ for scalar, $f$ for fermion, $g$ for gauge, denotes the field-dependent squared mass matrices appearing in the potential and~$\mu$ denotes the renormalization scale.
Let $s_i$ denote the spin and $n_i$ the number of
spin degrees of freedom; then,
\begin{equation}
	n_i = (-1)^{2 s_i} (2 s_i + 1),\qquad \text{that is,}\qquad n_s = 1,\quad n_f = -2, \quad n_g = 3\,.
\end{equation}

The constants $C_i$ depend on the renormalization scheme. Here and in the following, we will work in the \msbar scheme where
\begin{equation}
    C_s = C_f = \frac{3}{2}, \qquad C_g = \frac{5}{6}\,.
\end{equation}

To express the derivatives of the effective potential, it is useful to define new quantities in a basis where the mass matrices $M_i^2$ ($i=s,f,g)$ are diagonal. For simplicity, we drop the index $i$ from the mass matrices, keeping in mind that the following results are equally valid 
in the scalar, fermion and gauge sectors. 
Let $N$ denote the number of states, that is, the order of the mass matrix squared $M^2$; we 
define\footnote{Note that the symmetric matrices $\overline{\partial_\mu M^2}$ and $\overline{\partial_\mu\partial_\nu M^2}$ are in general 
not diagonal.}
\begin{align}
    U M^2 U^\trans &= \overline{M^2} = \diag\left(\lambda_{1}, \cdots, \lambda_{N}\right) \equiv D\,, \label{eq:diagM}\\
    U (\partial_\mu M^2) U^\trans &\equiv \overline{\partial_\mu M^2}\,, \label{eq:diagdM}\\
    U (\partial_\mu \partial_\nu M^2) U^\trans &\equiv \overline{\partial_\mu\partial_\nu M^2} \label{eq:diagd2M}\,.
\end{align}
Recall that~$U=U\left(\phi_a\right)$ is an orthogonal matrix of dimension $N \times N$, the eigenvalues of $M^2$ are denoted by $\lambda_1, \ldots, \lambda_N$,
and the bilinear index $\mu \in \{0,\ldots,3\}$.
With the eigenvalues $\lambda_I$ defined, the one-loop contributions $V^{(1)}_i$ to the effective potential coming from scalar, gauge and fermion contributions ($i=s,f,g$, respectively) can be rewritten as
\begin{equation} \label{eq:V1eigs}
    V^{(1)}_i = \frac{n_i}{4} \sum_{I=1}^N \lambda_I^2 \left[\log\frac{\lambda_I}{\mu^2} - C_i\right]\,.
\end{equation}
In a next step, we express the derivatives of the diagonalized mass matrix as:
\begin{align}
    \partial_\mu D &= (\partial_\mu U) M^2 U^\trans + U M^2 (\partial_\mu U^\trans) + U \left(\partial_\mu M^2\right) U^\trans \nonumber\\
    &= \left((\partial_\mu U) U^\trans\right) D + D \left(U (\partial_\mu U^\trans)\right) + U \left(\partial_\mu M^2\right) U^\trans \nonumber\\
    &= \left[W_\mu, D\right] + \overline{\partial_\mu M^2} \label{eq:diagDerivatives}\,.
\end{align}
In the last equation, we have defined $W_\mu = (\partial_\mu U) U^\trans$ and used the fact that $W_\mu$ are skew-symmetric matrices, since
\begin{equation} \label{eq:Vskew}
    \partial_\mu \left( U U^\trans \right) = 0 = W_\mu + W^\trans_\mu\,.
\end{equation}
As a consequence of the skew-symmetry of $W_\mu$ and the diagonal form of $D$, $[W_\mu, D]$ is a symmetric matrix with vanishing components on the diagonal.
Restricting~\eqref{eq:diagDerivatives} to its diagonal components yields
\begin{equation} \label{eq:firstEigDerivative}
    \partial_\mu \lambda_I = \left(\overline{\partial_\mu M^2}\right)^{II},
\end{equation}
while evaluating its off diagonal components allows to express the elements of $W_\mu$ as
\begin{equation}
    W_\mu^{IJ} = \begin{cases}
        \frac{\left(\overline{\partial_\mu M^2}\right)^{IJ}}{\lambda_I - \lambda_J} & \text{if}\quad I \neq J\\
        0 & \text{if}\quad I = J
    \end{cases}\,.
\end{equation}
If degenerate eigenvalues are present in the spectrum, it is always possible to choose a basis (fixed by the choice of the orthogonal matrix $U$) in which the above equation generalizes to
\begin{equation} \label{eq:VIJ}
    W_\mu^{IJ} = \begin{cases}
        \frac{\left(\overline{\partial_\mu M^2}\right)^{IJ}}{\lambda_I - \lambda_J} & \text{if}\quad \lambda_I \neq \lambda_J\\
        0 & \text{if}\quad \lambda_I = \lambda_J
    \end{cases}\,.
\end{equation}
In appendix~\ref{app:eigenvaluesSecondDerivatives}, we show that reiterating a similar procedure for the second derivatives gives the following formula:
\begin{equation} \label{eq:secondEigDerivative}
    \partial_\mu \partial_\nu \lambda_I = \left(\overline{\partial_\mu \partial_\nu M^2}\right)^{II} + 2 \sum_{\lambda_I \neq \lambda_J} \frac{\left(\overline{\partial_\mu M^2}\right)^{IJ} \left(\overline{\partial_\nu M^2}\right)^{JI}}{\lambda_I - \lambda_J}\,.
\end{equation}
Equipped with~\eqref{eq:firstEigDerivative} and~\eqref{eq:secondEigDerivative} we can now express the first and second derivatives of the effective potential contributions $V_i$. Starting with the first derivatives, we have
\begin{align}
    \partial_\mu V^{(1)}_i &= \frac{n_i}{4} \sum_{I=1}^N 2 (\partial_\mu \lambda_I) \lambda_I \left[\log\left(\frac{\lambda_I}{\mu^2}\right) -\left (C_i - \frac{1}{2}\right)\right] \nonumber\\
    &= \frac{n_i}{2} \sum_{I=1}^N \left(\overline{\partial_\mu M^2}\right)^{II} A_i \left(\lambda_I\right) \label{eq:dV1eigs},
\end{align}
where the functions $A_i$ are given by
\begin{equation} \label{eq:Afunctions}
    A_s(x) = A_f(x) \equiv A(x) = x \left[\log\left(\frac{x}{\mu^2}\right) - 1\right], \qquad A_g(x) = x \left[\log\left(\frac{x}{\mu^2}\right) - \frac{1}{3}\right]\,.
\end{equation}
Turning to the second derivatives, we can write
\begin{align}
    \partial_\mu\partial_\nu V^{(1)}_i &= \frac{n_i}{2} 
    \left\{\sum_{I=1}^N  (\partial_\mu \partial_\nu \lambda_I) \lambda_I \left[\log \left(\frac{\lambda_I}{\mu^2}\right) -\left (C_i - \frac{1}{2}\right)\right] + 
    \right.
    \nonumber\\
    & \qquad \qquad \left.
     (\partial_\mu \lambda_I) (\partial_\nu \lambda_I) \left[\log\left(\frac{\lambda_I}{\mu^2}\right) -\left (C_i - \frac{3}{2}\right)\right]\right\} \nonumber\\
    &= \frac{n_i}{2} \left\{\sum_{I=1}^N \left(\overline{\partial_\mu \partial_\nu M^2}\right)^{II} A_i \left(\lambda_I\right) + 
    \sum_{I=1}^N \sum_{J=1}^N \left(\overline{\partial_\mu M^2}\right)^{IJ} \left(\overline{\partial_\nu M^2}\right)^{JI} B_i \left(\lambda_I, \lambda_J\right) \right\} \label{eq:d2V1eigs}\,,
\end{align}
where the functions $B_i$ are given by
\begin{align} \label{eq:Bfunctions}
    B_s(x,y) &= B_f(x,y) \equiv B(x,y) = \frac{A(x) - A(y)}{x-y}, \qquad B_g(x,y) = \frac{A_g(x) - A_g(y)}{x-y}\,.
\end{align}
In the above definitions, it is understood that $B_i(x,y)$ is analytically continued along the line where $x = y$, namely:
\begin{align}
    B(x,x) &= \lim\limits_{y \to x} \frac{A(x) - A(y)}{x-y} = \frac{d A}{d x}(x) = \log \left(\frac{x}{\mu^2}\right)\,, \label{eq:Blim}\\
    B_g(x,x) &= \lim\limits_{y \to x} \frac{A_g(x) - A_g(y)}{x-y} = \frac{d A_g}{d x}(x) = \log \left(\frac{x}{\mu^2}\right) + \frac{2}{3}\,. \label{eq:Bglim}
\end{align}
The formulae in~\eqref{eq:dV1eigs} and~\eqref{eq:d2V1eigs} are the central result of this section. We now apply these equations to express the derivatives of the gauge, fermion and scalar contributions to the effective potential. While the first derivatives of $V_i^{(1)}$ are needed to solve analytically the one-loop stationary point equations~\eqref{eq:kSPE1}, the second derivatives will be useful for expressing the one-loop corrections to the physical spectrum in the $\hbar$-expansion (see Sec.~\ref{sec:oneLoopMasses}).

\subsection{Gauge contributions}
\label{sec:oneLoopGauge}

We begin with the gauge contributions: the gauge boson mass matrix 
$M_g^2\left(\phi_a\right)$
can be diagonalized as
\begin{equation}
    M_g^2 = \diag\left(M_W^2, M_W^2, M_Z^2, M_\gamma^2\right),
\end{equation}
where the field-dependent eigenvalues are conveniently expressed in terms of the bilinear fields as
\begin{equation}\label{eq:gaugeEigs}
	M_W^2 = \frac{1}{2} K_0 \left(g_+^2 - g_-^2\right), \qquad M_{Z, \gamma}^2 = \frac{1}{2}\left[K_0 g_{+}^2 \pm \sqrt{ g_{+}^4 \tvec{K}^2 + (K_0^2-\tvec{K}^2)\,g_{-}^4}\,\right],
\end{equation}
with
\begin{equation}
	g_+^2 \equiv \frac{g_1^2 + g_2^2}{2} \quad \text{and} \quad g_-^2 \equiv \frac{g_1^2 - g_2^2}{2}\,.
\end{equation}
In the charge-conserving case, where 
$K_0^2 =\tvec{K}^2$, 
corresponding to correct electroweak-symmetry breaking $\matheweakgroup\rightarrow\mathemgroup$, the above eigenvalues simplify to
\begin{align}
    M_W^2 &= \frac{1}{2}  \left(g_+^2 - g_-^2\right) K_0 = \frac{1}{2} g_2^2 K_0, &
     M_Z^2 &= g_+^2 K_0 = \frac{1}{2}\left(g_1^2 + g_2^2\right)  K_0,
   \\ 
    M_\gamma^2  &= 0\,.
\end{align}
According to~\eqref{eq:V1eigs}, the gauge contributions $V^{(1)}_g\left(\phi_a,\mu\right)$ to the one-loop potential are given by
\begin{equation}
    V^{(1)}_g = \frac{3}{4}\left\{2 M_W^4 \left[\log\left(\frac{M_W^2}{\mu^2}\right) - \frac{5}{6}\right] + M_Z^4 \left[\log\left(\frac{M_Z^2}{\mu^2}\right) - \frac{5}{6}\right] \right\}\,.
\end{equation}
Using~\eqref{eq:dV1eigs} and~\eqref{eq:d2V1eigs}, 
the first and second derivatives of $V^{(1)}_g$ with respect to the bilinear fields
can be written in the concise form (recall that the functions $A_g$ and $B_g$ were defined in~\eqref{eq:Afunctions} and~\eqref{eq:Bfunctions}, respectively):
\begin{align}
    \partial_\mu V^{(1)}_g &= \frac{3}{2} \Big\{2 g_{\mu W W}\,  A_g\left(M_W^2\right) + g_{\mu Z Z}\,  A_g\left(M_Z^2\right)\Big\}\,, \label{eq:dV1g}\\
\begin{split}
    \partial_\mu \partial_\nu V^{(1)}_g &= \frac{3}{2} \Big\{g_{\mu \nu Z Z}\,  A_g\left(M_Z^2\right) + 2 g_{\mu W W} \, g_{\nu W W} \, B_g\left(M_W^2, M_W^2\right) \\
    &\qquad + g_{\mu Z Z} \, g_{\nu Z Z} \, B_g\left(M_Z^2, M_Z^2\right) + g_{\mu \gamma \gamma} \, g_{\nu \gamma \gamma} \, B_g\left(0, 0\right)\Big\}\,, \label{eq:d2V1g}
\end{split}
\end{align}
where we have defined
\begin{align}
    g_{\mu AB} &\equiv \left(\overline{\partial_\mu M_g^2}\right)^{AB} = \left(\partial_\mu M_g^2\right)^{AB}, \label{gmuAB}\\
    g_{\mu \nu AB} &\equiv \left(\overline{\partial_\mu \partial_\nu M_g^2}\right)^{AB} = \left(\partial_\mu \partial_\nu M_g^2\right)^{AB}\,.\label{gmunuAB}
\end{align}
Note that the orthogonal matrix $U$ in \eqref{eq:diagdM} and~\eqref{eq:diagd2M} is the identity matrix, as $M_g^2$ is already diagonal 
in the initial basis. This explains the equalities in equations \eqref{gmuAB} and \eqref{gmunuAB}.
It is straightforward to show that the only non-vanishing $g_{\mu A B}$ and $g_{\mu \nu A B}$ coefficients 
appearing in~\eqref{eq:dV1g} and~\eqref{eq:d2V1g} can be expressed as
\begin{align}
    g_{\mu W W} &= \frac{M_W^2}{K_0} \delta^{\mu 0}\,, \label{gmuWW} \\
    g_{\mu Z Z} & = \frac{M_Z^2}{2 K_0^2} \left[K^\mu + \cos^2(2\theta_W) \gK^\mu\right]\,,\label{gmuZZ}\\
    g_{\mu \gamma \gamma} &= \frac{M_Z^2}{2 K_0^2} \sin^2(2 \theta_W) \gK^\mu\,,\label{gmugg}\\
    \begin{split}
        g_{\mu \nu Z Z} &=\frac{M_Z^2}{2 K_0^2} \sin^2(2\theta_W) 
\bigg[ \cos^2(2\theta_W) \frac{\gK^\mu \gK^\nu}{K_0^2} \\
&+ (1-\delta^{\mu 0})(1-\delta^{\nu 0}) \big(\delta^{\mu \nu} - \frac{K^\mu K^\nu}{K_0^2}\big)
\bigg]\,,
    \end{split}\label{gmunuZZ}\\
    g_{\mu \nu \gamma \gamma} ={}& - g_{\mu \nu Z Z}\,,\label{gmunugg}
\end{align}
where $\theta_W$ is the Weinberg angle,
\begin{equation}
    \cos (\theta_W) = \frac{M_W}{M_Z}\,.
\end{equation}
In particular, note that $g_{\mu \nu W W}$ vanishes.

At this point, we comment on the last term of~\eqref{eq:d2V1g}, which arises from the scalar-photon-photon interactions.
In light of the definition of $B_g$ in~\eqref{eq:Bfunctions} and~\eqref{eq:Bglim}, the quantity $B_g(0, 0)$ is logarithmically 
divergent at a charge-conserving minimum of the scalar potential. 
However, this divergence is spurious, indicating that the matrix of second derivatives with respect to the bilinear fields 
is unphysical.
As shown below, when computing the contributions of gauge boson loops to the scalar mass matrix, such divergences will properly cancel as expected.

The coefficients $g_{\mu A B}$ and $g_{\mu \nu A B}$ are closely related to the tree-level scalar-vector-vector and scalar-scalar-vector-vector vertices, respectively, denoted $g_{i A B}$ and $g_{i j A B}$ in the following. 
From~\eqref{gmuAB}, respectively~\eqref{gmunuAB}, recalling the connection of the scalar fields~$\phi_i$ to the bilinears $K^\mu$,~\eqref{eq:defgamma}, we find
\begin{align}
    g_{i A B} &= \left(\partial_i M_g^2\right)^{AB} = \Gamma^{\mu}_i g_{\mu A B}, \label{giAB}\\
    g_{i j A B} &= \left(\partial_i \partial_j M_g^2\right)^{AB} = \Delta^\mu_{ij} g_{\mu A B} + 
    \Gamma^{\mu}_i \Gamma^{\nu}_j g_{\mu \nu A B}\,.\label{gijAB}
\end{align}
The rotation of the scalar indices $i,j$ to the basis where the scalar mass matrix is diagonal is performed 
with the matrices $\bar{U}$, see~\eqref{eq:Ubar} and~\eqref{Gamma_Delta_bar}.
This yields the tree-level couplings between the mass eigenstates:
\begin{equation}\label{eq:gaugeCouplingsMassBasis}
    \bar{g}_{i A B} = \bar{\Gamma}^{\mu}_i g_{\mu A B}, \qquad \bar{g}_{i j A B} = \bar{\Delta}^\mu_{ij} g_{\mu A B} + \bar{\Gamma}^{\mu}_i \bar{\Gamma}^{\nu}_j g_{\mu \nu A B}\,.
\end{equation}
These definitions allow us to express the derivatives of $V_g^{(1)}$ with respect to the neutral scalar fields in a concise form,
\begin{align}
    \bar{D}_a V^{(1)}_g &= \frac{3}{2} \Big\{2 \bar{g}_{a W W}\, A_g\left(M_W^2\right) + \bar{g}_{a Z Z}\,  A_g\left(M_Z^2\right)\Big\}\,, \label{eq:DV1g}\\
\begin{split}
    \bar{D}_a \bar{D}_b V^{(1)}_g &= \frac{3}{2} \Big\{\bar{g}_{a b Z Z}\,  A_g\left(M_Z^2\right) + 2 \bar{g}_{a W W} \, \bar{g}_{b W W} \, B_g\left(M_W^2, M_W^2\right) \\
    &\qquad + \bar{g}_{a Z Z} \, \bar{g}_{b Z Z} \, B_g\left(M_Z^2, M_Z^2\right)\Big\}\,, \label{eq:D2V1g}
\end{split}
\end{align}
where
\begin{align}
    && \bar{g}_{a W W} &= \sqrt{\frac{2}{K_0}} M_W^2 \bar{k}^a, & \bar{g}_{a Z Z} &= \sqrt{\frac{2}{K_0}} M_Z^2 \bar{k}^a\,, && \label{eq:gbaXX} \\
    && \bar{g}_{a b W W} &= \frac{M_W^2}{K_0} \delta^{a b}, & \bar{g}_{a b Z Z} &= \frac{M_Z^2}{K_0} \delta^{a b}\,. &&\label{eq:gbabXX}
\end{align}
Following the discussion around~\eqref{eq:oneLoopSPE} and~\eqref{eq:MneutralDab}, the $\bar{D}$-derivatives in~\eqref{eq:DV1g} and~\eqref{eq:D2V1g} 
represent the contributions of the gauge bosons to the one-loop stationary point equation and to the one-loop neutral scalar mass matrix, respectively. Finally, the quantity $\partial_0 V^{(1)}_g$ required for compute $u^{(1)}$, see~\eqref{eq:u1}, can be obtained from~\eqref{eq:dV1g}:
\begin{equation} \label{par0V1g}
    \partial_0 V^{(1)}_g = \frac{3}{2} \left\{2 g_{0 W W}\,  A_g\left(M_W^2\right) + g_{0 Z Z}\,  A_g\left(M_Z^2\right)\right\},
\end{equation}
with
\begin{equation} \label{g0WW}
    g_{0 W W} = \frac{M_W^2}{K_0}, \qquad 
    g_{0 Z Z} = \frac{M_Z^2}{2 K_0} (1+\cos^2(2 \theta_W))\,.
\end{equation}

\subsection{Fermionic contributions}
\label{sec:oneLoopFermion}

The Yukawa interaction terms involving the two Higgs doublets $\varphi_1$ and $\varphi_2$ can be written in the standard form,
\begin{align}
    -\mathcal{L}_Y = \Big[
    \overline{Q}_L\big( y_u\,\widetilde{\varphi}_1 + \epsilon_u\, \widetilde{\varphi}_2 \big) u_R
    + \overline{Q}_L\big( y_d\,\varphi_1 + \epsilon_d\, \varphi_2 \big) d_R
    + \overline{L}\big( y_e\,\varphi_1 + \epsilon_e\, \varphi_2 \big) e_R  
\Big] + \mathrm{h.c.}\,. \label{eq:yukLag}
\end{align}
As usual, $Q_L$ denotes the left-handed quark doublets and $L$ the left-handed lepton doublets,
$u_R$, $d_R$ are the right-handed up- and down-type quark singlets, and $e_R$ the 
right-handed leptons. The corresponding Yukawa coupling matrices are denoted by $y_u$, $y_d$, $y_e$, as well as
$\epsilon_u$, $\epsilon_d$, $\epsilon_e$.
The conjugate doublets $\widetilde{\varphi}_a$ are given, as usual, by
\begin{equation} \label{phiconj}
    \left(\widetilde{\varphi}_a\right)^i = \varepsilon^{ij} \left(\varphi^*_a\right)_j\,, \qquad \text{with} \qquad \varepsilon = i \sigma_2\,,
\end{equation}
where the indices $i$ and $j$ refer to the weak-isospin component of the doublets $\varphi^a$, ($a=1,2$).  
Note that the position of the family index $a$ is relevant, since  $\varphi^a$
transforms under the fundamental representation of $U(2)_H$ (the group describing unitary mixing of the two doublets), while $\varphi^*_a$ and $\widetilde{\varphi}_a$ transform under the anti-fundamental representation of the family symmetry group.\\

For each fermion species $f = u,d,e$, the Yukawa matrices $y_f$ and $\epsilon_f$ can be unified into a single object $\mathcal{F} = \mathcal{U}, \mathcal{D}, \mathcal{E}$, transforming under the (anti-)fundamental representation of the family group. Namely, we define
\begin{subequations}
    \begin{align}
        \mathcal{U}^a = \begin{pmatrix} y_u \\  \epsilon_u \end{pmatrix}\ \ &, \ \ \mathcal{U}^\dagger_a = \begin{pmatrix} y_u^\dagger & \epsilon_u^\dagger \end{pmatrix}\,,\\ 
        \mathcal{D}_a = \begin{pmatrix} y_d & \epsilon_d \end{pmatrix}\ \ &, \ \ {\mathcal{D}^\dagger}^a = \begin{pmatrix} y_d^\dagger \\ \epsilon_d^\dagger \end{pmatrix}\,,\\
        \mathcal{E}_a = \begin{pmatrix} y_e & \epsilon_e \end{pmatrix}\ \ &, \ \ {\mathcal{E}^\dagger}^a = \begin{pmatrix} y_e^\dagger \\ \epsilon_e^\dagger \end{pmatrix}\,,
    \end{align}
\end{subequations}
to rewrite the Yukawa Lagrangian~\eqref{eq:yukLag} in a more compact form
\begin{align}
    -\mathcal{L}_Y = \Big[
    \overline{Q}_L\;\mathcal{U}^a\,\widetilde{\varphi}_a\; u_R
    + \overline{Q}_L\;\mathcal{D}_a\,\varphi^a\; d_R
    +  \overline{L}\;\mathcal{E}_a\,\varphi^a\;e_R
\Big] + \mathrm{h.c.}\,.
\end{align}
This expression is
manifestly invariant under a change of basis in the Higgs family space, that is, a unitary transformation $U$ of the two doublets, provided
\begin{align}
    \varphi^a \rightarrow U^a_{\ \, b} \, \varphi^b \quad \Rightarrow \quad \mathcal{D}_a &\rightarrow \mathcal{D}_b \left(U^\dagger\right)^b_{\ a}, &  \mathcal{E}_a &\rightarrow \mathcal{E}_b \left(U^\dagger\right)^b_{\ a}, & \mathcal{U}^a &\rightarrow U^a_{\ \, b} \, \mathcal{U}^b\,.
\end{align}
Although the Yukawa interactions are linear in the scalar fields (and the Yukawa matrices), physically relevant quantities such as the fermion mass matrix squared, $M_F^2$, depend quadratically on them. For such quantities, the bilinear formalism can be naturally extended to the basis-dependent objects $\mathcal{D}$, $\mathcal{E}$ and $\mathcal{U}$. In analogy with the definition of the bilinear fields 
in~\eqref{eq:KbarToKM} which we repeat here for clarity, 
\begin{equation}
    \twomat{K}^a_{\,\ b} = \varphi_b^\dagger \varphi^a  = \frac{1}{2} K^\mu \left(\sigma_\mu\right)^a_{\,\ b}\,,
\end{equation}
we may write
\begin{equation} \label{eq:Yudef}
 {\mathcal{U}_b}^\dagger \mathcal{U}^a \equiv \frac{1}{2} Y_u^\mu \left(\sigma_\mu\right)^a_{\,\ b}\,,
\end{equation}
as the definition of the bilinear up-type Yukawa coupling $Y_u$. 
Similarly, the four-component bilinear Yukawa couplings $Y_d$ and $Y_e$ can be defined through
\begin{equation}  \label{eq:Ydedef}
    \mathcal{D}_b {\mathcal{D}^\dagger}^a = \frac{1}{2} Y_d^\mu \left(\sigma_\mu\right)^a_{\,\ b}\,, \qquad 
    \mathcal{E}_b {\mathcal{E}^\dagger}^a = \frac{1}{2} Y_e^\mu \left(\sigma_\mu\right)^a_{\,\ b}
\end{equation}
with
\begin{equation}
    \left(\mathcal{D}_a (\mathcal{D}^\dagger)^b\right)^{ik} = \mathcal{D}_a^{ij} \left(\mathcal{D}^*\right)^{b}_{kj} = \left(\mathcal{D}^*\right)^{b}_{kj} \mathcal{D}_a^{ij}\,.
\end{equation}

Contracting both sides of~\eqref{eq:Yudef} and ~\eqref{eq:Ydedef} with $\left(\sigma^\nu\right)^b_{\,\ a}$ 
yields the expressions for $Y_{u,d,e}$:
\begin{align}
    Y_u^\nu & = \left(\sigma^\nu\right)^b_{\,\ a} {\mathcal{U}_b}^\dagger \mathcal{U}^a = \mathcal{U}^\dagger \sigma^\nu \mathcal{U}\, ,
    \\
    Y_d^\nu & = \left(\sigma^\nu\right)^b_{\,\ a} {\mathcal{D}_b} {\mathcal{D}^a}^\dagger = \mathcal{D} \sigma^\nu \mathcal{D}^\dagger\, ,
  \\
    Y_e^\nu & = \left(\sigma^\nu\right)^b_{\,\ a} {\mathcal{E}_b} 
    {\mathcal{E}^a}^\dagger = \mathcal{E} \sigma^\nu \mathcal{E}^\dagger\, .
  \end{align}
Explicitly, the components of $Y_f$ ($f=u,d,e$) are given by the hermitian matrices
\begin{equation} \label{Ymat}
    Y_u = \begin{pmatrix}
        y_u^\dagger y_u + \epsilon_u^\dagger \epsilon_u \\
        y_u^\dagger \epsilon_u
        +
        \epsilon_u^\dagger y_u
        \\
        -i\left(y_u^\dagger \epsilon_u - \epsilon_u^\dagger y_u\right)\\
        y_u^\dagger y_u - \epsilon_u^\dagger \epsilon_u
    \end{pmatrix}, \quad
    Y_d = \begin{pmatrix}
        y_d y_d^\dagger + \epsilon_d \epsilon_d^\dagger\\
        y_d \epsilon_d^\dagger + \epsilon_d y_d^\dagger \\
        -i\left(y_d \epsilon_d^\dagger - \epsilon_d y_d^\dagger\right)\\
        y_d y_d^\dagger - \epsilon_d \epsilon_d^\dagger
    \end{pmatrix}, \quad
    Y_e = \begin{pmatrix}
        y_e y_e^\dagger + \epsilon_e \epsilon_e^\dagger\\
        y_e \epsilon_e^\dagger + \epsilon_e y_e^\dagger \\
        -i\left(y_e \epsilon_e^\dagger - \epsilon_e y_e^\dagger\right)\\
        y_e y_e^\dagger - \epsilon_e \epsilon_e^\dagger
    \end{pmatrix}\,.
\end{equation}

Under a change of basis, the four-component bilinear Yukawa couplings transform in the same way as $\KT$, namely (for $f = u,d,e$)
\begin{equation} \label{RYuk}
    Y_f^0 \rightarrow Y_f^0, \qquad Y_f^a \rightarrow \overline{Y}_f^a = R(U)^{ab} \,  Y_f^b, \ a=1,2,3
\end{equation}
where $R(U)$ is an orthogonal $3\times3$ matrix defined in~\eqref{eq:bilinearRtransfo}.\\

From these transformation properties, we may define in particular the Yukawa couplings $\bar{\mathcal{U}}$, $\bar{\mathcal{D}}$ and $\bar{\mathcal{E}}$ obtained after performing the change of basis which diagonalizes the neutral mass matrix, 
described in~\eqref{eq:diagonalBasisNeutral} and \eqref{bilbasis}. Similarly, the four-component bilinear Yukawa couplings in such a basis are denoted
\begin{equation}
    \bar{Y}_f = 
    \begin{pmatrix}
        \bar{Y}_f^0 \\
        \bar{\tvec{Y}}_f
    \end{pmatrix} =
       \begin{pmatrix}
        Y_f^0 \\
        \bar{\tvec{Y}}_f
    \end{pmatrix}
\end{equation}
for $f=u,d,e$.

With these preparations, the fermion squared mass matrices can now be expressed in terms of bilinears - see~\cite{Maniatis:2020wfz}. 

The resulting fermion mass matrices squared read
\begin{equation}
\begin{split}
&M_{u/d}^2=
\frac{1}{4}
\left[
Y_\mu^u K^\mu + Y_\mu^d K^\mu
\pm
\sqrt{
(Y_\mu^u K^\mu - Y_\mu^d K^\mu)^2
+ 4 |\xi_{ud}|^2 \widetilde{\tvec{K}}^\trans \tilde{g} \widetilde{\tvec{K}}}
\right],\\
&
M_e^2=
\frac{1}{2} K_\mu Y_e^\mu\,.
\end{split}
\end{equation}
At the charge-conserving minimum, this simplifies to
\begin{equation}
M_u^2  =
\frac{1}{2} K_\mu Y_u^\mu\,,\qquad
M_d^2 = \frac{1}{2} K_\mu Y_d^\mu\,,\qquad
M_e^2 = \frac{1}{2} K_\mu Y_e^\mu\,.
\end{equation}
Here we have defined
\begin{equation}
\label{xiudsq}
|\xi_{ud}|^2 =
\left(y_u \epsilon_d^\dagger - \epsilon_u y_d^\dagger\right)
\left(\epsilon_d y_u^\dagger -y_d \epsilon_u^\dagger \right)\,.
\end{equation}

Analogous to the gauge boson contributions, the fermionic contributions $V^{(1)}_f\left(\phi_a,\mu\right)$ to the one-loop potential and its bilinear-field derivatives can be expressed as, including only contributions from the third generation,
\begin{align}
    V^{(1)}_f &= -\left\{m_\tau^4 \left[\log\left(\frac{m_\tau^2}{\mu^2}\right) - \frac{3}{2}\right] + 3 m_t^4 \left[\log\left(\frac{m_t^2}{\mu^2}\right) - \frac{3}{2}\right] + 3 m_b^4 \left[\log\left(\frac{m_b^2}{\mu^2}\right) - \frac{3}{2}\right] \right\}\,, \label{eq:V1f}\\
    \partial_\mu V^{(1)}_f &= -2\, \Big\{y_{\mu \boldsymbol{\tau \tau}} A(m_\tau^2) + 3 y_{\mu \mathbf{t t}} A(m_t^2) + 3 y_{\mu \mathbf{b b}} A(m_b^2) \Big\}\,, \label{eq:dV1f}\\
\begin{split} \label{eq:d2V1f}
    \partial_\mu \partial_\nu V^{(1)}_f &= -2\, \Big\{3y_{\mu \nu \mathbf{t t}} A(m_t^2) + 3y_{\mu \nu \mathbf{b b}} A(m_b^2) \\
    &\qquad 
    + y_{\mu \boldsymbol{\tau \tau}} y_{\nu \boldsymbol{\tau \tau}} B(m_\tau^2, m_\tau^2)
    + 3y_{\mu \mathbf{t t}} y_{\nu \mathbf{t t}} B(m_t^2, m_t^2)  + 3y_{\mu \mathbf{b b}} y_{\nu \mathbf{b b}} B(m_b^2, m_b^2) \Big\}\,,
\end{split}
\end{align}
where the coefficients $y_{\mu \mathbf{f}_1 \mathbf{f}_2}$ and $y_{\mu \nu \mathbf{f}_1 \mathbf{f}_2}$ are obtained by taking the first and second bilinear-field derivatives of the corresponding entries in the fermion mass matrix squared. Those are given, in terms of bilinear fields, by
{\allowdisplaybreaks
\begin{align}
\begin{split}
    y_{\mu \boldsymbol{\tau \tau}} &= \frac{1}{2} Y_\tau^\mu, \\
    y_{\mu \boldsymbol{t t}} &= \frac{1}{2} \left(Y_t^\mu + \frac{\left|\xi_{ud}\right|^2}{m_t^2 - m_b^2} \gK^\mu\right), \qquad 
    y_{\mu \boldsymbol{b b}} = \frac{1}{2} \left(Y_b^\mu - \frac{\left|\xi_{ud}\right|^2}{m_t^2 - m_b^2} \gK^\mu\right)\,,\\
    y_{\mu \nu \boldsymbol{t t}} &= \frac{\left|\xi_{ud}\right|^2}{2 (m_t^2 - m_b^2)} g^{\mu\nu} 
    - \frac{\left|\xi_{ud}\right|^2}{4 (m_t^2 - m_b^2)^2} \left[ \gK^\mu \left(Y_u - Y_d\right)^\nu 
    + \left(Y_u - Y_d\right)^\mu \gK^\nu \right] \\
    & - \frac{\left|\xi_{ud}\right|^4}{2 (m_t^2 - m_b^2)^3} \gK^\mu \gK^\nu\,,\\
    y_{\mu \nu \boldsymbol{b b}} &= - y_{\mu \nu \boldsymbol{t t}}\,.
\end{split}
\end{align}
}%
Note that $y_{\mu\nu \boldsymbol{\tau \tau}}$ vanishes.
In complete analogy to~\eqref{eq:gaugeCouplingsMassBasis}, we define the scalar-fermion-fermion and scalar-scalar-fermion-fermion coefficients $\bar{y}$ in the mass basis as
\begin{equation}\label{eq:fermionCouplingsMassBasis}
    \bar{y}_{i \mathbf{f}_1 \mathbf{f}_2} = \bar{\Gamma}_{i}^{\mu} y_{\mu \mathbf{f}_1 \mathbf{f}_2}, \qquad \bar{y}_{i j \mathbf{f}_1 \mathbf{f}_2} =  \bar{\Delta}^\mu_{ij} y_{\mu \mathbf{f}_1 \mathbf{f}_2} + \bar{\Gamma}_{i}^{\mu} \bar{\Gamma}_{j}^{\nu} y_{\mu \nu \mathbf{f}_1 \mathbf{f}_2}\,.
\end{equation}
This allows us to compute the first and second derivatives of $V^{(1)}_f$ with respect to the neutral scalar fields:
\begin{align}
    \bar{D}_a V^{(1)}_f &= -2 \, \Big\{\bar{y}_{a \boldsymbol{\tau \tau}} A(m_\tau^2) + 3 \bar{y}_{a \mathbf{t t}} A(m_t^2) + 3 \bar{y}_{a \mathbf{b b}} A(m_b^2) \Big\} \label{eq:DV1f}\,,
    \\
%
\begin{split} \label{eq:D2V1f}
    \bar{D}_a \bar{D}_b V^{(1)}_f &= -2 \, \Big\{3\bar{y}_{a b \mathbf{t t}} A(m_t^2) + 3\bar{y}_{a b \mathbf{b b}} A(m_b^2) \\
    &\qquad + \bar{y}_{a \boldsymbol{\tau \tau}} \bar{y}_{b \boldsymbol{\tau \tau}} B(m_\tau^2, m_\tau^2)
    + 3\bar{y}_{a \mathbf{t t}} \bar{y}_{b \mathbf{t t}} B(m_t^2, m_t^2) + 3\bar{y}_{a \mathbf{b b}} \bar{y}_{b \mathbf{b b}} B(m_b^2, m_b^2) \Big\}\,,
\end{split}
\end{align}
with
{\allowdisplaybreaks
\begin{align}
\begin{split}
    \bar{y}_{a \mathbf{ff}} &= \sqrt{\frac{K_0}{2}} \left(Y_f^0 \bar{k}^a + \bar{Y}_f^a\right),  (f =\tau,b,t)\, ,
    \label{ybaxx}\\
    \bar{y}_{a b \mathbf{ff}} &= \frac{1}{2 K_0} \left(Y_f^\trans \gT \KT\right) \delta^{ab} + \frac{1}{2}\left(\bar{k}^a \bar{Y}_f^b + \bar{Y}_f^a \bar{k}^b \right),  (f=b,t)\,.
\end{split}
\end{align}
}%
Finally, as required to compute $u^{(1)}$, the quantity $\partial_0 V^{(1)}_f$ is directly obtained from~\eqref{eq:dV1f} and can be put in the form
\begin{equation} \label{eq:d0V1f}
    \partial_0 V^{(1)}_f = -\left\{Y_\tau^0 A(m_\tau^2) + 3 Y_t^0 A(m_t^2) + 3 Y_b^0 A(m_b^2) + 3 K_0 \left|\xi_{ud}\right|^2 B(m_t^2, m_b^2) \right\}\,.
\end{equation}
    \subsection{Scalar contributions}
\label{sec:oneLoopScalar}

We now turn to the scalar contributions to the effective potential and follow the procedure from the previous two sections to derive expressions for $\partial_0 V_s^{(1)}$, $\bar{D}_a V_s^{(1)}$ and $\bar{D}_a \bar{D}_b V_s^{(1)}$. Recall that these quantities are needed to (i)~express $u^{(1)}$, (ii)~solve the one-loop stationary point equations and (iii)~compute the one-loop corrections to the neutral scalar masses, respectively. As explained below, the computation of the bilinear field derivatives of the scalar mass matrix (and, in turn, of $\partial_0 V_s^{(1)}$) is more involved than for the gauge and fermion sectors. On the other hand, the computation of $\bar{D}_a V_s^{(1)}$ and $\bar{D}_a \bar{D}_b V_s^{(1)}$ is rather straightforward since one only needs in practice to evaluate the derivatives of $V_s^{(1)}$ with respect to the neutral field components. We therefore begin with the evaluation of the $\bar{D}$-derivatives.\\

The structure of the mass matrix at a charge-conserving minimum has been discussed in Sec.~\ref{sec:gaugeInvariantFormalism}. Following these derivations and using \eqref{eq:V1eigs}, $V_s^{(1)}\left(\phi_a,\mu\right)$ can be written as
\begin{equation}
    V^{(1)}_s = \frac{1}{4}\left\{2 m_{H^\pm}^4 \left[\log\left(\frac{m_{H^\pm}^2}{\mu^2}\right) - \frac{3}{2}\right] + \sum_{a=1}^3 m_a^4 \left[\log\left(\frac{m_a^2}{\mu^2}\right) - \frac{3}{2}\right]\right\}\,.
\end{equation}
Then, using~\eqref{eq:dV1eigs}, we may express the first derivatives of $V_s^{(1)}$ with respect to the scalar fields $\phi^i$ ($i=1,\dots,8$) as
\begin{equation} \label{eq:dV1sGeneral}
    \partial_i V_s^{(1)} = \frac{1}{2} \sum_{j=1}^8 \left(\overline{\partial_i M_s^2}\right)_{jj} A\left(m_j^2\right)\,,
\end{equation}
where we remind the reader that the notation $\overline{\partial_i M_s^2}$ indicates that the derivatives of the scalar mass matrix have been rotated to the diagonal basis of $M_s^2$. Using the formalism introduced in Sec.~\ref{sec:gaugeInvariantFormalism}, we may write
\begin{align}
    \left(\partial_i M_s^2\right)_{jk} \equiv \lambda_{ijk} &= \partial_i \left(\Delta_{jk}^\mu \partial_\mu V^{(0)} + \Gamma^{\mu}_j \mathcal{M}_{\mu\nu} \Gamma^{\nu}_k\right)\\
    &= \left(\Delta_{ij}^\mu \Gamma^\nu_k + \Delta_{ik}^\mu \Gamma^\nu_j + \Delta_{jk}^\mu \Gamma^\nu_i\right) \mathcal{M}_{\mu\nu}\,.
\end{align}

After rotation to the diagonal basis, the above formula translates to

\begin{equation} \label{eq:lambdabijk}
    \bar{\lambda}_{ijk} = \left(\bar{\Delta}^\mu_{ij} \bar{\Gamma}_k^\nu + \bar{\Delta}^\mu_{ik} \bar{\Gamma}_j^\nu + \bar{\Delta}^\mu_{jk} \bar{\Gamma}_i^\nu\right) \mathcal{M}_{\mu \nu}\,,
\end{equation}
where the matrices $\bar{\Gamma}$ and $\bar{\Delta}^\mu$ have been defined in~\eqref{Gamma_Delta_bar}. The tensor $\bar{\lambda}_{ijk}$ defined in this way is the rank-3 tensor of tree-level cubic couplings among
the physical scalar states. All components of this tensor can be expressed in terms of gauge-invariant quantities within our formalism. The explicit form of $ \bar{\lambda}_{ijk}$ is given in Sec.~\ref{sec:hbar_nuts}.\\

Throughout this section, we adopt a notation where the eight scalar field components in the diagonal basis are rewritten as
\begin{equation}
    \bar{\phi} = \begin{pmatrix}
        G^0, & G^\pm_1, & G^\pm_2, H^\pm_1, H^\pm_2, H_1, & H_2, & H_3
    \end{pmatrix}\,.
\end{equation}
For readability, the indices $a,b,c,d \in \{1,2,3\}$ will be used as shorthand for the neutral components 
$H^a, H^b, H^c$ and $H^d$. 
Specializing~\eqref{eq:dV1sGeneral} to the neutral Higgs sector, we may finally write
(using $m_{H_1^\pm}=m_{H_2^\pm}=m_{H^\pm}$)
\begin{equation}
    \bar{D}_a V^{(1)}_s = \frac{1}{2} \left\{ \sum_{b=1}^3 \bar{\lambda}_{abb} A\left(m_b^2\right) + \sum_{p=1}^2 \bar{\lambda}_{a H^\pm_p H^\pm_p} A\left(m_{H^\pm}^2\right) \right\}\,,
\end{equation}
where the cubic couplings $\bar{\lambda}_{abb}$ and $\bar{\lambda}_{a H^\pm_p H^\pm_p}$ can be 
directly read from~\eqref{eq:lambdabijk}, respectively~\eqref{eq:scalarcouplings}. The above expression can be rewritten in the slightly more compact form
\begin{equation} \label{barDVs1}
    \bar{D}_a V^{(1)}_s = \frac{1}{2} \left\{ \sum_{b=1}^3 \bar{\lambda}_{abb} A\left(m_b^2\right) + \bar{\lambda}_{a H^\pm H^\pm} A\left(m_{H^\pm}^2\right) \right\}\,,
\end{equation}
where we have defined
\begin{equation}
    \bar{\lambda}_{a H^\pm H^\pm} = \sum_{p=1}^2 \bar{\lambda}_{a H^\pm_p H^\pm_p}\,.
\end{equation}
The above framework straightforwardly extends to the computation of the second neutral-scalar derivatives of $V_s^{(1)}$. In the diagonal basis, the quartic couplings can be computed using
\begin{equation}
    \bar{\lambda}_{ijkl} = \left(\bar{\Delta}^\mu_{ij} \bar{\Delta}^\nu_{kl} + \bar{\Delta}^\mu_{ik} \bar{\Delta}^\nu_{jl} + \bar{\Delta}^\mu_{il} \bar{\Delta}^\nu_{jk}\right) \mathcal{M}_{\mu \nu}\,,
\end{equation}
and, applying~\eqref{eq:d2V1eigs}, we express $\bar{D}_a \bar{D}_b V^{(1)}_s$ as
%
%
%
\begin{align}
\begin{split} \label{eq:D2V1s}
    \bar{D}_a \bar{D}_b V^{(1)}_s = \frac{1}{2} \Bigg\{ &\sum_{c=1}^3 \bar{\lambda}_{abcc} A\left(m_c^2\right) + \bar{\lambda}_{a b H^\pm H^\pm} A\left(m_{H^\pm}^2\right) \\
    +{}& \sum_{c,d=1}^3 \bar{\lambda}_{acd} \bar{\lambda}_{bcd} B\left(m_c^2, m_d^2\right) + \bar{\lambda}_{a H^\pm H^\pm} \bar{\lambda}_{b H^\pm H^\pm} B\left(m_{H^\pm}^2, m_{H^\pm}^2\right) \\
    +{}& 2 \sum_{c=1}^3 \bar{\lambda}_{ac G^0} \bar{\lambda}_{bc G^0} B\left(m_c^2, 0\right) 
    + \bar{\lambda}_{a H^\pm G^\pm} \bar{\lambda}_{b G^\pm H^\pm} B\left(m_{H^\pm}^2, 0\right)\\
    +{}& \bar{\lambda}_{a G^0 G^0} \bar{\lambda}_{b G^0 G^0} B\left(0, 0\right) + \bar{\lambda}_{a G^\pm G^\pm} \bar{\lambda}_{b G^\pm G^\pm} B\left(0, 0\right)  \Bigg\}\,,
\end{split}
\end{align}
where we have defined
\begin{align}
\begin{split}
    \bar{\lambda}_{a b H^\pm H^\pm} &= \sum_{p=1}^2 \bar{\lambda}_{a b H^\pm_p H^\pm_p}\,,\\
    \bar{\lambda}_{a H^\pm H^\pm} \bar{\lambda}_{b H^\pm H^\pm} &= 
    \sum_{p,q=1}^2 \bar{\lambda}_{a H^\pm_p H^\pm_q} \bar{\lambda}_{b H^\pm_q H^\pm_p}\,,\\
    \bar{\lambda}_{a H^\pm G^\pm} \bar{\lambda}_{b G^\pm H^\pm} &= 
    \sum_{p,q=1}^2 \bar{\lambda}_{a H^\pm_p G^\pm_q} \bar{\lambda}_{b G^\pm_q H^\pm_p}\,,\\
    \bar{\lambda}_{a G^\pm G^\pm} \bar{\lambda}_{b G^\pm G^\pm} &= 
    \sum_{p,q=1}^2 \bar{\lambda}_{a G^\pm_p G^\pm_q} \bar{\lambda}_{b G^\pm_q G^\pm_p}\,.
\end{split}
\end{align}
All cubic and quartic couplings appearing in the above formulae are given in Sec.~\ref{sec:hbar_nuts} in terms of gauge-invariant quantities. From~\eqref{eq:D2V1s}, we observe that the contributions arising from Goldstone-boson loops are proportional to $B(0,0)$, which is a logarithmically divergent and hence an ill-defined quantity. This divergence is spurious and indicates that the matrix of second field-derivatives of the effective potential is not, by itself, a physical quantity. As explained in more detail in the next section, these divergent contributions cancel exactly in the computation of the pole -- \textit{i.e.}~physical -- masses.\\

The last remaining quantity to be derived is the bilinear-field derivative $\partial_0 V^{(1)}_s$, needed to compute the one-loop correction to the parameter $u$ employing \eqref{eq:u1}.
More generally, $\partial_\mu V^{(1)}_s$ is obtained 
through~\eqref{eq:dV1eigs},
\begin{align}
    \partial_\mu V^{(1)}_s &= \frac{1}{2} \sum_{i=1}^8 \left(\overline{\partial_\mu M_s^2}\right)^{ii} A\left(m_i^2\right) \nonumber\\
    &= \frac{1}{2} \left\{ \sum_{a=1}^3 \left(\overline{\partial_\mu M_s^2}\right)^{aa} A\left(m_a^2\right) + \sum_{p=1}^2 \left(\overline{\partial_\mu M_s^2}\right)^{H^\pm_p H^\pm_p} A\left(m_{H^\pm}^2\right) \right\}\,. \label{eq:dmuVs2}
\end{align}
Defining 
\begin{equation} \label{lambda0}
\bar{\lambda}_{0aa} \equiv \left(\overline{\partial_0 M_s^2}\right)^{aa}, \qquad
\bar{\lambda}_{0H^\pm H^\pm} \equiv
\sum_{p=1}^2 \left(\overline{\partial_0 M_s^2}\right)^{H^\pm_p H^\pm_p},
\end{equation}
we write the derivative with respect to $K_0$ as
\begin{equation} \label{d0V1s}
\partial_0 V^{(1)}_s =
\frac{1}{2} \left\{ \sum_{a=1}^3 \bar{\lambda}_{0aa} A\left(m_a^2\right) + 
\bar{\lambda}_{0H^\pm H^\pm} A\left(m_{H^\pm}^2\right) \right\}\,.
\end{equation}
The results for the couplings $\bar{\lambda}_{0aa}$ and $\bar{\lambda}_{0H^\pm H^\pm}$ 
are given in equation \eqref{lambda0app} in App.~\ref{app:firstKder}.

\section{One-loop physical scalar masses}
\label{sec:oneLoopMasses}

Having derived all the necessary formulae to solve the one-loop stationary point equations, we now compute the corresponding corrections to the scalar spectrum. Generally speaking, in the $\hbar$-expansion framework, the most consistent way to obtain loop corrections to physical observables is to perform the $\hbar$-expansion of their all-order expressions. Applying this method to the scalar spectrum, the first step is to express the all-order scalar mass matrix after rotation to the diagonal basis,
\begin{equation}
    \bar{M}_s^2 = \bar{\Delta}^\mu \partial_\mu V + \bar{\Gamma} \mathcal{M} \bar{\Gamma}^\trans = \diag\begin{pmatrix}
        0, & 0, & 0, & m_{H^\pm}^2, & m_{H^\pm}^2, & m_1^2, & m_2^2, & m_3^2
    \end{pmatrix}\,.
\end{equation}
The main advantage of this method is that, even if all relevant quantities are obtained from a Taylor expansion around the tree-level minimum, the diagonal structure of the mass matrix is enforced order-by-order. This in particular prevents the appearance of massive Goldstone bosons, as generally expected when evaluating the scalar mass matrix away from the true minimum of the potential. Similarly, unphysical mixing between, \textit{e.g.}~physical and Goldstone states is thereby avoided. However, as observed in the previous section, evaluating the second-derivatives of the quantum potential at the tree-level vacuum inevitably introduces infrared divergences in the Goldstone sector. Of course, these divergences are spurious and should not affect the physical observables. While a consistent $\hbar$-expansion of the scalar mass matrix cannot prevent the occurrence of such divergences, we will explicitly show at the end of this section that they naturally cancel when switching from the \msbar to the on-shell scheme.\\

First focusing on the charged Higgs-boson mass, we have shown in Sec.~\ref{sec:gaugeInvariantFormalism} that, at any order in perturbation theory, $m_{H^\pm}^2$ is simply given by
\begin{equation}
    m_{H^\pm}^2 = 4 u K_0\,.
\end{equation}
An $\hbar$-expansion of this expression to first order yields
\begin{equation} \label{eq:mHp1aux}
    m_{H^\pm}^2 = 4 u^{(0)} K_0^{(0)} + \kappa \left[4 u^{(1)} K_0^{(0)} + 4 u^{(0)} K_0^{(1)} \right] + \mathcal{O}\left(\kappa^2\right)\,.
\end{equation}
While $u^{(1)}$ is directly given by~\eqref{eq:u1}, $K_0^{(1)}$ can be computed using~\eqref{eq:kgk}:
\begin{equation} \label{K10}
    K^{(1)}_0 = \frac{\tvec{K}^{(0)\,\trans} \tvec{K}^{(1)}}{K_0^{(0)}} = K_0^{(0)} \tvec{k}^{(0)\,\trans} \tvec{k}^{(1)}\,,
\end{equation}
where $\tvec{K}^{(1)}$ follows from the one-loop stationary point equation~\eqref{eq:oneLoopSPE} and where $\tvec{k}^{(1)} = \tvec{K}^{(1)}/K^{(0)}_0$. This yields the first key result of this section:
\begin{align}
\begin{split}
    m_{H^\pm}^2 &= \left(m_{H^\pm}^2\right)^{(0)} + \kappa \left(m_{H^\pm}^2\right)^{(1)} + \mathcal{O}\left(\kappa^2\right)\,,\\
    \left(m_{H^\pm}^2\right)^{(0)} &= 4 u^{(0)} K_0^{(0)}\,,\\
    \left(m_{H^\pm}^2\right)^{(1)} &= 4 \left[u^{(1)} + \tvec{k}^{(0)\,\trans}\tvec{k}^{(1)} u^{(0)}\right] K_0^{(0)}\,.
\end{split}
\end{align}

Next, we consider the neutral sector. The neutral mass matrix in the canonical basis is given by
\begin{align}
    \left(\widehat{\mathcal{M}}_\mathrm{neutral}^2\right)^{ab} &= D_a D_b V \label{eq:canonicalMneu1}\\
    &= \left[\widehat{\Delta}^\mu \partial_\mu V + \widehat{\Gamma}\mathcal{M}\widehat{\Gamma}^\trans\right]^{a b} \label{eq:canonicalMneu2}\\
    &= \left[\gamma_3\left(\mathcal{M} - 2 u \gT\right) \gamma_3^\trans\right]^{ab}\,, \label{eq:canonicalMneu3}
\end{align}
with $\mathcal{M}_{\mu\nu} = \partial_\mu \partial_\nu V$. As exemplified in the previous sections, multiple strategies can be followed to derive the various contributions to $\widehat{\mathcal{M}}_\mathrm{neutral}^2$ in terms of gauge-invariant quantities. At tree-level, \eqref{eq:canonicalMneu3} was used in Sec.~\ref{sec:gaugeInvariantFormalism} to obtain a basis-independent formula for the neutral mass matrix. 
At one loop order, the gauge and fermion contributions have been computed in Sec.~\ref{sec:oneLoopGauge}, \ref{sec:oneLoopFermion}, and \ref{sec:oneLoopScalar} using~\eqref{eq:canonicalMneu2} where $\partial_\mu V^{(1)}_{g,f}$ and $\partial_\mu \partial_\nu V^{(1)}_{g,f}$ were obtained from the diagonalized gauge boson, fermion, and scalar mass matrices. 

The all-order diagonalized mass matrix $\bar{\mathcal{M}}_\mathrm{neutral}^2$ is obtained from
\begin{equation}
    \bar{\mathcal{M}}_\mathrm{neutral}^2 = R \widehat{\mathcal{M}}_\mathrm{neutral}^2 R^\trans\,,
\end{equation}
with $R$ an orthogonal $3\times3$ matrix. Both $\widehat{\mathcal{M}}_\mathrm{neutral}^2$ and $R$ can be expanded perturbatively:
\begin{align}
\begin{split}
    \widehat{\mathcal{M}}_\mathrm{neutral}^2\big(\KT\big) ={}& \left(\widehat{\mathcal{M}}_\mathrm{neutral}^2\right)^{(0)}\big(\KT^{(0)}\big) \\
    &+ \kappa \left[\left(\widehat{\mathcal{M}}_\mathrm{neutral}^2\right)^{(1)}\big(\KT^{(0)}\big) + K^{(1)}_\mu \partial_\mu \left(\widehat{\mathcal{M}}_\mathrm{neutral}^2\right)^{(0)}\big(\KT^{(0)}\big) \right]\\
    &+ \mathcal{O}\left(\kappa^2\right)\, ,
\end{split}\\
R ={}& \left(\unitmatrix_3 + \kappa \, T^{(1)}\right) R^{(0)} + \mathcal{O}\left(\kappa^2\right)\,.
\end{align}
In the perturbative expansion of $R$, we have conveniently factored out the orthogonal matrix $R^{(0)}$ which diagonalizes the tree-level neutral mass matrix. Requiring the orthogonality condition $R R^\trans = \unitmatrix_3$ to hold order-by-order makes $T^{(1)}$ a skew-symmetric matrix. Further defining
\begin{align}
    \left(\bar{\mathcal{M}}_\mathrm{neutral}^2\right)^{(0)} = R^{(0)} \left(\widehat{\mathcal{M}}_\mathrm{neutral}^2\right)^{(0)}R^{(0)\,\trans}\,,\\
    \left(\bar{\mathcal{M}}_\mathrm{neutral}^2\right)^{(1)} = R^{(0)} \left(\widehat{\mathcal{M}}_\mathrm{neutral}^2\right)^{(1)}R^{(0)\,\trans}\,,
\end{align}
and
\begin{equation}
    \delta^{(1)} \bar{\mathcal{M}}_\mathrm{neutral}^2 = K^{(1)}_\mu R^{(0)} \partial_\mu \left(\widehat{\mathcal{M}}_\mathrm{neutral}^2\right)^{(0)}R^{(0)\,\trans}\,,
\end{equation}
the $\hbar$-expansion of the all-order diagonalized mass matrix $\bar{\mathcal{M}}_\mathrm{neutral}^2$ reads (all quantities in the right-hand side are understood to be evaluated at the tree-level minimum $\KT^{(0)}$)
\begin{align}
\begin{split} \label{eq:hbarMneutral}
    \bar{\mathcal{M}}_\mathrm{neutral}^2 = \left(\bar{\mathcal{M}}_\mathrm{neutral}^2\right)^{(0)} + \kappa &\bigg\{\left(\bar{\mathcal{M}}_\mathrm{neutral}^2\right)^{(1)} + \delta^{(1)} \bar{\mathcal{M}}_\mathrm{neutral}^2 + \left[T^{(1)}, \left(\bar{\mathcal{M}}_\mathrm{neutral}^2\right)^{(0)}\right]\bigg\}\,.
\end{split}
\end{align}
The tree-level mas matrix $\left(\bar{\mathcal{M}}_\mathrm{neutral}^2\right)^{(0)}$ is directly obtained from~\eqref{eq:MneutralTreeLevel}, 
\begin{align}
    \left(\bar{\mathcal{M}}_\mathrm{neutral}^2\right)^{(0)} &= 4 K_0^{(0)} \left[ \big(\eta_{00} - u^{(0)}\big) \bar{\tvec{k}}^{(0)} \bar{\tvec{k}}^{(0)\,\trans} + \bar{\tvec{\eta}} \bar{\tvec{k}}^{(0)\,\trans} + \bar{\tvec{k}}^{(0)} \bar{\tvec{\eta}}^\trans + \left(\bar{E} + u^{(0)} \unitmatrix_3\right)\right] \nonumber\\
    &= \diag\begin{pmatrix} \left(m_1^2\right)^{(0)}, & \left(m_2^2\right)^{(0)}, & \left(m_3^2\right)^{(0)}\end{pmatrix}\,,
\end{align}
and, in turn, providing an expression for the shift $\delta^{(1)} \bar{\mathcal{M}}_\mathrm{neutral}^2$ is straightforward. First defining
\begin{equation} \label{delta1}
    \bar{\tvec{\delta}}^{(1)} \equiv \bar{\tvec{k}}^{(1)} - \frac{K^{(1)}_0}{K^{(0)}_0} \bar{\tvec{k}}^{(0)}\,,
\end{equation}
and using the definition of $\tvec{f}_\pm$ in~\eqref{eq:fpmdef}, we may write
\begin{equation}
    \delta^{(1)} \bar{\mathcal{M}}_\mathrm{neutral}^2 = \frac{K^{(1)}_0}{K^{(0)}_0} \left(\bar{\mathcal{M}}_\mathrm{neutral}^2\right)^{(0)} + \frac{1}{2} \left(\bar{\tvec{f}}_\pm^\trans \bar{\tvec{\delta}}^{(1)} + \bar{\tvec{\delta}}^{(1)\,\trans} \bar{\tvec{f}}_\pm\right)\,.
\end{equation}
The contributions from the one-loop potential read
\begin{equation}
    \left(\bar{\mathcal{M}}_\mathrm{neutral}^2\right)^{(1)} = \bar{D}_a \bar{D}_b V^{(1)} = \bar{D}_a \bar{D}_b V^{(1)}_g + \bar{D}_a \bar{D}_b V^{(1)}_f + \bar{D}_a \bar{D}_b V^{(1)}_s\,,
\end{equation}
with $\bar{D}_a \bar{D}_b V^{(1)}_{g,f,s}$ given in~\eqref{eq:D2V1g}, \eqref{eq:D2V1f} and~\eqref{eq:D2V1s}, respectively. The entries of the skew-symmetric matrix $T^{(1)}$ are chosen such that the diagonal structure of the mass matrix is preserved order-by-order, yielding for $a \neq b$
\begin{equation} \label{eq:T1expr}
    \left(T^{(1)}\right)^{ab} = \frac{1}{m_a^2 - m_b^2}\left[\bar{D}_a \bar{D}_b V^{(1)} + \frac{1}{2} \left(\bar{f}_\pm^a \bar{\delta}^{(1),\,b} + \bar{f}_\pm^b \bar{\delta}^{(1),\,a}\right)\right]\,.
\end{equation}
Finally, the one-loop corrected neutral masses are simply obtained from the diagonal elements of~\eqref{eq:hbarMneutral} to which the commutator involving $T^{(1)}$ does not contribute:
\begin{equation} \label{scalarmasses}
\begin{split}
    m_a^2 &= \left(m_a^2\right)^{(0)} + \kappa \left(m_a^2\right)^{(1)} + \mathcal{O}\left(\kappa^2\right)\,,\\
    \left(m_a^2\right)^{(1)} &= \bar{D}_a \bar{D}_a V^{(1)} + \frac{K^{(1)}_0}{K^{(0)}_0} \left(m_a^2\right)^{(0)} + \bar{f}_\pm^a \bar{\delta}^{(1),\,a}\,.
\end{split}
\end{equation}
Before closing this section, we comment on the infrared divergences occurring in the one-loop mass spectrum through the Goldstone contributions in $\bar{D}_a\bar{D}_b V^{(1)}_s$. We first recall 
(see \textit{e.g.}~\cite{Denner:1991kt}) that the, all order, scalar pole masses are obtained by solving the secular equation
\begin{equation} \label{eq:poleSecular}
    \det \left[M_s^2 + \Pi(p^2) - \Pi(0) - p^2 \unitmatrix \right] = 0\,,
\end{equation}
where $M_s^2$ stands for the matrix of second field-derivatives of the effective potential and $\Pi(p^2)$ is the matrix of scalar self-energies evaluated at momentum $p^2$. To any order in perturbation theory, the zero-momentum self-energies actually coincide with the second derivatives of the effective potential,
\begin{equation}
    \left[\Pi(0)\right]^{ab} = \left(M_s^2\right)^{ab} = \partial_a \partial_b V\,,
\end{equation}
resulting in a cancellation in the secular equation~\eqref{eq:poleSecular}. As we will now explicitly show, this cancellation ensures that any infrared-divergent Goldstone contribution appearing in $M_s^2$ will drop out of the computations when going to the on-shell scheme. Adapting the results in \cite{Denner:1991kt},
the scalar-loop contributions to the one-loop scalar self-energy matrix are given in the diagonal basis by
\begin{equation}
    \left[\bar{\Pi}^{(1)}_s(p^2)\right]^{ij} = \frac{1}{2} \Big\{ \bar{\lambda}_{ijkk} A\big(m_k^2\big) + \bar{\lambda}_{ikl}\bar{\lambda}_{jkl} B\big(p^2, m_l^2, m_k^2\big)\Big\}\,,
\end{equation}
where $i,j$ span the eight scalar field components. The function $A(x)$ was defined in~\eqref{eq:Afunctions} while $B(p^2, x, y)$ is given is the \msbar scheme by the regularized loop integral
\begin{equation}
    B\big(p^2, x,y\big) = \int_0^1 \log\left(\frac{x t + y (1-t) + p^2 t (1-t)}{\mu^2}\right)\, dt\,.
\end{equation}
The two-argument function $B(x,y)$ previously defined in~\eqref{eq:Bfunctions} corresponds in fact to the zero-momentum limit of the function $B\left(p^2, x, y\right)$:
\begin{equation}
    B\big(0, x,y\big) = \int_0^1 \log\left(\frac{x t + y (1-t)}{\mu^2}\right)\, dt = \frac{A(x) - A(y)}{x-y} = B(x,y)\,.
\end{equation}
As mentioned before, massless Goldstone-boson loops generate infrared divergences at zero-momentum. On the other hand, $B(p^2, 0, 0)$ is finite and well-behaved, since
\begin{equation}
    B\big(p^2, 0, 0\big) = \int_0^1 \log\left(\frac{p^2 t (1-t)}{\mu^2}\right)\, dt = \log \left( \frac{p^2}{\mu^2} \right) - 2\,.
\end{equation}
Recall that, according to~\eqref{eq:poleSecular}, any momentum-dependent function entering the expression of the scalar self-energies is in fact amputated from its zero-momentum limit when computing the pole masses through the secular equation. Therefore, any infrared divergence in the Goldstone sector appearing in the \msbar scalar mass matrix will cancel in~\eqref{eq:poleSecular}. The pole masses thus obtained are, as expected, free from such divergences. 

\section{The $\hbar$ expansion in a nutshell}
\label{sec:hbar_nuts}

Here, we collect all expressions needed to study the $\hbar$ expansion in a general THDM. 
An explicit example shall be presented in Sec.~\ref{sec:application}.
The steps to compute the $\hbar$ expansion are summarized as follows: first, express the THDM in terms of bilinears as outlined in Sec.~\ref{sec:bilinear}. In particular, the tree-level potential is written in~\eqref{eq:pot} with parameters~\eqref{eq:para3}. Equivalently, the potential has been given in~\eqref{eq:potK4} with parameters~\eqref{eq:para4} in four-vector bilinear notation. The tree-level charged Higgs-boson masses are then computed by~\eqref{eq:allOrderMcharged}, or~\eqref{mHpm} and the neutral squared mass matrix~\eqref{eq:allOrderMneutral}, or explicitly by~\eqref{eq:MneutralTreeLevel}. The neutral squared mass matrix can be diagonalized by the similarity transformation~\eqref{eq:diagonalBasisNeutral} with an orthogonal matrix $R$. We thus have at this stage the complete scalar spectrum determined, that is, $m_{H^\pm}^{(0)}$ as well as the three neutral scalars~$(m_a^2)^{(0)}$ with the superscript added here to indicate the perturbation level, that is, tree level. 
We recall in this context that the diagonalization corresponds to a basis change, in terms of the bilinear vector $\tvec{K}$ with rotation matrix~$R$; see~\eqref{bilbasis}.
In the basis, where the scalar squared mass matrix is diagonal, all basis-dependent expressions are denoted with a bar symbol.
The tree level bilinear vector in the diagonal basis reads therefore $\bar{K}^{(0)}_\mu = R_{\mu \nu} K^{(0)}_\nu$ and
the dimensionless bilinear in the diagonal basis is~$\bar{\tvec{k}}^{(0)} = \bar{\tvec{K}}^{(0)}/K_0^{(0)}$.
In this diagonal basis the expansion of the vector $\bar{K}_\mu$ is accordingly 
$\bar{K}_\mu = \bar{K}^{(0)}_\mu + \kappa \bar{K}^{(1)}_\mu + \mathcal{O}\left(\kappa^2\right)$. 

We now go ahead and determine the one-loop term $\bar{K}^{(1)}_\mu$.
As shown in detail in Sec.~\ref{sec:hbarExpansion}, this can be conveniently achieved by solving~\eqref{eq:oneLoopSPE}
for $\bar{K}^{(1)}_a$, 
\begin{equation} \label{eq:oneLoopSPEa}
    \sqrt{2 K_0^{(0)}} \bar{D}_a V^{(1)} + \left(m_a^2\right)^{(0)} \bar{K}^{(1)}_a = 0\,.
\end{equation}
We then get immediately from~\eqref{K10}
\begin{equation} \label{K01n}
\bar{\tvec{k}}^{(1)} = \frac{ \bar{\tvec{K}}^{(1)} } {K_0^{(0)} }, \qquad
K_0^{(1)} = K_0^{(0)} \bar{\tvec{k}}^{(0) \trans}
\bar{\tvec{k}}^{(1)}\,.
\end{equation}
Now we obtain the shift of the Lagrange multiplier~\eqref{eq:u1} in the expansion 
$u = u^{(0)} + \kappa u^{(1)} + \mathcal{O}\left(\kappa^2\right)$ from
\begin{equation} \label{eq:u1a}
    u^{(1)} = \frac{1}{4 K_0^{(0)}}\left[ 2 \partial_0 V^{(1)} + \frac{1}{2} \tvec{f}_\pm^\trans \tvec{k}^{(1)} \right]\,,
\end{equation}
where
\begin{equation}
    \tvec{f}_\pm = 8 K_0^{(0)} \left(\eta_{00}\tvec{k}^{(0)} + \tvec{\eta}\right) - 2 \left(m_{H^\pm}^2\right)^{(0)} \tvec{k}^{(0)} \label{eq:fpmdefa}\,.
\end{equation}
Note that the scalar product $\tvec{f}_\pm^\trans \tvec{k}^{(1)} = \tvec{\bar{f}}_\pm^\trans \tvec{\bar{k}}^{(1)}$ with 
$\tvec{\bar{f}}_\pm = R \tvec{f}_\pm$ in~\eqref{eq:u1a} is basis invariant.

From the shift of the Lagrange multiplier in~\eqref{eq:u1a} as well as $K_0^{(1)}$ from~\eqref{K01n}
we determine the squared mass of the pair of charged Higgs bosons, \eqref{mHpm}, $m_{H^\pm}^2 = 4 u K_0$, that is,
\begin{equation} \label{mHpmhbar}
\begin{split}
m_{H^\pm}^2  = & 4 u K_0 =
\left(m_{H^\pm}^2 \right)^{(0)} + \kappa  \left(m_{H^\pm}^2 \right)^{(1)} + \mathcal{O}\left(\kappa^2\right)
\\ =  &
 4 u^{(0)} K_0^{(0)} + 
\kappa \left(4 u^{(0)} K_0^{(1)} + 4  u^{(1)} K_0^{(0)} \right) + \mathcal{O}\left(\kappa^2\right)\,.
\end{split}
\end{equation}
The shift of the scalar squared masses finally follows from~\eqref{scalarmasses},
\begin{equation} \label{scalarmasses2}
\begin{split}
    m_a^2 &= \left(m_a^2\right)^{(0)} + \kappa \left(m_a^2\right)^{(1)} + \mathcal{O}\left(\kappa^2\right)\,,\\
    \left(m_a^2\right)^{(1)} &= \bar{D}_a \bar{D}_a V^{(1)} + \frac{K^{(1)}_0}{K^{(0)}_0} \left(m_a^2\right)^{(0)} + \bar{f}_\pm^a \bar{\delta}^{(1),\,a}\,,
\end{split}
\end{equation}
with \eqref{delta1}
\begin{equation} \label{delta12}
    \bar{\tvec{\delta}}^{(1)} \equiv \bar{\tvec{k}}^{(1)} - \frac{K^{(1)}_0}{K^{(0)}_0} \bar{\tvec{k}}^{(0)}\,.
\end{equation}

To get the solutions $\bar{K}_\mu^{(1)}$, $u^{(1)}$, and $\left(m_a^2\right)^{(1)}$ we see from~\eqref{eq:oneLoopSPEa}, \eqref{eq:u1a},
 and \eqref{scalarmasses2} 
that the derivatives $\bar{D}_a V^{(1)}$, $\bar{D}_a \bar{D}_b V^{(1)}$, and
$\partial_0 V^{(1)}$ are required.
In Secs.~\ref{sec:oneLoopGauge}, \ref{sec:oneLoopFermion}, and \ref{sec:oneLoopScalar}, respectively, the gauge, fermion, and scalar contributions to these derivatives have been given, where the 
gauge, fermionic, and scalar parts are
 \begin{equation}
 V^{(1)} = V^{(1)}_g +  V^{(1)}_f + V^{(1)}_s\,.
 \end{equation}
For the convenience of the reader, we repeat the required expressions here.
\\
 
\noindent{\bf Gauge contributions}\\

Collecting~\eqref{eq:DV1g} and \eqref{eq:D2V1g},
\begin{align}
    \bar{D}_a V^{(1)}_g &= \frac{3}{2} \Big\{2 \bar{g}_{a W W}\, A_g\left(M_W^2\right) + \bar{g}_{a Z Z}\,  A_g\left(M_Z^2\right)\Big\}\,, \label{eq:DV1ga}\\
\begin{split}
    \bar{D}_a \bar{D}_b V^{(1)}_g &= \frac{3}{2} \Big\{\bar{g}_{a b Z Z}\,  A_g\left(M_Z^2\right) + 2 \bar{g}_{a W W} \, \bar{g}_{b W W} \, B_g\left(M_W^2, M_W^2\right) \\
    &\qquad + \bar{g}_{a Z Z} \, \bar{g}_{b Z Z} \, B_g\left(M_Z^2, M_Z^2\right)\Big\}\,, \label{eq:D2V1ga}
\end{split}
\end{align}
with~\eqref{eq:gbaXX}, \label{eq:gbabXX2}
\begin{align}
    && \bar{g}_{a W W} &= \sqrt{\frac{2}{K_0}} M_W^2 \bar{k}^a, & \bar{g}_{a Z Z} &= \sqrt{\frac{2}{K_0}} M_Z^2 \bar{k}^a\,, && \\
    && \bar{g}_{a b W W} &= \frac{M_W^2}{K_0} \delta^{a b}, & \bar{g}_{a b Z Z} &= \frac{M_Z^2}{K_0} \delta^{a b}\,. &&
\end{align}
Furthermore, see \eqref{par0V1g},
\begin{equation}
    \partial_0 V^{(1)}_g = \frac{3}{2} \left\{2 g_{0 W W}\,  A_g\left(M_W^2\right) + g_{0 Z Z}\,  A_g\left(M_Z^2\right)\right\}\,,
\end{equation}
with~\eqref{g0WW}, recalling, $\cos(\theta_W)=M_W/M_Z$,
\begin{equation} \label{g0WW2}
    g_{0 W W} = \frac{M_W^2}{K_0}, \qquad 
    g_{0 Z Z} = \frac{M_Z^2}{2 K_0} (1 + \cos^2(2 \theta_W))\,.
\end{equation}

\noindent{\bf Fermionic contributions}\\

We have, collecting~\eqref{eq:DV1f}, \eqref{eq:D2V1f},
\begin{align}
    \bar{D}_a V^{(1)}_f &= -2 \, \Big\{\bar{y}_{a \boldsymbol{\tau \tau}} A(m_\tau^2) + 3 \bar{y}_{a \mathbf{t t}} A(m_t^2) + 3 \bar{y}_{a \mathbf{b b}} A(m_b^2) \Big\} \label{eq:DV1f7},\\
\begin{split} \label{eq:D2V1f7}
    \bar{D}_a \bar{D}_b V^{(1)}_f &= -2 \, \Big\{3\bar{y}_{a b \mathbf{t t}} A(m_t^2) + 3\bar{y}_{a b \mathbf{b b}} A(m_b^2) \\
    &\qquad + \bar{y}_{a \boldsymbol{\tau \tau}} \bar{y}_{b \boldsymbol{\tau \tau}} B(m_\tau^2, m_\tau^2)
    + 3\bar{y}_{a \mathbf{t t}} \bar{y}_{b \mathbf{t t}} B(m_t^2, m_t^2) + 3\bar{y}_{a \mathbf{b b}} \bar{y}_{b \mathbf{b b}} B(m_b^2, m_b^2) \Big\}\,,
\end{split}
\end{align}
with~\eqref{ybaxx}
{\allowdisplaybreaks
\begin{align}
\begin{split}
    \bar{y}_{a \boldsymbol{\tau \tau}} &= \sqrt{\frac{K_0}{2}} \left(Y_\tau^0 \bar{k}^a + \bar{Y}_\tau^a\right), \\
    \bar{y}_{a \mathbf{t t}} &= \sqrt{\frac{K_0}{2}} \left(Y_t^0 \bar{k}^a + \bar{Y}_t^a\right), \\
    \bar{y}_{a \mathbf{b b}} &= \sqrt{\frac{K_0}{2}} \left(Y_b^0 \bar{k}^a + \bar{Y}_b^a\right), \\
    \bar{y}_{a b \mathbf{t t}} &= \frac{1}{2 K_0} \left(Y_t^\trans \gT \KT\right) \delta^{ab} + \frac{1}{2}\left(\bar{k}^a \bar{Y}_t^b + \bar{Y}_t^a \bar{k}^b \right), \\
    \bar{y}_{a b \mathbf{b b}} &= \frac{1}{2 K_0} \left(Y_b^\trans \gT \KT\right) \delta^{ab} + \frac{1}{2}\left(\bar{k}^a \bar{Y}_b^b + \bar{Y}_b^a \bar{k}^b \right)\,,
\end{split}
\end{align}
}%
and~\eqref{eq:d0V1f}
\begin{equation} \label{eq:d0V1f2}
    \partial_0 V^{(1)}_f = -\left\{Y_\tau^0 A(m_\tau^2) + 3 Y_t^0 A(m_t^2) + 3 Y_b^0 A(m_b^2) + 3 K_0 \left|\xi_{ud}\right|^2 B(m_t^2, m_b^2) \right\}\,.
\end{equation}
The fermionic couplings $Y_t$, $Y_b$, $Y_\tau$ can be found in~\eqref{Ymat}. The squared phase $|\xi_{ud}|^2$ is defined in~\eqref{xiudsq}. The expressions with a bar symbol~\eqref{RYuk} correspond to these vectors in the basis where the scalar squared mass matrix is diagonal, that is 
$\bar{Y}_t^a = R_{ac} Y_t^c$, $\bar{Y}_b^a = R_{ac} Y_b^c$, $\bar{Y}_\tau^a = R_{ac} Y_\tau^c$ and the zero components unchanged. The rotation matrix $R$  is the diagonalization matrix of the scalar squared mass matrix, given in~\eqref{eq:diagonalBasisNeutral}. Note, however, that expressions of the form $Y_f^\trans \gT \KT = \bar{Y}_f^\trans \gT \overline{\KT}$ are basis invariant.\\

\noindent{\bf Scalar contributions}\\

The scalar contributions are given in~\eqref{barDVs1}, \eqref{eq:D2V1s}, and \eqref{d0V1s},

\begin{align}
  \bar{D}_a V^{(1)}_s &= \frac{1}{2} \left\{
    \sum_{b=1}^3 \bar{\lambda}_{abb} A\left(m_b^2\right)
    + \bar{\lambda}_{a H^\pm H^\pm} A\left(m_{H^\pm}^2\right)
  \right\} \;, \notag \displaybreak[1] \\
%
  \bar{D}_a \bar{D}_b V^{(1)}_s &= \frac{1}{2} \Bigg\{
    \sum_{c=1}^3 \bar{\lambda}_{abcc} A\left(m_c^2\right)
    + \bar{\lambda}_{ab H^\pm H^\pm} A\left(m_{H^\pm}^2\right) \notag \\
  &\quad + \sum_{c,d=1}^3 \bar{\lambda}_{acd} \bar{\lambda}_{bcd} B\left(m_c^2, m_d^2\right)
    + \bar{\lambda}_{a H^\pm H^\pm} \bar{\lambda}_{b H^\pm H^\pm} B\left(m_{H^\pm}^2, m_{H^\pm}^2\right) \notag \\
  &\quad + 2 \sum_{c=1}^3 \bar{\lambda}_{ac G^0} \bar{\lambda}_{bc G^0} B\left(m_c^2, 0\right)
    + \bar{\lambda}_{a H^\pm G^\pm} \bar{\lambda}_{b G^\pm H^\pm} B\left(m_{H^\pm}^2, 0\right) \notag \\
  &\quad + \bar{\lambda}_{a G^0 G^0} \bar{\lambda}_{b G^0 G^0} B(0, 0)
    + \bar{\lambda}_{a G^\pm G^\pm} \bar{\lambda}_{b G^\pm G^\pm} B(0, 0)
  \Bigg\}\;, \notag \displaybreak[1] \\
  \label{eq:d0V1sfirsta}
  \partial_0 V^{(1)}_s &= \frac{1}{2} \left\{
    \sum_{a=1}^3 \lambda_{0aa} A\left(m_a^2\right)
    + \lambda_{0 H^\pm H^\pm} A\left(m_{H^\pm}^2\right)\;.
  \right\}
\end{align}
The couplings in the scalar sector follow from our discussion in Sec.~\ref{sec:oneLoopScalar}
and are given explicitly in \cite{Sartore:2022sxh} (see there the formulae (4.20) and appendix~B), as well
as~\eqref{lambda0}:
\begin{align} \label{eq:scalarcouplings}
&\bar{\lambda}_{abc} = \frac{1}{\sqrt{2 K_0}}\bigg\{\left(\delta^{ab} - \bar{k}^a \bar{k}^b\right) \bar{f}_\pm^c + \delta^{ab} \bar{k}^c \left( m_a^2 + m_b^2 - m_c^2\right)  \bigg\} + (a \leftrightarrow c) + (b \leftrightarrow c)\,,
\nonumber    \\
& \bar{\lambda}_{aG^0G^0} = \frac{1}{\sqrt{2 K_0}}  \bar{k}^a m_a^2\,,
\nonumber    \\
& \bar{\lambda}_{aH^\pm H^\pm} = \frac{\bar{f}^a}{\sqrt{2 K_0}}\,,
\nonumber    \\
& \bar{\lambda}_{aG^\pm G^\pm} = \frac{1}{\sqrt{2 K_0}} \bar{k}^a m_a^2\,,
\nonumber    \\
&  \bar{\lambda}_{abG^0} = \frac{1}{\sqrt{2 K_0}} \varepsilon_{abc} \bar{k}^c \big(m_b^2-m_a^2 \big)\,,
\nonumber    \\
&  \bar{\lambda}_{aG^\pm H^\pm} = \frac{1}{\sqrt{2 K_0}} 
        \big(m_a^2 - m_{H^\pm}^2 \big)
        \big(\bar{x}^a + i \bar{y}^a \big)\,,
\nonumber    \\
&  \bar{\lambda}_{abcd} = g_{abcd}+ g_{cdab} + g_{acbd} + g_{bdac} + g_{adbc} + g_{bcad}\,,
    \\
&    \bar{\lambda}_{abH^\pm H^\pm} = \frac{1}{2 K_0}
        \bigg(
        \delta^{ab} 
                \big(  
            \bar{\tvec{k}}^\trans \bar{\mathcal M}^2_{\text{neutral}} \bar{\tvec{k}} - 16 K_0 \bar{\tvec{\eta}}^\trans \bar{\tvec{k}} 
                \big) 
           \nonumber \\ 
            & 
        \phantom{ \bar{\lambda}_{abH^\pm H^\pm} =} + \bar{k}^a \bar{k}^b \big ( 8 K_0 (\eta_{00} + 
            \bar{\tvec{\eta}}^\trans \bar{\tvec{k}}  ) - (m_a^2 + m_b^2) \big)
        + 8 K_0 \big( \bar{k}^a \bar{\eta}^b + \bar{k}^b \bar{\eta}^a \big)
        \bigg)\,,
 \nonumber       \\
  &
\bar{\lambda}_{0H^\pm H^\pm}
    =8\left(\eta_{00}-\tvec{k}^\trans\tvec{\eta}\right)
    +\sum_{a=1}^3\frac{\bar{f}^a}{2K_0}\left(\frac{\bar{f}^a}{m_{H^\pm}^2-m_a^2}-\bar{k}_a\right)\, ,
    \nonumber\\
    &\bar{\lambda}_{0aa}=-2\left(\eta_{00}+\tvec{k}^\trans\tvec{\eta}\right)+
    \left(1-\bar{k}_a^2\right)\left[4\eta_{00}
    -\frac{m_{H^\pm}^2}{2K_0}\left(3-\frac{m_{H^\pm}^2}{m_a^2}\right)\right]
   \nonumber\\
   &\phantom{\bar{\lambda}_{0aa}=} +\frac{m_a^2}{K_0}
   +\frac{\bar{f}^a}{2K_0}\left(\frac{\bar{f}^a}{m_a^2-m_{H^\pm}^2}+\bar{k}_a\right)\, .
   \nonumber
\end{align}

\noindent
The auxiliary functions are
\begin{equation}
\begin{split}
   &\bar{f}^a = 8 K_0 \left(\eta_{00} \bar{k}^a + \bar{\eta}^a\right) - \bar{k}^a m_a^2\,,\\
    &\bar{f}_\pm^a = 8 K_0 \left(\eta_{00} \bar{k}^a + \bar{\eta}^a\right) - 2 \bar{k}^a m_{H^\pm}^2 \,,\\
    &g_{abcd}=
    \frac{\delta^{ab}}{2 K_0}
        \bigg(
        \frac{\delta^{cd}}{2}
        \bar{\tvec{k}}^\trans \bar{\mathcal M}^2_{\text{neutral}} \bar{\tvec{k}}
        +
         \bar{k}^c \bar{k}^d \big( m_a^2 + m_b^2 - m_c^2 - m_d^2 \big)
         \bigg)
         \\
        &\qquad 
        + 4 \big(
        \delta^{ab} - \bar{k}^a \bar{k}^b \big)
        \left( \bar{k}^c \bar{k}^d 
        \left(\eta_{00} - \frac{m_{H^\pm}^2}{4 K_0} \right)
        + \bar{\eta}^c \bar{k}^d + \bar{\eta}^d \bar{k}^c
        - \bar{\tvec{\eta}}^\trans \bar{\tvec{k}}
             \delta^{cd}
            \right)\,.
\end{split}
\end{equation}

The three-component vectors $\tvec{x}$ and $\tvec{y}$ which appear 
in~\eqref{eq:scalarcouplings} correspond to the orthogonal 
rotation of the vector ${\tvec{k}}$ into the three direction,
\begin{equation} \label{RH}
    R_H = \begin{pmatrix}
        - \tvec{y}^\trans\\
        \tvec{x}^\trans\\
        \tvec{{k}}^\trans
    \end{pmatrix} = 
    \begin{pmatrix}
        -y_1 & -y_2 & -y_3\\
        x_1 & x_2 & x_3\\
        {k}_1 & {k}_2 & {k}_3
    \end{pmatrix}, \qquad R_H {\tvec{k}} = \begin{pmatrix}
        0 \\ 0 \\ 1
    \end{pmatrix}\,,
\end{equation}
with 
\begin{equation} \label{RHcond}
    \tvec{x}^\trans \tvec{{k}} = \tvec{y}^\trans \tvec{{k}} = \tvec{x}^\trans \tvec{y} = 0 ,\quad
    \text{and} \quad \tvec{x}^\trans \tvec{x} = \tvec{y}^\trans \tvec{y} = \tvec{{k}}^\trans \tvec{{k}} = 1\,, \quad
    \tvec{x}\times\tvec{y} = \tvec{{k}}\,.
\end{equation}

\subsection*{Scalar integrals}

The scalar one-point integrals have been defined in \eqref{eq:Afunctions}:
\begin{equation} \label{eq:Afunctionsa}
    A_s(x) = A_f(x) \equiv A(x) = x \left[\log \left(\frac{x}{\mu^2}\right) - 1\right], \qquad 
    A_g(x) = x \left[\log \left(\frac{x}{\mu^2}\right) - \frac{1}{3}\right]\, .
\end{equation}
Moreover, the scalar two-point integrals have been introduced in \eqref{eq:Bfunctions}.
They read for the case $x \neq y$:
\begin{align} \label{eq:Bfunctionsa}
    B_s(x,y) &= B_f(x,y) \equiv B(x,y) = \frac{A(x) - A(y)}{x-y}, \qquad B_g(x,y) = \frac{A_g(x) - A_g(y)}{x-y}\,,
\end{align}
and for the case of equal arguments as well as for the massless Goldstone bosons they are given by
\begin{align}
    B(x,x) &= \log \left(\frac{x}{\mu^2}\right)\,, \label{eq:Blima}\\
    B_g(x,x) &= \log \left(\frac{x}{\mu^2}\right) + \frac{2}{3}\,, \label{eq:Bglima}\\
    B_s(0,0) &= 0\,.
\end{align}

\section{Application of the $\hbar$ expansion}
\label{sec:application}

We now apply the previously derived formalism to the simple case of a CP-conserving THDM.
This illustrates how any THDM can be analyzed beyond tree level.

Requiring an explicit, that is, non softly broken, $\mathbb{Z}_2$ symmetry at the level of the scalar and Yukawa sectors further reduces the number of parameters of the model. As a result, taking as input the value of the following parameters
\begin{equation}
    v^2, \quad m_h^2, \quad m_H^2, \quad m_A^2, \quad m_{H^\pm}^2, \quad \tan\beta, \quad \cos(\alpha-\beta)
\end{equation}
will allow us to fully determine the parameters of the scalar sector, thus providing a straightforward way to identify and study realistic benchmarks. 

\subsection{The Higgs potential at tree level}
Let us first write down the scalar potential of the model in terms of conventional parameters and fields,
\begin{equation} \label{potZ2}
    \begin{split}
    V^{(0)}_{\text{THDM}}(\varphi_1, \varphi_2) &= 
m_{11}^2 \varphi_1^\dagger \varphi_1
+ m_{22}^2 \varphi_2^\dagger \varphi_2
+ \frac{\lambda_1}{2} \left(\varphi_1^\dagger \varphi_1\right)^2
+ \frac{\lambda_2}{2} \left(\varphi_2^\dagger \varphi_2\right)^2
\\
&
+ \lambda_3 \left(\varphi_1^\dagger \varphi_1\right)\left(\varphi_2^\dagger \varphi_2\right)
+ \lambda_4 \left(\varphi_1^\dagger \varphi_2\right)\left(\varphi_2^\dagger \varphi_1\right)
+ \frac{\lambda_5}{2} 
\left[
\left(\varphi_1^\dagger \varphi_2 \right)^2 + 
\left(\varphi_2^\dagger \varphi_1 \right)^2
\right]\,.
\end{split}
\end{equation}
Compared to the most general THDM, \eqref{eq:Vconv}, we have therefore in this model the restrictions
\begin{equation} \label{eq:Vapppara}
m_{12}^2=0, \quad
\im(\lambda_5) =0, \quad
\lambda_6 = \lambda_7 =0.
\end{equation} 
We assume that the parameters are chosen such that the potential is stable, that is, bounded from below.  
We further assume that the model has the correct electroweak symmetry breaking with a vacuum given by 
\begin{equation} \label{appvev}
    \langle\varphi_1\rangle = \begin{pmatrix}
        0\\
        v_1^0
    \end{pmatrix}, \qquad
    \langle\varphi_2\rangle = \begin{pmatrix}
        0\\
        v_2^0
    \end{pmatrix}, \quad \text{with } v_1^{0} = \frac{v}{\sqrt{2}} \cos(\beta), \quad v_2^{0} = \frac{v}{\sqrt{2}} \sin(\beta),
    \end{equation}
with $v = 246\,\mathrm{GeV}$.
See~\cite{Maniatis:2006fs} for the conditions ensuring 
stability and electroweak symmetry-breaking in the general THDM.
The Yukawa sector will be taken either of type I or type II, in the one-generation approximation,
\begin{align} \label{Yukapp1}
    \textbf{Type I:}\quad -\mathcal{L}_Y &= \epsilon_t \overline{Q}_L \widetilde{\varphi}_2 u_R + \epsilon_b \overline{Q}_L \varphi_2 d_R + \epsilon_\tau \overline{L} \varphi_2 e_R + \mathrm{h.c.}\,,\\
    \label{Yukapp2}
    \textbf{Type II:}\quad -\mathcal{L}_Y &= \epsilon_t \overline{Q}_L \widetilde{\varphi}_2 u_R + y_b \overline{Q}_L \varphi_1 d_R + y_\tau \overline{L} \varphi_1 e_R + \mathrm{h.c.}\,,
\end{align}
where all Yukawa couplings above can be made real by suitable phase redefinition. The conjugate field~$\widetilde{\varphi}_2$ is defined in~\eqref{phiconj}. \\

Now we translate the conventional parameters of the potential~\eqref{potZ2} into the bilinear parameters. This can be easily done recalling~\eqref{eq:para3} employing~\eqref{eq:Vapppara}:
\begin{multline} \label{apppara}
    \xi_0= \frac{1}{2} (m_{11}^2 + m_{22}^2), \quad
    \tvec{\xi} = \begin{pmatrix}0 \\ 0 \\ 
    \frac{1}{2}(m_{11}^2 - m_{22}^2)\end{pmatrix}\,, \quad
    \eta_{00}=\frac{\lambda_1+\lambda_2}{8}+\frac{\lambda_3}{4}\,, \quad
    \tvec{\eta} = \begin{pmatrix}0 \\ 0 \\ \frac{\lambda_1-\lambda_2}{8}\end{pmatrix}\,,\\
    E = \frac{1}{4}\diag \begin{pmatrix}\lambda_4+\lambda_5, & \lambda_4-\lambda_5, & \frac{1}{2}(\lambda_1-\lambda_2)-\lambda_3\end{pmatrix} = \diag \begin{pmatrix}E_{11}, & E_{22}, & E_{33}\end{pmatrix}\,.
\end{multline}
In four-component notation this reads, \eqref{eq:para4},
\begin{equation} \label{eq:para4a}
    \tilde{\tvec{\xi}}=\begin{pmatrix} \xi_0 \\ \tvec{\xi}\end{pmatrix},
    \qquad
    \tilde{E}=\begin{pmatrix} \eta_{00} & \tvec{\eta}^\trans\\
    \tvec{\eta} & E \end{pmatrix}.
\end{equation}
We form the four-component vector $\KT = \begin{pmatrix}K_0,& K_1,& K_2,& K_3\end{pmatrix}^\trans$ 
with the expressions for the bilinears for any THDM given in~\eqref{eq:Kphi}, that is,
\begin{align}
\label{eq:Kphi2}
&K_0 = \varphi_1^\dagger \varphi_1 + \varphi_2^\dagger \varphi_2,
&&K_1 = \varphi_1^\dagger \varphi_2 + \varphi_2^\dagger \varphi_1, \nonumber \\
&K_2 = i\big( \varphi_2^\dagger \varphi_1 - \varphi_1^\dagger \varphi_2 \big), 
&&K_3 = \varphi_1^\dagger \varphi_1 - \varphi_2^\dagger \varphi_2. 
\end{align}
The potential then takes the form
\begin{equation} \label{eq:potK42}
    V^{(0)}_{\text{THDM}}(\KT) =  \KT^\trans \tilde{\tvec{\xi}} + 
     \KT^\trans \tilde{E}  \KT.
\end{equation}

In the Yukawa sector, as discussed in detail in Sec.~\ref{sec:oneLoopFermion}, we get for the Hermitian Yukawa matrices comparing~\eqref{Yukapp1} and \eqref{Yukapp2}  with \eqref{Ymat} and \eqref{xiudsq},
\begin{align} 
    && \textbf{Type I:} && Y_t &= \begin{pmatrix}
        \epsilon_t^2\\ 0 \\ 0 \\ -\epsilon_t^2
    \end{pmatrix}\,, & Y_b &= \begin{pmatrix}
        \epsilon_b^2\\ 0 \\ 0 \\ -\epsilon_b^2
    \end{pmatrix}\,, & Y_\tau &= \begin{pmatrix}
        \epsilon_\tau^2\\ 0 \\ 0 \\ -\epsilon_\tau^2
    \end{pmatrix}\,, & |\xi_{ud}|^2 &= 0\,,&&\label{Ytype1} \\[.15cm]
    && \textbf{Type II:} && Y_t &= \begin{pmatrix} 
        \epsilon_t^2\\ 0 \\ 0 \\ -\epsilon_t^2
    \end{pmatrix}\,, & Y_b &= \begin{pmatrix}
        y_b^2\\ 0 \\ 0 \\ y_b^2
    \end{pmatrix}\,, & Y_\tau &= \begin{pmatrix}
        y_\tau^2\\ 0 \\ 0 \\ y_\tau^2
    \end{pmatrix}\,, & |\xi_{ud}|^2 &= \epsilon_t^2 y_b^2\,.&& \label{Ytype2}
\end{align}
The tree-level stationary-point equations
for a charge-conserving minimum, \eqref{eqEW}, read
\begin{equation} \label{vaca}
    0 = \xiT + 2 \left(\ET - u \gT\right) \KT = \begin{pmatrix}
        \xi_0 + 2\left(\eta_{00} - u\right) K_0 + 2 \eta_3 K_3\\
        2\left(E_{11} + u\right) K_1\\
        2\left(E_{22} + u\right) K_2\\
        \xi_3 + 2 K_3 \left(E_{33}+u\right) + 2 \eta_3 K_0
    \end{pmatrix}\,.
\end{equation}
The relation~\eqref{mHpm}, valid in any THDM,  
\begin{equation} \label{mHpma}
m_{H^\pm}^2 = 4 u K_0\;,
\end{equation}
uniquely fixes $u$.

Expressing $\KT$ in terms of the conventional vacuum expectation values, from~\eqref{appvev} and \eqref{eq:Kphi2} we get
\begin{equation} \label{appvac}
    \KT = \begin{pmatrix}
        K_0\\
        K_1\\
        0 \\
        K_3\\
    \end{pmatrix} = \begin{pmatrix}
        (v^0_1)^2 + (v^0_2)^2\\
        2 v^0_1 v^0_2\\
        0 \\
        (v^0_1)^2 - (v^0_2)^2\\
    \end{pmatrix} = \frac{v^2}{2} \begin{pmatrix}
        1\\
        \sin(2\beta)\\
        0\\
        \cos(2\beta)
    \end{pmatrix}.
\end{equation}

We observe from~\eqref{vaca} that for $E_{11} \neq -u$  we must have $K_1=0$ and similar for $E_{22} \neq -u$ we must have  $K_2=0$.\ Note that since the first two components of $\tvec{\xi}$ and $\tvec{\eta}$ vanish, we can by a basis transformation, \eqref{eq:b1}, \eqref{eq:b2} interchange the parameters $E_{11}$ and $E_{22}$. 
From~\eqref{vaca} and \eqref{appvac} we conclude that the case $E_{11} \neq -u$ or $E_{22} \neq -u$ corresponds to a $\mathbb{Z}_2$-conserving, inert vacuum, where one of the vacuum expectation values in~\eqref{appvev} vanishes.

The remaining quartic couplings can be fixed by examining the neutral mass matrix~\eqref{eq:MneutralTreeLevel} with 
$\tvec{k} = \tvec{K}/K_0 = \begin{pmatrix} k_1,& 0,& k_3 \end{pmatrix}^\trans = \begin{pmatrix} \sin(2\beta),& 0,& \cos(2\beta) \end{pmatrix}^\trans$,
\begin{align}\label{eq:CPconsNeutralMassMat}
    \widehat{\mathcal{M}}_\mathrm{neutral}^2 &= 4 K_0 \left[ (\eta_{00}-u) \tvec{k} \tvec{k}^\trans + \tvec{\eta} \tvec{k}^\trans + \tvec{k} \tvec{\eta}^\trans + \left(E + u \unitmatrix_3\right) \right]\\
    &= \begin{pmatrix}
        4 K_0 \left[ \left(\eta_{00}-u\right) k_1^2  + E_{11} + u \right]  & 0 & 4 K_0 \left[\left(\eta_{00}-u\right) k_1 k_3 + k_1 \eta_3\right] \\
         0 & m_A^2 & 0 \\
         4 K_0 \left[\left(\eta_{00}-u\right) k_1 k_3 + k_1 \eta_3\right] & 0 & 4 K_0 \left[ \left(\eta_{00}-u\right) k_3^2  + 2 \eta_3 k_3 + E_{33} + u \right]
    \end{pmatrix}, \nonumber
\end{align}
with 
\begin{equation} \label{appmA2}
m_A^2 = 4 K_0 (E_{22} + u)\,.
\end{equation}
Since the scalar squared mass matrix is real and symmetric it can be diagonalized, that is,
\begin{equation}\label{eq:massDiagonalisation}
    R \widehat{\mathcal{M}}_\mathrm{neutral}^2 R^\trans = \diag\begin{pmatrix}
        m_h^2, & m_A^2, & m_H^2
    \end{pmatrix}
\end{equation}
with
\begin{equation} \label{basisa}
    R = \begin{pmatrix}
        \cos(\alpha+\beta) & 0 & \sin(\alpha+\beta) \\
        0 & 1 & 0 \\
        -\sin(\alpha+\beta) & 0 & \cos(\alpha+\beta)
    \end{pmatrix}\,.
\end{equation}

In the case of an inert vacuum with $k_1 = 0$, the mass matrix~\eqref{eq:CPconsNeutralMassMat} is already diagonal and $\alpha$ and $\beta$ are no longer free parameters. Requiring a viable fermion sector, we see from~\eqref{RYuk} that keeping the basis transformation~\eqref{basisa} diagonal requires  
\begin{equation} \label{viabferma}
    |\alpha| = |\beta| = \frac{\pi}{2}
\end{equation}
and two quartic couplings remain undetermined. More precisely, \eqref{eq:CPconsNeutralMassMat} yields
\begin{equation}
    E_{11} = \frac{m_H^2 - m_{H^\pm}^2}{2 v^2}, \quad \eta_{00} + E_{33} - 2 \eta_3 = \frac{\lambda_2}{4} = \frac{m_h^2}{2 v^2}
\end{equation}
and the quartic couplings $\lambda_1$ and $\lambda_3$
would be free parameters of the model. This case, which would require imposing additional constraints, is not considered here. Instead we focus on the case of a non-inert vacuum, where the parameters have to fulfill the condition $E_{11}=-u = - \frac{m_{H^\pm}^2}{2v^2}$. Taking also into account the constraint of a viable Yukawa sector~\eqref{viabferma}, we may finally summarize the values taken by all the couplings of the model in terms of the input parameters:
\begin{gather} 
    u = \frac{m_{H^\pm}^2}{2 v^2}, 
    \quad \KT = \frac{v^2}{2} \begin{pmatrix}
        1\\
        \sin(2\beta)\\
        0\\
        \cos(2\beta)
    \end{pmatrix}, 
    \quad \overline{\KT} = \frac{v^2}{2} \begin{pmatrix}
        1\\
        \sin(\alpha + 3\beta)\\
        0\\
        \cos(\alpha + 3\beta)
    \end{pmatrix},\nonumber\\ 
    E_{11} = -\frac{m_{H^\pm}^2}{2 v^2}, \quad E_{22} = \frac{m_A^2 - m_{H^\pm}^2}{2 v^2}, \nonumber\\
\begin{aligned}
    E_{33} &= \frac{1}{4v^2 \sin^2(2\beta)}\left[ m_h^2 + m_H^2 + m_{H^\pm}^2 (\cos(4\beta)-1) + (m_h^2-m_H^2) \cos(2\alpha + 6 \beta) \right],\\
    \eta_{00} &= \frac{1}{4v^2 \sin^2(2\beta)}\left[ m_h^2 + m_H^2 - m_{H^\pm}^2 (\cos(4\beta)-1) + (m_h^2-m_H^2) \cos(2\alpha + 2\beta) \right],\label{paraadiag}\\
    \eta_3 &= \frac{-1}{4v^2 \sin^2(2\beta)}\left[ (m_h^2 + m_H^2) \cos(2\beta)+ (m_h^2-m_H^2) \cos(2\alpha + 4\beta) \right],
\end{aligned}\\
\begin{aligned}
    \textbf{Type I:}\quad \epsilon_t &= \frac{\sqrt{2} m_t}{v \sin(\beta)} , \quad \epsilon_b = \frac{\sqrt{2}m_b}{v \sin(\beta)} , \quad \epsilon_\tau  = \frac{\sqrt{2}m_\tau}{v \sin(\beta)} , \quad \left|\xi_{ud}\right|^2 = 0\,, 
        \\
    \textbf{Type II:}\quad \epsilon_t &= \frac{\sqrt{2}m_t}{v \sin(\beta)} , \quad y_b = \frac{\sqrt{2}m_b}{v \cos(\beta)} , \quad y_\tau  = \frac{\sqrt{2}m_\tau}{v \cos(\beta)} , \quad 
    |\xi_{ud}|^2 = \frac{16 m_t^2 m_b^2}{v^4 \sin^2\left(2\beta\right)}\,.\nonumber
\end{aligned}
\end{gather}

\subsection{Scalar mass spectrum at next-to leading order}
Here we compute the $\hbar$ expansion at next-to leading order of the scalar spectrum in the example of the CP-conserving model of type~I. As input parameters we 
choose the tree-level values~\cite{ParticleDataGroup:2022pth},
\begin{equation} \label{numpara}
m_h^{(0)}=~125.25~\text{GeV}, \quad
m_A^{(0)}=~190~\text{GeV}, \quad
m_H^{(0)} =~300~\text{GeV}, \quad
m_{H^\pm}^{(0)} =~200~\text{GeV},
\end{equation}
and set $\cos(\alpha-\beta)=1$, use the standard value, $v=246~\text{GeV}$, and for the fermion masses
$m_b =~4.18~\text{GeV}$,
$m_t =~172.5~\text{GeV}$, and
$m_\tau =~1.777~\text{GeV}$. Note that from now on, we indicate the perturbation order as an upper index. We vary the value of $\tan(\beta)$ in our calculations, that is, the tangents of the mixing angle of the two Higgs-boson doublets.
From the input parameters we first compute all the bilinear parameters, as outlined in 
\eqref{paraadiag}. From the methods in~\cite{Maniatis:2006fs} we can check stability (see their theorem 1) and that the model has the correct, that is, electroweak symmetry breaking
$\matheweakgroup \to \mathemgroup$ (see 
their
theorems 2 and 3). 

With these preparations, we compute the one-loop approximation of the vacuum, the parameter~$u$, as well as the scalar mass spectrum as outlined in Sec.~\ref{sec:hbar_nuts}.

\begin{figure}
\includegraphics[width=\textwidth]{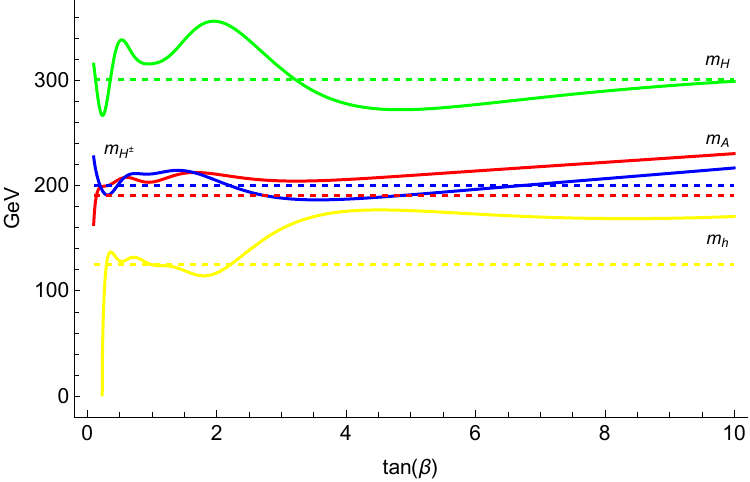}
\caption{\label{plotscalarmasses} Spectrum of the scalar masses in the CP-conserving model of type~I at the one-loop order. Shown are the masses of the three electrically neutral Higgs bosons, $m_h$, $m_A$, $m_H$ as well as the mass of the pair of charged Higgs bosons $m_{H^\pm}$, depending on the parameter $\tan(\beta)$. The dashed lines denote the corresponding tree level masses. 
The energy scale is set to the $Z$-boson mass.
}
\end{figure}

 The result for the scalar mass spectrum in the case of a model of type~I is shown in Fig.~\ref{plotscalarmasses}. The tree-level masses are given in~\eqref{numpara}.
Note that for small values of $\tan(\beta)$ the Yukawa couplings become very large giving large corrections to the tree level masses~\eqref{numpara}.

\section{Conclusions}
\label{sec:conclusion}

Keeping gauge symmetry manifest has been shown to enable a concise study of stability and electroweak symmetry breaking in the general THDM. In particular, the physical structure of the model is not obscured by unphysical gauge redundancies. The essential step is to introduce bilinear expressions~\cite{Nagel:2004sw, Maniatis:2006fs, Nishi:2006tg} which have been shown to provide a one-to-one correspondence with the Higgs-boson doublet fields, modulo unobservable gauge degrees of freedom

Moreover, basis transformations, that is, unitary mixing of the two doublets and general symmetries appearing as subgroups of $O(3)$ for the bilinears, can be studied geometrically and  transparently. 
For instance, basis transformations are simple rotations of $SO(3)$ and CP transformations correspond to point or plane 
reflections of the bilinears. 

The formalism has been developed firstly to describe the potential of the THDM. Recently, this bilinear formalism has been shown to be applicable also to the gauge sector and to the fermion sector of the THDM, that is, to the complete model~\cite{Sartore:2022sxh}. In particular, the scalar squared mass matrices have been computed in a model-independent manner. It has been shown that the masses of the pair of charged Higgs bosons depend only on the vacuum of the Higgs potential.
This result turns out to be valid for any THDM - to any perturbation order and even for effective models. 

In this work, we have considered quantum corrections to the THDM. 
This was achieved by adapting the $\hbar$ loop expansion~\cite{Brodsky:2010zk} to the THDM - maintaining  gauge invariance manifest at every step through the use of bilinears.
We have presented iterative solutions of the stationary-point equations starting from the tree-level solution. 
The loop corrections to the scalar potential have to be evaluated at the tree-level vacuum. We have given analytic expression for the stationary-point equations and for the quantum corrections to the scalar mass spectrum. In particular, this approach avoids gauge dependence and circumvents the infrared singularities which typically arise for the Goldstone bosons. 
The Goldstone modes remain massless, and any mixing with the physical sector is avoided order by order. 

The main results were summarized in a dedicated section,
and the method was illustrated using  a specific CP symmetric model.
The outlined methods are applicable to any THDM.
We hope that the results developed in this paper will form a basis for future phenomenological applications 
of THDMs beyond tree level
providing new insights due to the gauge invariant formalism.

\acknowledgments

This work is supported by ANID Fondecyt projects 1200641 and 1250132.


\clearpage

\appendix
\section{Second derivatives of mass eigenvalues}
\label{app:eigenvaluesSecondDerivatives}

Here we want to derive~\eqref{eq:secondEigDerivative}, the second derivative of the squared mass eigenvalues. 
We start with the diagonalized form of the squared mass matrix~\eqref{eq:diagM}
\begin{equation} \label{eq:Da}
    D = U M^2 U^\trans = \diag(\lambda_1, \ldots, \lambda_n)
\end{equation}
with orthogonal matrix~$U$. The
second derivative of the diagonal matrix~$D$ reads
\begin{equation} \label{eq:ddMq}
\begin{split}
\partial_\mu \partial_\nu D = 
&
(\overline{\partial_\mu \partial_\nu  M^2}) +
D U (\partial_\mu \partial_\nu U^\trans) +
(\overline{\partial_\nu M^2}) U (\partial_\mu U^{\trans}) +
(\partial_\nu U) U^\trans (\overline{\partial_\mu M^2}) 
\\ & 
+
(\overline{\partial_\mu M^2}) U (\partial_\nu U^\trans) +
(\partial_\mu U) U^\trans (\overline{\partial_\nu M^2}) +
(\partial_\mu \partial_\nu U) U^\trans D 
\\ &
 +
(\partial_\nu U) U^\trans D U (\partial_\mu U^\trans) +
(\partial_\mu U) U^\trans D U (\partial_\nu U^\trans)\,.
\end{split}
\end{equation}
Here we have used the abbreviations~\eqref{eq:diagM}--\eqref{eq:diagd2M} and orthogonality of the matrix~$U$, that is, 
$U^\trans U = U U^\trans = \unitmatrix$.
Now we use the abbreviation
\begin{equation}
W_\mu=(\partial_\mu U) U^\trans
\end{equation}
giving also
$W_\mu^\trans= U (\partial_\mu U^\trans)$. 
For the second partial derivatives of $U$ we get
\begin{equation}
    \partial_\mu \partial_\nu U = (\partial_\nu W_\mu) U - W_\mu W^\trans_\nu U\,.
\end{equation}
This allows us to replace all partial derivatives of $U$ and $U^\trans$. Recalling that $W_\mu$ is skew symmetric, that is, $W_\mu^\trans = -W_\mu$, we get
\begin{equation} \label{eq:ddMq2}
\begin{split}
\partial_\mu \partial_\nu D = &
(\overline{\partial_\mu \partial_\nu  M^2}) +
D (\partial_\nu W_\mu^\trans) + 
(\partial_\nu W_\mu) D +
W_\nu (\overline{\partial_\mu M^2}) 
\\ & +
W_\mu (\overline{\partial_\nu M^2}) -
(\overline{\partial_\nu M^2}) W_\mu -
(\overline{\partial_\mu M^2}) W_\nu 
\\ &  +
D W_\nu W_\mu - 
W_\nu D W_\mu -
W_\mu D W_\nu +
W_\mu W_\nu D\,.
\end{split}
\end{equation}
We compute the diagonal part and write the indices of the matrices explicitly. In particular we write
$\partial_\mu \partial_\nu D^{II} = \partial_\mu \partial_\nu \lambda_I$; see~\eqref{eq:Da}. 
Note that there is no summation over repeated indices here if not explicitly written.
We see that 
the expressions $D^{II} (\partial_\nu W_\mu^\trans)^{II}$ and similar $(\partial_\nu W_\mu)^{II} D^{II}$ vanish.
We now insert the explicit expressions for 
$(W_\mu)^{IJ}= (\overline{\partial_\mu M^2})^{IJ}/(\lambda_I - \lambda_J)$ 
for 
$\lambda_I \neq \lambda_J$ and 
$(W_\mu)^{IJ}=0$ for 
$\lambda_I = \lambda_J$,  
found in~\eqref{eq:VIJ}, and get~\eqref{eq:secondEigDerivative}
\begin{equation} \label{eq:ddMq4}
\partial_\mu \partial_\nu \lambda_I =
(\overline{\partial_\mu \partial_\nu  M^2})^{II}       
+ 2
\sum_{\lambda_I \neq \lambda_J} 
\frac{(\overline{\partial_\mu M^2})^{IJ} (\overline{\partial_\nu M^2})^{JI}}
{\lambda_I-\lambda_J}.
\end{equation}

\section{First bilinear derivative of the scalar mass matrix}
\label{app:firstKder}

For the scalar contribution of the one-loop corrections~\eqref{d0V1s} we have to determine the partial derivatives of the scalar mass matrices with respect to~$K_0$, \eqref{lambda0},
\begin{equation} \label{lambda0_app}
\bar{\lambda}_{0aa} \equiv \left(\overline{\partial_0 M_s^2}\right)^{aa},
\quad a \in \{1,2,3\}, 
\qquad
\bar{\lambda}_{0H^\pm H^\pm} \equiv
\sum_{p=1}^2 \left(\overline{\partial_0 M_s^2}\right)^{H^\pm_p H^\pm_p}\;.
\end{equation}
The overbar symbol indicates the similarity rotation which diagonalizes the mass matrix; see \eqref{eq:diagM} and \eqref{eq:diagdM}.  The index~$a$ corresponds to the three neutral Higgs bosons and $p$ to the two charged Higgs bosons.

The connection of the scalar mass matrix squared defined in terms of the eight field components, $M_s^2= \partial_i \partial_j V$ with $i, j \in \{1, \ldots, 8\}$, to the bilinear form $\mathcal{M}_{\mu\nu} = \partial_\mu \partial_\nu V$ with $\mu, \nu \in \{0, \ldots, 3\}$, has been given in~\eqref{eq:msExprMat},
\begin{equation} \label{eq:msExprMat_app}
    M_s^2 = \Delta^\nu \partial_\nu V + \Gamma \mathcal{M} \Gamma^\trans.
\end{equation}
In order to compute the coefficients~\eqref{lambda0_app} we need the partial derivative\footnote{Note that there is no term $\Gamma (\partial_\mu \mathcal{M}) \Gamma^T$ because we suppose that the potential $V$ is
quadratic in $K_\mu$.},
\begin{equation} \label{eq:partmsExprMat_app}
    \partial_\mu M_s^2 = \Delta^\nu \mathcal{M}_{\mu\nu} + (\partial_\mu \Gamma) \mathcal{M} \Gamma^\trans + \Gamma \mathcal{M} (\partial_\mu \Gamma^\trans).
\end{equation}
In Sec.~\ref{sec:gaugeInvariantFormalism} we have determined the squared scalar mass matrix in the canonical basis, indicated by a hat symbol:
\begin{equation} \label{eq:Ms2hat_app}
    \widehat{M^2_s}
    = U_c M^2_s U_c^T
 = \widehat{\Delta}^\nu \partial_\nu V + \widehat{\Gamma} \mathcal{M} \widehat{\Gamma}^\trans
\end{equation}
where $U_c$ is the rotation matrix to the canonical basis
and, using \eqref{eq:partmsExprMat_app},
\begin{equation}
\label{eq:delMshat_app}
\begin{aligned}
    \widehat{\partial_\mu M^2_s}&=U_c( \partial_\mu M^2_s) U_c^\trans   
    =\widehat{\Delta}^\nu\mathcal{M}_{\mu\nu}+
    \widehat{\partial_\mu\Gamma}\mathcal{M}\widehat{\Gamma}^\trans+\widehat{\Gamma}\mathcal{M}\widehat{\partial_\mu\Gamma}^\trans.
\end{aligned}
\end{equation}
We now calculate the expressions for 
$\widehat{\Gamma}$, $\widehat{\Delta}^\nu$, and $\widehat{\partial_\mu\Gamma}$ needed in~\eqref{eq:Ms2hat_app} and~\eqref{eq:delMshat_app}.
The expression found for the connection in the canonical basis~\eqref{eq:canonicalGamma} with $\gamma$ in~\eqref{eq:gamma} reads
\begin{equation} \label{eq:Ghat_app}
    \widehat{\Gamma} =
    \begin{pmatrix}
    0_{4\times4} \\ \gamma
    \end{pmatrix}\, \quad \text{with }
    \gamma = \sqrt{2 K_0} \begin{pmatrix}
       \alpha & \tvec{0}^\trans \\
       \tvec{k} & \unitmatrix_3
    \end{pmatrix}\,.
\end{equation}
Here, we have defined
\begin{equation}
\alpha 
\equiv \sqrt{\frac{K_0^2 - \tvec{K}^\trans \tvec{K}}{K_0^2} }
= \sqrt{1 - \tvec{k}^\trans \tvec{k}} \ge 0\;.
\end{equation}
The charge-conserving case corresponds to $\alpha=0$ with $K_0>0$, but at this stage we make no assumptions with respect to electroweak symmetry breaking. 

The general expression $\widehat{\Delta}^\nu$, without any assumptions on the electroweak-symmetry breaking behaviour, has been computed
in the canonical bases in~\cite{Sartore:2022sxh} and reads
\begin{equation} \label{eq:DeltahatB}
    \widehat{\Delta}^\mu = \begin{pmatrix}
        A_{44}^\mu & C_{44}^\mu \\[.15cm]
        \left(C_{44}^\mu\right)^\trans & B_{44}^\mu \\
    \end{pmatrix}\,,
\end{equation}
where
\begin{align}
    A_{44}^0 &= \unitmatrix_4, & A_{44}^a &= k_a \unitmatrix_4 - \frac{1}{|\tvec{k}|^2} \begin{pmatrix}
        0 & \alpha k_a \tvec{k}^\trans \\
        \alpha k_a \tvec{k} \quad & 2 k_a \tvec{k} \tvec{k}^\trans\!\! -\!\! \left(\left[\tvec{e}_a\times\tvec{k}\right] \tvec{k}^\trans + \tvec{k} \left[\tvec{e}_a\times\tvec{k}\right]^\trans\right)
    \end{pmatrix}\label{eq:A44}\\[.2cm] 
    B_{44}^0 &= \unitmatrix_4, & B_{44}^a &= - k_a \unitmatrix_4 + \begin{pmatrix}
        0 & \alpha\, \tvec{e}_a^\trans \\
        \alpha\, \tvec{e}_a \quad & \tvec{e}_a \tvec{k}^\trans + \tvec{k} \tvec{e}_a^\trans
    \end{pmatrix}\,,
    \label{eq:B44}\\[.2cm]
    C_{44}^0 &= 0_{4\times4}, & C_{44}^a &= \frac{1}{|\tvec{k}|} \begin{pmatrix}
        0 & \left[\tvec{e}_a \times \tvec{k}\right]^\trans \\[.15cm]
        |\tvec{k}|^2 \tvec{e}_a - k_a \tvec{k} \quad 
        & \alpha \left(k_a \unitmatrix_3 - \tvec{e}_a \tvec{k}^\trans\right)
    \end{pmatrix}\label{eq:C44}\,.
\end{align}
The missing piece in~\eqref{eq:delMshat_app}
is~$\widehat{\partial_\mu\Gamma}$, that is, 
$\partial_\mu\Gamma$ in the canonical basis. 
From the definition of the connection, \eqref{eq:defgamma}, that is, 
$\Gamma^\lambda_j = \partial_j K^\lambda = \Delta^\lambda_{ji} \phi_i =  \Delta^\lambda_{ij} \phi_i$, we get 
\begin{equation}
\partial_i \Gamma^\lambda_j = \Gamma^\nu_i \partial_\nu \Gamma^\lambda_j = \Delta^{\lambda}_{ij}.
\end{equation}
Contracting this equation with $\Gamma^\mu_i$ we find
\begin{align}
\Gamma^\mu_i \Gamma^\nu_i \partial_\nu \Gamma^\lambda_j &= 
(\Gamma^\trans \Gamma)^{\mu\nu} \partial_\nu \Gamma^\lambda_j
= \Gamma^\mu_i \Delta^\lambda_{ij} 
= \Gamma^\alpha_j T_\alpha^{\mu\lambda} - \Gamma^\lambda_i \Delta^\mu_{ij}\, ,
\end{align}
where the last equality has been derived in equation (A.6) of \cite{Sartore:2022sxh} and
$T_\alpha^{\mu\lambda}$ is the 3rd-rank tensor defined in~\eqref{eq:ga22}.
Writing $\Gamma^T \Gamma =: \Gamma^2$ we hence obtain
\begin{equation}
(\Gamma^2)^{\mu \nu} \partial_\nu \Gamma^\lambda_j = 
\Gamma^\alpha_j T^{\mu\lambda}_\alpha -\Delta^\mu_{ji} \Gamma^\lambda_i \, ,
\end{equation}
which can be written in the following more compact form:
\begin{equation}
(\Gamma^2)^{\mu \nu} \partial_\nu \Gamma
=
\Gamma T^\mu -\Delta^\mu \Gamma \, .
\end{equation}
Multiplying with the inverse of $(\Gamma^2)^{\mu \nu}$ we eventually have
\begin{equation}
    \partial_\mu \Gamma^\lambda_j = (\Gamma^{-2})_{\mu\nu} (\Gamma^\alpha_j T^{\nu\lambda}_\alpha
    -\Delta^\nu_{ji} \Gamma^\lambda_i)
\end{equation}
or short
\begin{equation}
\partial_\mu \Gamma=
(\Gamma^{-2})_{\mu\nu}
(\Gamma T^\nu - \Delta^\nu \Gamma)\;.
\end{equation}
This gives
in the canonical basis\footnote{Note that $\hat \Gamma^\trans \hat \Gamma = \Gamma^\trans \Gamma$.}
 \begin{equation} \label{eq:parGcan}
\widehat{\partial_\mu\Gamma}=U_c\partial_\mu\Gamma=\left(\Gamma^{-2}\right)_{\mu\nu}\left(\widehat{\Gamma} T^\nu-\widehat{\Delta}^\nu \widehat{\Gamma} \right).
\end{equation}
We need to compute the component $\widehat{\partial_0\Gamma}$ and with the explicit expressions in~\eqref{eq:Ghat_app}, \eqref{eq:Deltahat}, as well as \eqref{eq:Texp}, \eqref{eq:Gamma2expr}, and with the inverted matrix
\begin{equation} \label{Gamma2inv}
\Gamma^{-2} = \frac{1}{2 K_0 \alpha^2} \begin{pmatrix} 
       1 & -\tvec{k}^\trans \\
       -\tvec{k}\; & \alpha^2 \unitmatrix_3 + \tvec{k} \tvec{k}^\trans
    \end{pmatrix}
\end{equation}
we get\footnote{The second equation on the right side is only valid to order $\mathcal{O}(\alpha)$
using that $|\tvec{k}| = 1 - \alpha^2/2 + {\cal O}(\alpha^3)$, as well as 
$1/|\tvec{k}| = 1 + \alpha^2/2 + {\cal O}(\alpha^3)$.
}
\begin{equation}\label{eq: d0gammacan}
\begin{aligned}
    \widehat{\partial_0\Gamma}&=\frac{1}{\sqrt{2K_0}|\alpha\tvec{k}|}\begin{pmatrix}
        0_{1\times1} & 0_{1\times3} \\
        0_{3\times1} & \tvec{k}^\trans\tvec{k}\unitmatrix_3-\tvec{k}\tvec{k}^\trans \\
        |\tvec{k}| & \tvec{k}^\trans|\tvec{k}| \\
        0_{3\times1} & |\alpha\tvec{k}|\unitmatrix_3
    \end{pmatrix}=\frac{1}{\sqrt{2K_0}\alpha}\begin{pmatrix}
        0_{1\times1} & 0_{1\times3} \\
        0_{3\times1} & \unitmatrix_3-\tvec{k}\tvec{k}^\trans \\
        1_{1\times1} & \tvec{k}^\trans \\
        0_{3\times1} & \alpha \unitmatrix_3
    \end{pmatrix}+\mathcal{O}(\alpha).
\end{aligned}
\end{equation}

 We see that this matrix is singular in the limit of $\alpha=0$. We therefore cannot set $\alpha$ to zero in the intermediate steps. Instead, we expand all expressions with respect to $\alpha$. In this way, we see that the singularities drop out in the final expressions. 
Since we are ultimately interested in the charge-conserving case, we only need to keep terms up to order $\alpha$.

Now we are in a position to compute \eqref{eq:delMshat_app}, inserting \eqref{eq:Ghat_app}, \eqref{eq:Deltahat}, and
\eqref{eq: d0gammacan} we get explicitely for the zero component of the partial derivative
\begin{equation} \label{part0Msq}
    \widehat{\partial_0M^2_s}=2\begin{pmatrix}
        \hat{A}_{1\times1} & \hat{B}^\trans_{1\times3} & 0_{1\times1} & \hat{D}^\trans_{1\times3} \\
        \hat{B}_{3\times1} & \hat{C}_{3\times3} & \hat{E}_{3\times1} & \hat{F}^\trans_{3\times3} \\
        0_{1\times1} & \hat{E}^\trans_{1\times3} & \hat{G}_{1\times1} & \hat{H}^\trans_{1\times3} \\
        \hat{D}_{3\times1} & \hat{F}_{3\times3} & \hat{H}_{3\times1} & \hat{I}_{3\times3} \\
    \end{pmatrix},
\end{equation}
where
\begin{equation}
\begin{aligned}
    \hat{A}&=\eta_{00}+\tvec{k}^\trans\tvec{\eta}, \\
    \hat{B}&=-\frac{|\alpha|\tvec{k}^\trans\tvec{\eta}}{1-\alpha^2}\tvec{k}, \\
    \hat{C}&=\left(\eta_{00}+\tvec{k}^\trans\tvec{\eta}\right)\unitmatrix_3+\frac{\tvec{k}\left(\tvec{\eta}\times\tvec{k}\right)^\trans+\left(\tvec{\eta}\times\tvec{k}\right)\tvec{k}^\trans-2\left(\tvec{k}^\trans\tvec{\eta}\right)\tvec{k}\tvec{k}^\trans}{1-\alpha^2}, \\
    \hat{D}&=\frac{\tvec{\eta}\times\tvec{k}}{\sqrt{1-\alpha^2}}, \\
    \hat{E}&=2\big[\left(1-\alpha^2\right)\tvec{\eta}-\left(\tvec{k}^\trans\tvec{\eta}\right)\tvec{k}\big], \\
    \hat{F}&=\alpha^2\tvec{k}^\trans\tvec{\eta}\unitmatrix_3+\frac{\left(1-2\alpha^2\right)\tvec{k}\tvec{\eta}^\trans-\left(\tvec{k}^\trans\tvec{\eta}\right)\tvec{k}\tvec{k}^\trans+\left(1-\alpha^2\right)\tvec{E}-(\tvec{E}\tvec{k})\tvec{k}^\trans}{\sqrt{1-\alpha^2}|\alpha|}, \\
    \hat{G}&=3\left(\eta_{00}-\tvec{k}^\trans\tvec{\eta}\right), \\
    \hat{H}&=\frac{\left(\eta_{00}-\tvec{k}^\trans\tvec{\eta}\right)\tvec{k}+\left(1+2\alpha^2\right)\tvec{\eta}-\tvec{E}\tvec{k}}{|\alpha|}, \\
    \hat{I}&=\left(\eta_{00}-\tvec{k}^\trans\tvec{\eta}\right)\unitmatrix_3+2\left(\tvec{E}+\tvec{k}\tvec{\eta}^\trans+\tvec{\eta}\tvec{k}^\trans\right).
\end{aligned}
\end{equation}

The expansion of $\widehat{M^2_s}$ in~\eqref{eq:Ms2hat_app} to order $\alpha$ can now be computed with the help of the explicit expressions for $\widehat{\Delta}^\nu$ and $\widehat{\Gamma}$ as given in~\eqref{eq:Deltahat}, respectively~\eqref{eq:Ghat_app}, 
\begin{equation} \label{eq:Mshatexp}
   \widehat{M^2_s} = \widehat{M^2_{s,0}} +
   \alpha \widehat{M^2_{s,1}} + {\cal{O}}(\alpha^2)\;.
\end{equation}

We need the squared scalar mass matrices in the diagonal basis in order to compute the  couplings in~\eqref{lambda0_app}. The diagonalization can be performed with the orthogonal matrix $\bar{U}_\alpha$ up to terms of order $\alpha^2$,
\begin{equation}\label{eq:Msbarapp}
\overline{M^2_s} = \bar{U}_\alpha \widehat{M^2_s} 
\bar{U}_\alpha^\trans\;.
\end{equation}
Expanding the orthogonal matrix $\bar{U}_\alpha$ we may write,
\begin{equation} \label{eq:Uexp}
\bar{U}_\alpha = \bar{U} + \alpha \bar{U}_1 + {\cal{O}}(\alpha^2) =
\left( \unitmatrix_8 + \alpha S_1 \right) \bar{U} + {\cal{O}}(\alpha^2),
\end{equation}
where $\bar{U}$ is the matrix diagonalizing $\widehat{M}^2_s$ at order $\alpha^0$, corresponding to the charge-conserving case.  
Orthogonality of~$\bar{U}_\alpha$ implies that the matrix $S_1$ is skew symmetric. 
The matrix $\widehat{\partial_\mu M^2_s}~\eqref{eq:delMshat_app}$ inherits the singularity of $\partial_\mu \widehat{\Gamma}$  discussed below~\eqref{eq:Ghat_app}, and with respect to $\alpha$ can be expanded as
\begin{equation} \label{eq:delhatMs2exp}
    \widehat{\partial_\mu M^2_s} =
   \alpha^{-1}  {\widehat{\partial_\mu M^2_{s}}}_{,-1} +
    {\widehat{\partial_\mu M^2_{s}}}_{,0} + 
   {\cal{O}}(\alpha)\;.
\end{equation}
We apply the same similarity transformation, diagonalizing $\widehat{M}^2_{s,0}$, to the derivative expressions,
\begin{equation}
    \overline{\partial_\mu M^2}_{\!\!\!s,0} =
    \bar{U} {\widehat{\partial_\mu M^2}}_{\!\!\!s,0} \bar{U}^\trans, \qquad
    \overline{\partial_\mu M^2}_{\!\!\!s,-1} =
    \bar{U}  {\widehat{\partial_\mu M^2_{s}}}_{,-1} \bar{U}^\trans \, .
\end{equation}
With these abbreviations we can write 
\begin{equation} \label{eq:dbarMs}
   \overline{\partial_\mu M^2}_{\!\!\!s} =
   \bar{U}_\alpha \widehat{\partial_\mu M^2_{s}} \bar{U}^\trans_\alpha =
   \overline{\partial_\mu M^2}_{\!\!\!s,0}
   + \left[S_1, 
   \overline{\partial_\mu M^2}_{\!\!\!s,-1} \right] + 
   \alpha^{-1} 
   \overline{\partial_\mu M^2}_{\!\!\!s,-1} + {\cal{O}}(\alpha)\;.
\end{equation}
From the explicit expansion 
\eqref{eq:delhatMs2exp} which can be derived from \eqref{eq:delMshat_app} using the 
known expressions for $\widehat{\Delta}$, $\widehat{\Gamma}$ and 
$\widehat{\partial_\mu \Gamma}$
we find that all entries of the four $(8 \times 8)$-matrices $\overline{\partial_\mu M^2}_{\!\!\!s,-1}$ vanish, except the elements
\begin{equation}
\left(
\overline{\partial_\mu M^2}_{\!\!\!s,-1} \right)^{a H^\pm_p} =
\left(
\overline{\partial_\mu M^2}_{\!\!\!s,-1} \right)^{H^\pm_p a} \quad
\text{and} \quad
\left(
\overline{\partial_\mu M^2}_{\!\!\!s,-1} \right)^{a G^\pm_p} =
\left(
\overline{\partial_\mu M^2}_{\!\!\!s,-1} \right)^{G^\pm_p a}\;,
\end{equation}
where $a \in \{1,2,3\}$ and $p \in \{1,2\}$. In particular all diagonal elements of 
$\overline{\partial_\mu M^2}_{\!\!\!s,-1}$ vanish. With view on~\eqref{lambda0_app} this means in turn that the term of order $\alpha^{-1}$ in~\eqref{eq:dbarMs} will not contribute to $\partial_\mu V_s^{(1)}$, as expected. It remains to determine the relevant terms of the skew-symmetric matrix~$S_1$.
Writing the expansion terms of the squared mass matrix after the similarity transformation
\begin{align} \label{diagMs}
    &\overline{M^2}_{\!\!\!s,0} = 
    \bar{U} \widehat{M}^2_{\!s,0} \bar{U}^\trans=
    \diag(0,0,0, m_{H^\pm}^2, m_{H^\pm}^2, m_1^2, m_2^2, m_3^2 ),\\
    &\overline{M^2}_{\!\!\!s,1} = 
    \bar{U} \widehat{M}^2_{\!s,1} \bar{U}^\trans ,
\end{align}
we find using \eqref{eq:Mshatexp}, \eqref{eq:Msbarapp}, and \eqref{eq:Uexp}:
\begin{equation} \label{eq:Msbarexpansion}
   \overline{M^2}_{\!\!\!s} =
   \overline{M^2}_{\!\!\!s,0} + 
   \alpha \left(
   \overline{M^2}_{\!\!\!s,1} + 
   \left[S_1, \overline{M^2}_{\!\!\!s,0} \right]
   \right) +
   {\cal{O}}(\alpha^2)\;.
\end{equation}
As has been shown in~\cite{Sartore:2022sxh}, the diagonalization matrix~$\bar{U}$ has the form,
\begin{equation}
\label{eq:barU}
\bar{U} = 
\diag(1, R_H, 1, R)
\end{equation}
with $(3\times 3)$-matrices $R_H$ and $R$. $R$ corresponds to a basis transformation, where we write in particular $R \tvec{k}= \tvec{\bar{k}}$, whereas $R_H$ has the property
$R_H \tvec{k}= (0,0,1)^\trans$.

Since $\overline{M^2}_{\!\!\!s}$ is diagonal, the off-diagonal entries in~\eqref{eq:Msbarexpansion} must vanish order by order, yielding in particular
\begin{equation}
    0 =
    \left(\overline{M^2}_{\!\!\!s,1} \right)^{ij}
    +
    \left[S_1, \overline{M^2}_{\!\!\!s,0} \right]^{ij} =
    \left(\overline{M^2}_{\!\!\!s,1} \right)^{ij}
    +
    (m_j^2 - m_i^2) S_1^{ij},
\end{equation}
for all $i,j \in \{1, \ldots, 8\}$ with $i \neq j$. If at least one of the tree-level scalar eigenvalues is non-zero, we find 
\begin{equation}
    S_1^{ij} = \frac{\left(\overline{M^2}_{\!\!\!s,1} \right)^{ij}}{m_i^2 - m_j^2}\;.
\end{equation}

With this expression for $S_1$ we can therefore give the expressions~\eqref{eq:dbarMs} for $a \in \{1,2,3\}$ and $p \in \{1,2\}$,
\begin{align}
&\left(\overline{ \partial_\mu M^2}_{\!\!\!s} \right)^{aa} =
\left(\overline{ \partial_\mu M^2}_{\!\!\!s,0} \right)^{aa} +
2 \sum_{p=1}^{2} \bigg[
\frac{
\left(\overline{M^2}_{\!\!\!s,1} \right)^{a H^\pm_p}
\left(\overline{ \partial_\mu M^2}_{\!\!\!s,{-1}} \right)^{a H^\pm_p}
}
{m_a^2 - m_{H^\pm}^2}
+
\frac{
\left(\overline{M^2}_{\!\!\!s,1} \right)^{a G^\pm_p}
\left(\overline{ \partial_\mu M^2}_{\!\!\!s,{-1}} \right)^{a G^\pm_p}
}
{m_a^2}
\bigg],\\
&\left(\overline{ \partial_\mu M^2}_{\!\!\!s} \right)^{H_p^\pm H_p^\pm} =
\left(\overline{ \partial_\mu M^2}_{\!\!\!s,0} \right)^{H_p^\pm H_p^\pm} +
2 \sum_{a=1}^{3}
\frac{
\left(\overline{M^2}_{\!\!\!s,1} \right)^{a H^\pm_p}
\left(\overline{ \partial_\mu M^2}_{\!\!\!s,{-1}} \right)^{a H^\pm_p}
}
{m_{H^\pm}^2 - m_a^2}\;.
\end{align}

We also have to express $\widehat{\partial_0 M_s^2}$, \eqref{part0Msq},
in the basis diagonalizing 
$\widehat{M}^2_{\!s,0}$, \eqref{diagMs}, that is,
\begin{equation} \label{bardel0Msq}
\overline{\partial_0 M_s^2} =
\bar{U} \widehat{\partial_0 M_s^2} \bar{U}^\trans
\end{equation}
with $\bar{U}$ given in~\eqref{eq:barU}.

We get the following explicit expressions
at the charge conserving minimum:
\begin{alignat}{2}
&\left(\overline{M^2_{s,1}}\right)^{G_0G_p^\pm}
&&=0, \nonumber\\
&\left(\overline{M^2_{s,1}}\right)^{G_0H_1^\pm}
&&=2 u K_0, \nonumber\\
&\left(\overline{M^2_{s,1}}\right)^{aG_p^\pm}
&&= - 2 u K_0 R_{ai} (R_H)_{pi}, \\
&\left(\overline{M^2_{s,1}}\right)^{aH_1^\pm}
&&= 0, \nonumber\\
&\left(\overline{M^2_{s,1}}\right)^{aH_2^\pm}
&&= 4 K_0 (\bar{\eta}_a + \eta_{00} \bar{k}_a)
- 2 u K_0 \bar{k}_a,\nonumber
\end{alignat}
and, using $R_H$ in equation \eqref{RH},
\begin{alignat}{2}
&\left(\overline{\partial_0M^2_{s,0}}\right)^{G_0G_0}
&&=2\left(\eta_{00}+\tvec{k}^\trans\tvec{\eta}\right), \nonumber\\
&\left(\overline{\partial_0M^2_{s,0}}\right)^{G_0a}
&&=2(\overline{\tvec{\eta}\times\tvec{k}})_a, \nonumber\\
&\left(\overline{\partial_0M^2_{s,0}}\right)^{G^\pm_pG^\pm_l}
&&=2\left(\eta_{00}+\tvec{k}^\trans\tvec{\eta}\right)\delta^{pl}, \nonumber \\
&\left(\overline{\partial_0M^2_{s,0}}\right)^{G^\pm_pH^\pm_1}
&&=2\left(R_H\right)_{pi}\left(\tvec{\eta}\times\tvec{k}\right)_i,\nonumber \\
&\left(\overline{\partial_0M^2_{s,0}}\right)^{H^\pm_1H^\pm_1}
&&=2\big[\eta_{00}-\tvec{k}^\trans\tvec{\eta}+2\left(R_H\right)_{3i}\left(\tvec{\eta}\times\tvec{k}\right)_i\big]=2\left(\eta_{00}-\tvec{k}^\trans\tvec{\eta}\right), \nonumber\\
&\left(\overline{\partial_0M^2_{s,0}}\right)^{G^\pm_pH^\pm_2}
&&=4\left(R_H\right)_{pi}\eta_i, \\
&\left(\overline{\partial_0M^2_{s,0}}\right)^{H^\pm_1H^\pm_2}
&&=4\big[\left(R_H\right)_{3i}\eta_i-\tvec{k}^\trans\tvec{\eta}\big]=0, \nonumber\\
&\left(\overline{\partial_0M^2_{s,0}}\right)^{H^\pm_2H^\pm_2}
&&=6\left(\eta_{00}-\tvec{k}^\trans\tvec{\eta}\right), \nonumber\\
&\left(\overline{\partial_0M^2_{s,0}}\right)^{ab}
&&=2\left(\eta_{00}-\tvec{k}^\trans\tvec{\eta}\right)\delta^{ab}+4\left(\bar{E}_{ab}+\bar{k}_a\bar{\eta}_b+\bar{\eta}_a\bar{k}_b\right), \nonumber
\end{alignat}
as well as
\begin{alignat}{2}
&\left(\overline{\partial_0M^2_{s,-1}}\right)^{aG^\pm_p}
&&=2\big[\bar{k}_a\left(R_H\right)_{pi}\eta_i+R_{ai}E_{ij}\left(R_H\right)_{pj}\big], \nonumber\\
&\left(\overline{\partial_0M^2_{s,-1}}\right)^{aH^\pm_1}
&&=2\big[\bar{k}_a\left(R_H\right)_{3i}\eta_i-\left(\tvec{k}^\trans\tvec{\eta}\right)\bar{k}_a+\bar{E}_{ai}R_{ij}\left(R_H\right)_{3j}-\bar{E}_{ai}\bar{k}_i\big]=0, \nonumber\\
&\left(\overline{\partial_0M^2_{s,-1}}\right)^{aH^\pm_2}
&&=2\big[\left(\eta_{00}-\tvec{k}^\trans\tvec{\eta}\right)\bar{k}_a+\bar{\eta}_a-\bar{E}_{ai}\bar{k}_i\big].\nonumber
\end{alignat}
We find eventually for the 
couplings in~\eqref{lambda0_app},
\begin{equation}
\label{lambda0app}
\begin{aligned}
    \bar{\lambda}_{0H^\pm H^\pm}
    &=8\left(\eta_{00}-\tvec{k}^\trans\tvec{\eta}\right)+\sum_{a=1}^3\frac{\bar{f}^a}{2K_0}\left(\frac{\bar{f}^a}{m_{H^\pm}^2-m_a^2}-\bar{k}_a\right)\, ,
    \\
    \bar{\lambda}_{0aa}&=-2\left(\eta_{00}+\tvec{k}^\trans\tvec{\eta}\right)+
    \left(1-\bar{k}_a^2\right)\left[4\eta_{00}-\frac{m_{H^\pm}^2}{2K_0}\left(3-\frac{m_{H^\pm}^2}{m_a^2}\right)\right]
   \\
   &\phantom{=} +\frac{m_a^2}{K_0}
   +\frac{\bar{f}^a}{2K_0}\left(\frac{\bar{f}^a}{m_a^2-m_{H^\pm}^2}+\bar{k}_a\right)\, .
\end{aligned}
\end{equation}

\bibliographystyle{JHEP}
\bibliography{references}

\end{document}